\documentclass[twocolumn,prb,showpacs,preprintnumbers,amsmath,amssymb,superscriptaddress]{revtex4-1}
\usepackage{graphicx}
\usepackage{setspace}

\usepackage{color}
\usepackage{ulem} 

\usepackage{amsmath,amsthm,amssymb}
\usepackage{mathrsfs}

\usepackage[
colorlinks=true, citecolor=blue, urlcolor=blue, linkcolor=blue,
setpagesize=false, bookmarks=false
]{hyperref}

\newcommand{\hc}{\hat{c}}

\newcommand{\ha}{\hat{a}}
\newcommand{\hb}{\hat{b}}

\newcommand{\hH}{\hat{H}}
\newcommand{\hn}{\hat{n}}
\newcommand{\hX}{\hat{X}}
\newcommand{\hP}{\hat{P}}
\newcommand{\hrho}{\hat{\rho}}
\newcommand{\brho}{\boldsymbol{\rho}}

\newcommand{\al}{\alpha}
\newcommand{\be}{\beta}

\newcommand{\bchi}{\boldsymbol{\chi}}
\newcommand{\bT}{{\bf T}}
\newcommand{\bsig}{\boldsymbol{\sigma}}

\begin{document}
\title{Collective Modes in Excitonic Insulators: \\Effects of Electron-Phonon Coupling and Signatures in Optical Response }

\author{Yuta Murakami}
\affiliation{Department of Physics, Tokyo Institute of Technology, Meguro, Tokyo 152-8551, Japan}
\author{Denis Gole\v{z}}
\affiliation{Center for Computational Quantum Physics, Flatiron Institute, New York, New York 10010, USA}
\author{Tatsuya Kaneko}
\affiliation{Department of Physics, Columbia University, New York, New York 10027, USA}
\author{Akihisa Koga}
\affiliation{Department of Physics, Tokyo Institute of Technology, Meguro, Tokyo 152-8551, Japan}
\author{Andrew J. Millis}
\affiliation{Center for Computational Quantum Physics, Flatiron Institute, New York, New York 10010, USA}
\affiliation{Department of Physics, Columbia University, New York, New York 10027, USA}
\author{Philipp Werner}
\affiliation{Department of Physics, University of Fribourg, 1700 Fribourg, Switzerland}
\date{\today}

\begin{abstract}
We consider a two-band spinless model describing an excitonic insulator (EI) on the two-dimensional square lattice with anisotropic hopping parameters and electron-phonon (el-ph) coupling, inspired by the EI candidate Ta$_2$NiSe$_5$.
We systematically study the nature of the collective excitations in the ordered phase which originates from the interband Coulomb interaction and the el-ph coupling. 
When the ordered phase is stabilized only by the Coulomb interaction~(pure EI phase), its collective response exhibits a massless phase mode in addition to the amplitude mode. 
We show that in the BEC regime, the signal of the amplitude mode becomes less prominent 
and that the anisotropy in the phase mode velocities is relaxed compared to the model bandstructure. 
Through coupling to the lattice, the phase mode acquires a mass and the signal of the amplitude mode becomes less prominent. Importantly, character of the softening mode at the boundary between the normal semiconductor phase and the ordered phase depends on the parameter condition. In particular, we point out that even for el-ph coupling smaller than the Coulomb interaction the mode that softens to zero at the boundary can have a phonon character. 
We also discuss how the collective modes can be observed in the optical conductivity.
Furthermore, we study the effects of nonlocal interactions on the collective modes and show the possibility of realizing a coexistence of an in-gap mode and an above-gap mode split off from the single amplitude mode in the system with the local interaction only.
\end{abstract}

\maketitle

\section{Introduction}
The spontaneous condensation of composite fermions triggers intriguing macroscopic phenomena and collective motions. 
One typical example is the superconducting phase, which has been studied for a long time. 
This phase is characterized by supercurrents and exhibits a characteristic collective mode, the Higgs amplitude mode.\cite{Anderson1958,Anderson1963,littlewood1982,Matsunaga2013,Matsunaga2014,Measson2014,Benfatto2014,pekker2015,Murakami2016,Katsumi2018,shimano2020,Schwarz2020}  A closely related example is the excitonic insulating (EI) phase.\cite{jerome1967,kohn1967,keldysh1968,halperin1968,halperin1968RMP} It was proposed more than 50 years ago that the weakly screened Coulomb interaction in semiconductors (semimetals) with small bandgaps (band overlaps) 
could lead to the spontaneous formation of stable electron-hole bound states below a critical temperature~$T_c$. This  excitonic ordered state can potentially show superfluid-like transport and  collective motions. 
Exciton condensation has been realized in semiconductor double-layer systems,\cite{butov2002,eisenstein2004,wang2019} where the spatial separation between the electrons and holes provides stability and enables a rather straightforward measurement of perfect Coulomb drag.\cite{nandi2012}

On the other hand, the EI phase in bulk materials has been elusive for decades. Recently, several 
transition metal chalcogenides~(TMCs) are attracting interest as EI candidates. \cite{cercellier2007,Monney2011PRL,Monney2012,Monney2015PRL,kogar2017,kaneko2018,chen2018} The low dimensionality of TMCs results in a weak screening of the Coulomb interaction and a large exciton binding energy. \cite{chernikov14,he2014,ugeda2014,mueller2018}
A remarkable property of these materials is that they may exhibit condensation at non-cryogenic temperatures. 
The potential to study intriguing collective phenomena originating from the macroscopic condensation at elevated temperatures makes TMCs a unique platform. 

A well-studied EI candidate is the quasi one-dimensional layered chalcogenide Ta$_2$NiSe$_5$ (TNS),\cite{sunshine1985,disalvo1986} which exhibits a small direct gap.\cite{wakisaka2009,kaneko2013,*kaneko2013e} It shows a phase transition at 328~K accompanied by a structural distortion.\cite{disalvo1986} Experimental indications of the existence of an EI phase in TNS have been obtained, for instance, using equilibrium and time-resolved angle-resolved photoemission spectroscopy (ARPES),\cite{seki2014,mor2017,okazaki2018} as well as anomalies in the temperature dependence of transport~\cite{lu2017,nakano2019} and phonon properties of the material.\cite{werdehausen2018,nakano2018,larkin2018}
Another indication is the BCS-BEC (semimetal-semiconductor) crossover driven by physical or chemical pressure.~\cite{lu2017}

A strict definition of the EI phase relies on the conservation of charges within each of the involved bands (continuous $U(1)$ symmetries). The breaking of the $U(1)$ symmetry leads to a massless Goldstone mode, which can bring interesting coherent phenomena such as super-transport.
However, the crucial issue regarding the EI phase in real materials is that the system may weakly break the $U(1)$ symmetry, either by coupling to phonons~\cite{kaneko2013,zenker2014,murakami2017} or by a direct hybridization term.~\cite{mazza2019,Watson2019} 
In particular, such electron-phonon (el-ph) coupling can cooperate with the Coulomb interaction to stabilize the ordered phase.
These terms breaking the continuous symmetry make the phase mode massive and suppress the potential super-transport from the exciton condensate.~\cite{zenker2014,kaneko2015,murakami2017}
Therefore, practically important questions are (i) whether the excitonic or el-ph mechanism is dominant in these materials, and (ii) how the properties of the ordered phase in realistic materials differ from the pure EI phase. 
In order to address these questions, it is useful to study the nature of the collective excitations.

In this work, we present a systematic analysis of the properties of the collective excitations for the ordered phase driven by the excitonic mechanism (interband Coulomb interaction) and the el-ph coupling, using a  two-band model with strongly anisotropic dispersion in two dimensions which is inspired by TNS. The collective modes are studied by evaluating linear response functions within the random-phase approximation (RPA) or equivalently solving the time-dependent mean-field (tdMF) theory under small perturbations. 

We start by considering collective excitations in the pure EI phase driven by the local Coulomb interaction. We demonstrate the existence of the massless phase mode and reveal the different properties of the amplitude modes in the BCS and BEC regimes [Fig.~\ref{fig:chi_pure_EI}]. While the band dispersion is strongly anisotropic, we will point out the anisotropy in terms of the velocity of the phase mode is suppressed in the BEC regime [Fig.~\ref{fig:velocity}]. 
Then we discuss the effects of the el-ph coupling on the collective modes. 
We show that the el-ph coupling cooperates with the Coulomb interaction to stabilize the ordered phase. Moreover, it makes the phase mode massive, and suppresses the peak structure in the response functions associated with the amplitude mode even in the BCS regime [Fig.~\ref{fig:chi11_type1_lam}]. 
We discuss the origin of this suppression and argue that it can be used as a measure of the relative contribution of the el-ph coupling and Coulomb interaction to the ordered phase.  Importantly, we identify the parameter regime, where even if the strength of the el-ph coupling is weaker than the Coulomb interaction, the character of the mode that softens to zero at the boundary between the ordered phase and the normal semiconductor phase can be phonon-like [Fig.~\ref{fig:chi11_type1_Utot}]. This poses a question on how to determine the dominant mechanism in this regime. 

We furthermore discuss the manifestation of the collective excitations in the optical conductivity under the assumption of a finite dipolar moment between the bands. We show that a massive phase mode can be observed in this case and  that it can serve as a direct experimental measure of the $U(1)$ symmetry breaking of the Hamiltonian [Fig.~\ref{fig:Chi_JJ}]. We also study the potential effects of the non-local interactions  and show that a new in-gap mode can emerge, which is reminiscent of the multiple bound states in the hydrogen atom [Fig.~\ref{fig:chi_q_w_nonlocU}]. 

The paper is organized as follows. In Sec.~\ref{sec:formulation}, we introduce our two-band model with local Coulomb interactions and el-ph coupling, the tdMF theory and the corresponding response functions.
In Sec.~\ref{sec:pure_EI}, we study the collective modes in the pure EI, while 
Sec.~\ref{sec:effect_elph} presents a systematic study of the effects of the el-ph coupling on the collective modes, and a discussion of their observability in the optical conductivity.  The effects of longer-range interactions are studied in 
Sec.~\ref{sec:nonlocU}.
The conclusions are presented in Sec.~\ref{sec:conclusion}.

\section{Formulation}\label{sec:formulation}
\subsection{Models}
In this paper, we consider a two-band spinless fermion model coupled to phonons on the two-dimensional square lattice with anisotropic hopping parameters,
\begin{align}
\hat{H}&=\hat{H}_{\rm kin}+\hat{H}_{\rm int}+\hat{H}_{\rm el-ph}+\hat{H}_{\rm ph} .\label{eq:Hamiltonian}
\end{align}

The first term includes the kinetic term and the band energies 
\begin{align}
\hat{H}_{\rm kin}&=-\sum_{\langle i,j\rangle ,a=0,1} J_a({\bf r}_{ij}) \hc^\dagger_{i,a} \hc_{j,a}+\sum_{i,a} D_a \hc^\dagger_{i,a}\hc_{i,a} \ ,
\end{align}
where, $\langle i,j\rangle$ indicates a pair of nearest-neighbor sites, and $a=0,1$ refers to the conduction band and valence band, respectively. 
$\hc^\dagger$ is the creation operator of electrons, $J_a({\bf r}_{ij})$ is the hopping parameter, ${\bf r}_{ij}={\bf r}_i - {\bf r}_j$ is the spatial vector connecting site $j$ to site $i$, and $D_a$ is the energy of band $a$.
The dispersion of the free electron is given by $\epsilon_a({\bf k}) \equiv -\sum_l J_a({\bf r}_l) e^{-i{\bf k}\cdot{\bf r}_l}$.
For simplicity, we first consider the case where the electrons interact with local interactions
\begin{align}
&\hH_{\rm int}=U\sum_i \hat{n}_{i,0}\hat{n}_{i,1},
\end{align}
where $U$ is the local interband Coulomb interaction and $\hat{n}_{i,a} = \hc^\dagger_{i,a}\hc_{i,a}$. Without el-ph coupling the system has $U(1)$ symmetries and the number of particles in the valence band and the conduction band are separately conserved. At low enough temperatures, the system can break the $U(1)$ symmetry (the symmetry about the relative phase between two bands), realizing an EI phase.\cite{ihle2008,seki2011,zenker2012} 
When the maximum (minimum) of the valence band (conduction band) in the normal state is at the Gamma point (${\bf k}={\bf 0}$), the system keeps the translational invariance in the EI phase, 
and we can introduce the order parameter of the phase as $\phi=\langle \hc^\dagger_{i,0} \hc_{i,1}\rangle$.
In the following, we consider this case, which is relevant for TNS.

We furthermore introduce an el-ph coupling of the form
\begin{equation}
\begin{split}
\hH_{\rm el-ph}&=g\sum_i (\hb_i^\dagger+\hb_i)(\hc^\dagger_{i,1}\hc_{i,0}+\hc^\dagger_{i,0}\hc_{i,1}),\\ 
\hH_{\rm ph}&=\omega_0\sum_i\hb_i^\dagger \hb_i,
\end{split}
\end{equation}
where $\omega_0$ is the phonon frequency, $g$ is the el-ph coupling constant, and $\hb^\dagger$ is the phonon creation operator.
This phonon reduces the $U(1)$ symmetries down to a $Z_2$ symmetry and cooperates with the Coulomb interaction to trigger the phase transition, where a real value of order parameter $\phi$ is favored~\cite{murakami2017}.  We will use $\lambda\equiv\frac{2g^2}{\omega_0}$ as a measure of the strength of the el-ph coupling. 

For the following analysis, we introduce the single-particle density matrix as 
\begin{align}
\hrho_{ia,jb}&\equiv \hc^\dagger_{jb}\hc_{ia},
\end{align}
and we use the symbol $\hat{{\brho}}$ when we regard the density matrix as a matrix with indices $(i,a)$ and $(j,b)$.
In the two-band model,  a useful parametrization of the local density matrix is
\begin{align}
\hrho_{\nu j} \equiv \hat{\Psi}_j^\dagger \boldsymbol{\sigma}_\nu \hat{\Psi}_j,
\end{align}
with $\nu=0,1,2,3$.
Here, $\hat{\Psi}^\dagger_j=[\hc^\dagger_{j,0},\hc^\dagger_{j,1}]$ and $\boldsymbol{\sigma}_\nu$ ($\nu=1,2,3$) is a Pauli matrix, while $\boldsymbol{\sigma}_0$ is the identity matrix. 
We also introduce the coordinate $X$ and momentum $P$ of the phonons as 
\begin{align}
& \hX_i=\hb_i+\hb^\dagger_i, \;\; \hP_i=\frac{1}{i}(\hb_i-\hb^\dagger_i),
\end{align}
where their commutator is $[\hX_i,\hP_i]=2i$.
In the following, we express all operators with a hat (e.g. $\hrho_{\nu j}$) and the time-dependent expectation values
without a hat (e.g.~$\rho_{\nu j}$).
\subsection{Mean-field theory}
In this section we introduce the time-dependent mean-field theory (tdMF) for the model \eqref{eq:Hamiltonian}.
We consider the situation where the system is initially in equilibrium and it is excited by  $\hH_{\rm ex}(t)$. 
Within the tdMF theory, the time evolution of the system is described by the MF Hamiltonian, 
\begin{align}
&\hH^{\rm MF}_{\rm tot}[\brho,X](t) = \hH_{\rm el}^{\rm MF}[\brho,X](t) + \hH_{\rm ph}^{\rm MF}[\brho](t) + \hH_{\rm ex}(t).
\end{align}
Here, we use the arguments $[\brho,X]$ and $[\brho]$ to explicitly show the dependency of the MF Hamiltonian on 
the expectation value of the density matrix and the phonon displacements.
 $\hH_{\rm el}^{\rm MF}[\brho,X](t)$ and $\hH_{\rm ph}^{\rm MF}[\brho](t)$ are obtained by decoupling
 the el-el interaction term and the el-ph coupling in the original Hamiltonian as 
\begin{align}
&\hH^{\rm MF}_{\rm el}[\brho,X](t) = \hH_{\rm kin} + \hH^{\rm H}(t) + \hH^{\rm F}(t) + \hH^{\rm MF,el}_{\rm el-ph}(t)  ,\nonumber\\
&\hH^{\rm MF}_{\rm ph}[\brho](t) =\omega_0\sum_i\hb^\dagger_i \hb_i + \sum_{l,\nu} g_{\nu}\rho_{\nu l}(t) \hX_l, \label{eq:MF}
\end{align}
where $g_\nu=\delta_{1,\nu}g$.
The MF terms in $\hH^{\rm MF}_{\rm el}$ can be expressed as 
\begin{equation}
\begin{split}
\hH^{\rm H}[\brho](t) & =U\sum_{l,a}\rho_{l\bar{a},l\bar{a}}(t) \hn_{la}, \\
\hH^{\rm F}[\brho](t)& = -U\sum_{l,a}\rho_{l\bar{a},la}(t) \hc^\dagger_{l\bar{a}}\hc_{la},\\
\hH^{\rm MF,el}_{\rm el-ph}[X](t) & = \sum_{l,\nu} g_\nu X_l(t)\hrho_{\nu l}. 
\end{split}
\end{equation}
Here, $\hH^{\rm H}$ and $\hH^{\rm F}$ correspond to the Hartree and the Fock contributions, respectively.
We also note that $\hH^{\rm H}(t) +\hH^{\rm F}(t) = \sum_{l,\nu} U_\nu \rho_{\nu l}(t) \hrho_{\nu l} $ with $[U_0,U_1,U_2,U_3]\equiv [\frac{U}{2},-\frac{U}{2},-\frac{U}{2},-\frac{U}{2}]$.

The equilibrium state is determined as follows.
The equilibrium MF Hamiltonians become
\begin{subequations}
\begin{align}
\hH_{\rm el}^{\rm MF}&=\frac{1}{2}\sum_{\bf k} \hat{\Psi}_{\bf k}^\dagger
\begin{bmatrix}
C_{\bf k}+B^z_{\bf k} & B^x_{\bf k}-iB^y_{\bf k}\\
 B^x_{\bf k}+iB^y_{\bf k} & C_{\bf k}-B^z_{\bf k}
\end{bmatrix}
 \hat{\Psi}_{\bf k},\label{eq:MF_ele}\\
\hH^{\rm MF}_{\rm ph} &=\omega_0\sum_i\hat{b}^\dagger_i\hat{b}_i+g(\phi+\phi^*)\sum_i  \hat{X}_i,\label{eq:H_mf_ph_eq}
\end{align}
\end{subequations}
with $\hat{\Psi}^\dagger_{\bf k}=\frac{1}{\sqrt{N}}\sum_{\bf k} e^{i{\bf k}\cdot {\bf r}_i} \hat{\Psi}^\dagger_i$ and 
\begin{align}
C_{\bf k}&=(\epsilon_{0}({\bf k})+\epsilon_{1}({\bf k}))+(D_0+D_1)+U n_{\rm tot}.
\end{align}
Here, $n_{\rm tot} \equiv n_0+ n_1 $.
From Eq.~\eqref{eq:H_mf_ph_eq}, we obtain the expectation values
\begin{align}
P_i&=0,\,\,\,\,\,X_i=-\frac{4g}{\omega_0}{\rm Re}\phi.
\end{align}
Using these values, 
$B^{(x,y,z)}_{\bf k}$ in Eq.~\eqref{eq:MF_ele} is expressed as
\begin{subequations} \label{eq:MF_self}
\begin{align}
B_{\bf k}^x&=-2(U+2\lambda){\rm Re} \phi,\label{eq:MF_self_a}\\
B_{\bf k}^y&=-2U{\rm Im} \phi,\\
B_{\bf k}^z&=(\epsilon_{0}({\bf k})-\epsilon_{1}({\bf k})) + D_{01}-U \Delta n, \label{eq:MF_self_c}
\end{align}
\end{subequations}
with $D_{01}\equiv D_0-D_1$ and $\Delta n \equiv n_0-n_1$. 
The eigenvalues of $\hH_{\rm el}^{\rm MF}$ are given by $E_{\pm} ({\bf k}) = \frac{1}{2}(\pm B_{\bf k}+ C_{\bf k})$ with
$B_{\bf k}= \sqrt{( B_{\bf k}^{x})^2+( B_{\bf k}^{y})^2+(B_{\bf k}^{z})^2}$.
This provides the single-particle dispersion in the MF theory.
Thus, the MF self-consistency relation becomes 
\begin{equation} \label{eq:MF2}
\begin{split}
&\phi=\frac{1}{N}\sum_{\bf k}\frac{B_{\bf k}^x+i B_{\bf k}^y}{2 B_{\bf k}}
[f(E_{+}({\bf k}),T)-f(E_{-}({\bf k}),T)],\\
&\Delta n=\frac{1}{N}\sum_{\bf k} \frac{B_{\bf k}^z}{B_{\bf k}}
[f(E_{+}({\bf k}),T)-f(E_{-}({\bf k}),T)],\\
&n_{\rm tot}=\frac{1}{N}\sum_{{\bf k}} [f(E_{+}({\bf k}),T)+f(E_{-}({\bf k}),T)],
\end{split}
\end{equation}
and we solve these equations to obtain the equilibrium MF values, $\phi$, $\Delta n$ and $n_{\rm tot}$.
Here, $T$ is the temperature and $f(\epsilon,T)$ is the corresponding Fermi distribution function.
The phase of $\phi$ can be arbitrary when $\lambda=0$, but $\phi\in \mathbb{R}$ is favored for nonzero $\lambda$.
Therefore, we assume that $\phi\in \mathbb{R}$ in the following.
Note that Equation~\eqref{eq:MF_self_a} tells that $U+2\lambda$ can be regarded as the effective interaction that drives the system into the ordered phase.
Thus, one measure of the relative contribution of the el-ph coupling and the Coulomb interaction to the order phase is the relative strength of $U$ and $2\lambda$.

When we discuss the effects of the el-ph coupling on the collective modes, we should compare the systems with the same single-particle properties.
From Eqs.~\eqref{eq:MF_self} and \eqref{eq:MF2}, if two sets of parameters $(D_{01},U,\lambda,D_{0}+D_1)$ and $(D'_{01},U',\lambda',D'_0+D'_1)$ satisfy
\begin{equation} \label{eq:condition}
\begin{split}
U+2\lambda & =U'+2\lambda',\\
D_{01}-U \Delta n &= D'_{01}-U' \Delta n, \\
(D_0+D_1)+U n_{\rm tot} & = (D'_0+D'_1)+U' n_{\rm tot}, 
\end{split} 
\end{equation}
these sets yield the same single-particle dispersions ($E_{\pm} ({\bf k})$) and MF parameters, $\phi$, $\Delta n$ and $n_{\rm tot}$. In other words, for a given reference set of parameters $(D_{01,{\rm ref}},U_{\rm ref},\lambda=0,D_{0,{\rm ref}}+D_{1,{\rm ref}})$ and arbitrary choice of $\lambda'$, one can always find $(D'_{01},U',D'_0+D'_1)$ such that $B_{\bf k}$, $C_{\bf k}$ and $E_{\pm} ({\bf k})$ are identical for these two sets. We will use this to discuss the effects of the el-ph coupling on the collective modes.
Note that this means that the single-particle spectra in the MF level is irrelevant to the origin of the ordered phase.  

For the evaluation of the linear response functions in the next section, we introduce the single-particle Green's function in equilibrium. The lesser and greater components of the Green's functions are defined as $G^<_{a,b}(t,t';{\bf k})\equiv i\langle \hat{c}^\dagger_{{\bf k},b}(t') \hat{c}_{{\bf k},a}(t)\rangle$ and $G^>_{a,b}(t,t';{\bf k})\equiv -i\langle \hat{c}_{{\bf k},a}(t) \hat{c}^\dagger_{{\bf k},b}(t')\rangle$ and we can regard them as $2\times2$ matrices in the band index space.
Assuming an equilibrium condition and a real order parameter, they can be expressed as 
\begin{equation}\label{eq:G_eq}
\begin{split}
{\bf G}^{<}(t;{\bf k}) &=i \sum_{\alpha=\pm}  f(E_{\alpha}({\bf k}),T)e^{-iE_{\alpha}({\bf k})t}{\bf W}_{\alpha}({\bf k}),\\
{\bf G}^{>}(t;{\bf k})&= -i \sum_{\alpha=\pm} f(-E_{\alpha}({\bf k}),T)e^{-iE_{\alpha}({\bf k})t}{\bf W}_{\alpha}({\bf k}),
\end{split}
\end{equation}
where 
\begin{align}
{\bf W}_{\pm}({\bf k})&=\frac{1}{2}\Big[\pm\frac{B^x_{\bf k}}{B_{\bf k}}\boldsymbol{\sigma}_1\pm\frac{B^z_{\bf k}}{B_{\bf k}}\boldsymbol{\sigma}_3+\boldsymbol{\sigma}_0\Big]. 
\end{align}

As for the time evolution, in general, one can write down the equations of motion of $\brho(t)$, $X(t)$ and $P(t)$
with respect to $\hH^{\rm MF}_{\rm tot}[\brho,X]$.~\cite{murakami2017,Murakami2020PRB} 
However, in the linear response regime, we do not need to explicitly compute the time evolution. 
We can derive the expression of the linear response function
consistent with the tdMF by regarding the deviation of the MF terms \eqref{eq:MF} from the equilibrium MF Hamiltonian $\hH^{\rm MF}_{\rm eq}$ as an additional external field [see the next section and  Appendix.~\ref{sec:general_rpa}].

\subsection{Linear response functions}
\begin{figure}[t]
     \centering
\includegraphics[width=50mm]{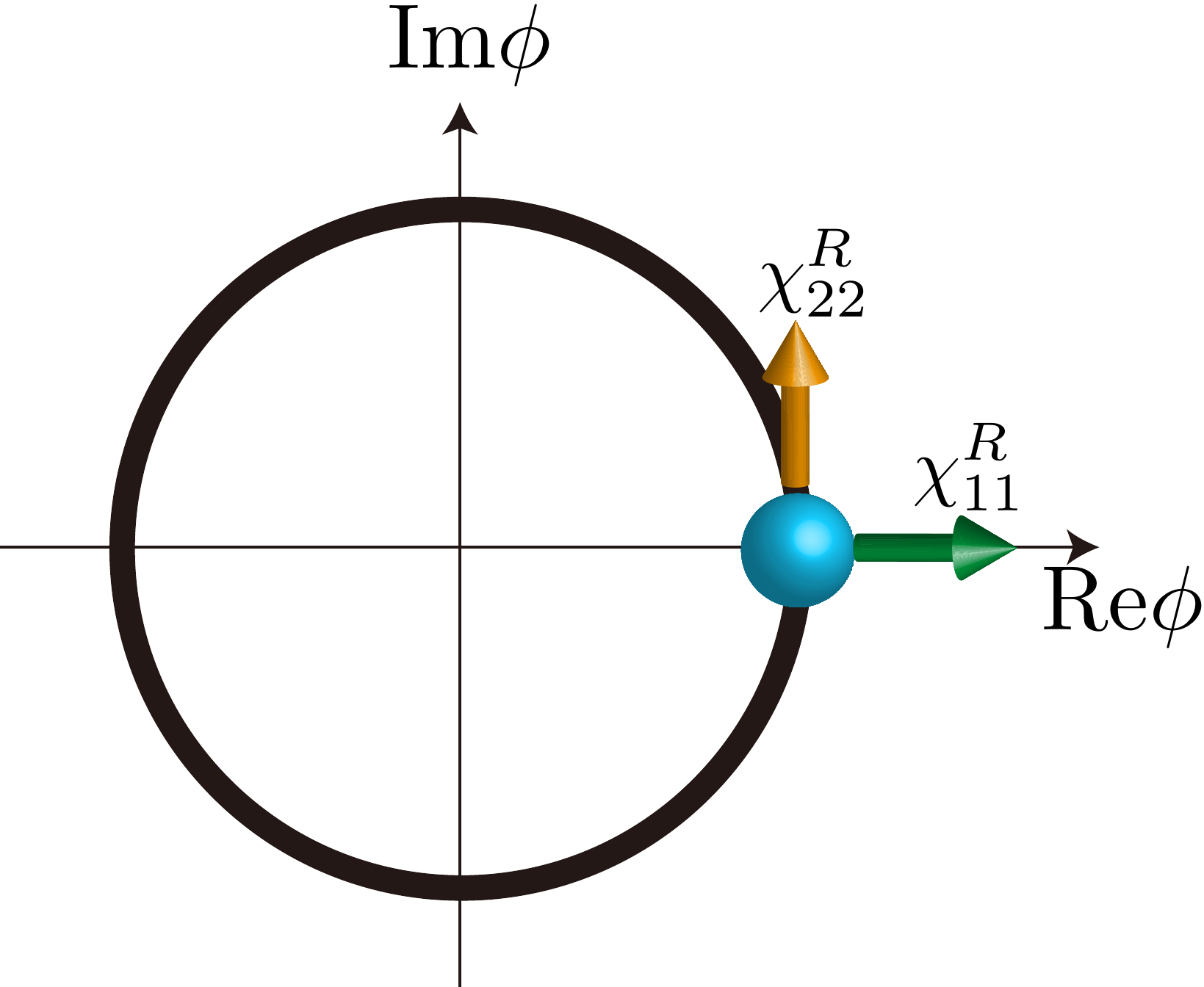}
\caption{Schematic picture of the dynamics of the order parameter corresponding to $\chi^R_{11}$ and $\chi^R_{22}$ at ${\bf q=0}$ in the complex plane of the order parameter $\phi$. }
\label{fig:chi_schematic}
\end{figure}

In this section, we present the expressions for the linear response functions evaluated by the tdMF theory introduced above.
In particular, we are interested in the linear response functions
\begin{align} \label{eq:chi_munu}
\chi_{\mu\nu}^R(t-t';{\bf r}_{i j})\equiv -i\theta(t-t') \langle [\hrho_{\mu i}(t),\hrho_{\nu j}(t')]\rangle.
\end{align}
We assume translational invariance in space and time as the system is in equilibrium in the absence of the external perturbation. 
The response function \eqref{eq:chi_munu} is obtained by exciting the system with a small perturbation $\hH_{\rm ex}(t) =   \delta F^{\rm el}_{\rm ex,\nu j} (t) \hrho_{\nu j}$ and measuring the evolution of $\hrho_{\mu i}$.
Since we take the order parameter $\phi$ to be real, $\chi^R_{11}$ corresponds to the dynamics in the amplitude direction, while $\chi^R_{22}$ corresponds to the dynamics in the phase direction [see Fig.~\ref{fig:chi_schematic}]. 
We can also introduce the Fourier transformation of $\chi$ as $\chi^R(\omega;{\bf q})\equiv \sum_l \int dt \chi^R(t;{\bf r}_l) e^{i\omega t-i{\bf q}\cdot{\bf r}_l}$.

Considering the time evolution induced by a weak excitation within the tdMF and focusing on the linear components (see Appendix~\ref{sec:general_rpa} for details),
we obtain 
\begin{equation}\label{eq:RPA_chi_Uloc}
\begin{split}
\bchi^R(\omega;{\bf q})&=\bchi^R_{0}(\omega;{\bf q})+\bchi^R_{0}(\omega;{\bf q}) \boldsymbol{\Theta}(\omega;{\bf q}) \bchi^R(\omega;{\bf q}) ,
\end{split}
\end{equation}
with $\boldsymbol{\Theta}=\boldsymbol{\Theta}^{\rm el} + \boldsymbol{\Theta}^{\rm ph}$ and 
\begin{equation}\label{eq:theta}
\begin{split} 
\boldsymbol{\Theta}^{\rm el}(\omega;{\bf q})&= {\rm Diag} [\frac{U}{2}, -\frac{U}{2}, -\frac{U}{2},-\frac{U}{2}], \\
\boldsymbol{\Theta}^{\rm ph}(\omega;{\bf q})&= {\rm Diag} [0, g^2 D^R_0(\omega), 0,0].
\end{split}
\end{equation}
Here, $\bchi^R_{0}$ is the response evaluated by keeping the MF Hamiltonian the same as in equilibrium (without updating the mean fields in the time evolution), 
and ${\rm Diag}[]$ indicates a diagonal matrix.
In terms of the Feynman diagrams, $\bchi^R_{0}$ corresponds to a bubble diagram ($\chi^R_{0}(t)=\theta(t)[G^<(-t)G^>(t)-G^>(-t)G^<(t)]$)), whose specific expression is 
\begin{align} \label{eq:chi0}
\chi_{0,\mu\nu}^R(\omega;{\bf q})&=\frac{1}{N}\sum_{\bf k} \Bigg\{\sum_{\alpha,\beta=\pm}{\rm Tr}[{\bf W}_\alpha({\bf k-q})\bsig_{\mu} {\bf W}_\beta({\bf k})\bsig_{\nu}]\nonumber \\
&\;\;\;\times\frac{f(E_\alpha({\bf k-q}),T)-f(E_\beta({\bf k}),T)}{\omega+ 0^+-(E_\beta({\bf k})-E_\alpha({\bf k-q}))}\Bigg\}.
\end{align}
Here, $0^+$ is an infinitesimally small positive value.
$D^R_0(\omega)$ is the free phonon Green's functions,
\begin{align}
D_0^R(\omega) & =\frac{2\omega_0}{(\omega+i0^+)^2-\omega_0^2}.
\end{align}
In terms of Feynman diagrams, the expression of the linear response function Eq.~\eqref{eq:RPA_chi_Uloc} consists of the ring diagrams and the ladder diagrams for the Coulomb interaction and the ring diagrams for the el-ph coupling (the random phase approximation (RPA)), cf. Ref.~\onlinecite{zenker2014}.
The detailed derivations of the susceptibilities are presented in a more general form in Appendix~\ref{sec:general_rpa} [see also Ref.~\onlinecite{Murakami2020PRB} for the case of normal states].

\subsection{Optical conductivity}\label{sec:opt_response}
In this section, we introduce the expression for the optical conductivity and explain how it is related to the response functions of the order parameters discussed above.
The optical response represents one of the most versatile experimental probes for equilibrium~\cite{basov2011} and nonequilibrium~\cite{giannetti2016} materials properties. 
For instance, detailed insights into the collective response of superconductors have been obtained from the non-linear terahertz optical response.~\cite{Matsunaga2014,Tsuji2015,Cea2016,Tsuji2016,Murotani2019,shimano2020}
In order to study the optical conductivity (linear response) of our two band model, we need to define how the system is coupled to the external field. 
Here, we consider a minimal gauge-invariant description in the restricted two-band space, which includes the interband acceleration~(Peierls term) and dipolar excitation,~\cite{golez2019multiband}
\begin{align} \label{eq:ham_optic}
\hH(t) =&-\sum_{\langle i,j\rangle, a= 0,1} J_a ({\bf r}_{ij},t)\hc^\dagger_{ia} \hc_{ja}-{\bf E}(t)\cdot \hat{\bf P}.
\end{align}
Here, ${\bf E}(t)$ is the electric field, $J_a ({\bf r}_{ij},t) = J_a ({\bf r}_{ij}) \exp[iq{\bf r}_{ij}\cdot {\bf A}(t)]$, ${\bf A}(t)$ is the vector potential with ${\bf E}(t) = -\partial_t {\bf A}(t)$, and $q$ is the electron charge.
The polarization operator $\hat{\bf P}$ is defined as $\hat{\bf P}  = \sum_{i,a} {\bf d}_a \hat{c}^\dagger_{ia}\hc_{i\bar{a}}$,
where ${\bf d}_a$ is a dipole matrix and, for simplicity, we assume that it is local~(momentum independent). 
Then, the current of the system with the coupling \eqref{eq:ham_optic} to the field consists of the intraband current (${\bf J}_{\rm intra}$) and the interband current (${\bf J}_{\rm inter}$),
\begin{equation}
\begin{split}
{\bf J}_{\rm intra}(t) &=i q \sum_{\langle i,j\rangle,a} {\bf r}_{ij} J_a({\bf r}_{ij}, t) \rho_{ja,ia}(t),\\
{\bf J}_{\rm inter}(t) &=  \partial_t \langle \hat{\bf P}(t) \rangle =   \partial_t   \sum_{i,a} {\bf d}_{a} \rho_{i\bar{a},ia}(t).
\end{split}
\end{equation}
In particular, we note that, in the linear response regime, the intraband current consists of the paramagnetic current and the diamagnetic current, which is proportional to the vector potential, and the operator for the paramagnetic current is $\hat{{\bf J}} \equiv iq \sum_{\langle i,j\rangle, a} {\bf r}_{ij}J_a ({\bf r}_{ij})\hc^\dagger_{ia} \hc_{ja}$.

By considering the linear response of these currents to the applied electromagnetic fields ${\bf A}$ and ${\bf E}$,
we obtain the expression of the optical conductivity;
\begin{align} 
{\boldsymbol \sigma}(\omega) =
\Bigg[i\omega {\boldsymbol \chi}_{PP}^R(\omega) 
-{\boldsymbol \chi}_{JP}^R(\omega)+{\boldsymbol \chi}_{PJ}^R(\omega) 
+\frac{-{\boldsymbol \chi}_{JJ}^R(\omega)+{\bf C}}{i\omega}
\Bigg] \label{eq:opt_cond}
\end{align}
with 
\begin{equation}\label{eq:Chi_opt}
\begin{split}
[{\boldsymbol \chi}_{PP}^R(t)]_{\alpha\beta} &= -i \theta(t)\langle [\hat{P}_\alpha(t),\hat{P}_\beta(0)] \rangle, \\
[{\boldsymbol \chi}_{JP}^R(t)]_{\alpha\beta} &= -i \theta(t)\langle [\hat{J}_\alpha(t),\hat{P}_\beta(0)] \rangle, \\
[{\boldsymbol \chi}_{PJ}^R(t)]_{\alpha\beta} &= -i \theta(t)\langle [\hat{P}_\alpha(t),\hat{J}_\beta(0)] \rangle, \\
[{\boldsymbol \chi}_{JJ}^R(t)]_{\alpha\beta} &= -i \theta(t)\langle [\hat{J}_\alpha(t),\hat{J}_\beta(0)] \rangle, 
\end{split}
\end{equation}
and the diamagnetic term 
\begin{align}
C_{\alpha\beta} =-q^2\sum_{\langle i,j\rangle,a} ({\bf r}_{ij})_\alpha ({\bf r}_{ij})_\beta J_a({\bf r}_{ij})  \rho_{ja,ia}.
\end{align}
Here, $\alpha,\beta=x,y$ indicate the spatial directions.
These quantities can be evaluated in such a way that they are consistent with the 
tdMF theory [see Appendix~\ref{sec:general_rpa} and \ref{sec:opt_cond2}].

With this simple coupling to the external field, one can make several exact statements on the optical conductivity without explicit calculations. First, ${\boldsymbol \chi}_{PP}^R$ is up to a constant the same as $\bchi^R(\omega;{\bf q=0})$ discussed in the previous section. For example, if ${\bf d}_a = d{\bf e}_x$ with $d\in{\bf R}$ and ${\bf e}_x$ is the unit vector in the $x$ direction, $[{\boldsymbol \chi}_{PP}^R]_{xx} = d^2 \chi^R_{11}(\omega;{\bf q=0})$. Second, ${\boldsymbol \chi}_{JP}^R$ and ${\boldsymbol \chi}_{PJ}^R$ turn out to be zero, because of the inversion symmetry (${\bf r}_i \leftrightarrow -{\bf r}_i$) of the Hamiltonian \eqref{eq:Hamiltonian}. Finally, the vertex correction to ${\boldsymbol \chi}_{JJ}^R$ is zero due to the inversion symmetry and the local self-energy of the MF theory of  the Hamiltonian \eqref{eq:Hamiltonian}, as in the dynamical mean-field theory.~\cite{georges1996} Therefore, ${\boldsymbol \chi}_{JJ}^R$ can simply be expressed as a bubble diagram with MF Green's functions, and it is fully determined by the single-particle properties. 
Therefore, it does not provide any information on whether the EI state originates from the Coulomb interaction or the el-ph coupling.

\section{Results}
\begin{figure}[t]
     \centering
\includegraphics[width=60mm]{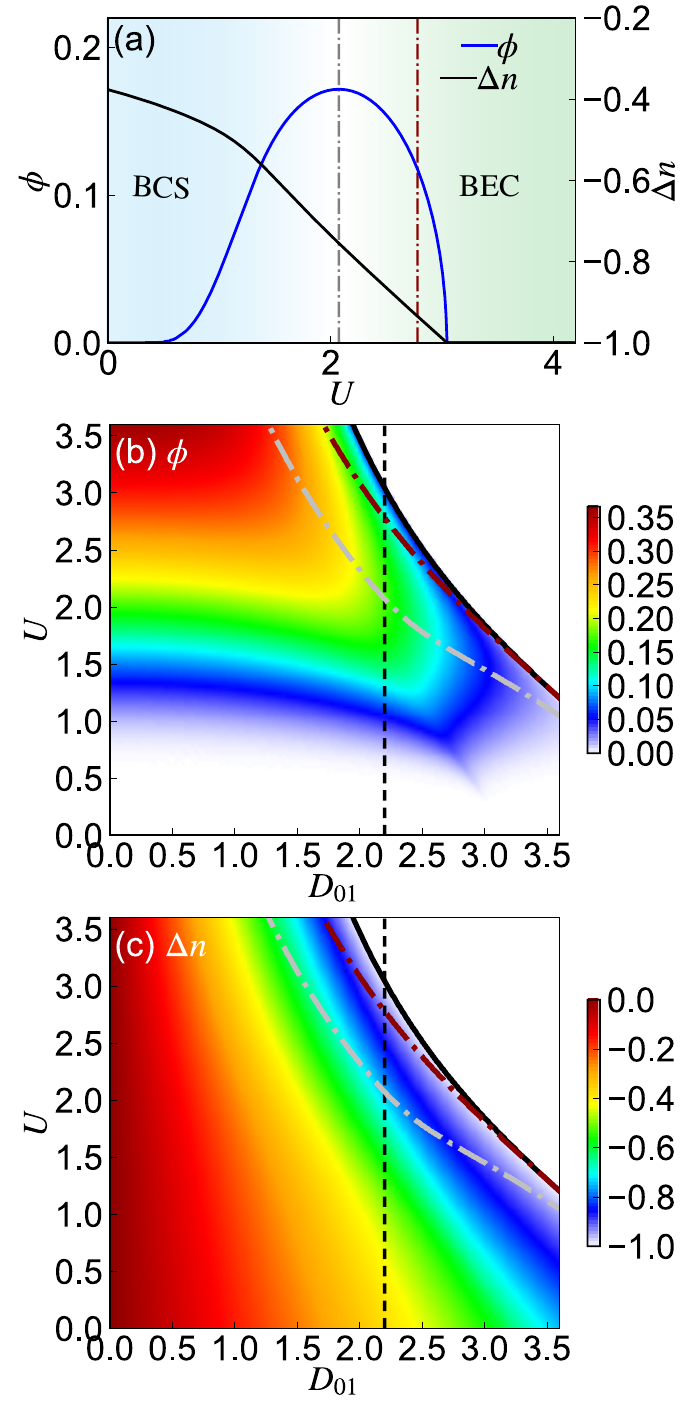}
\caption{(a) The order parameter $\phi$ $(\in \mathbb{R})$ and  the difference of the band occupation $\Delta n$ $(=n_0-n_1)$ as a function of the local interaction $U$ for a fixed energy difference between the bands $D_{01}=2.2$. 
The red dash-dotted line indicates the boundary between the $B_{\bf 0}^z<0$ and $B_{\bf 0}^z>0$ regimes ($U_{\rm SS}$), and the gray dash-dotted line indicates the maximum of $\phi$ for each $D_{01}$ ($U_{\rm BB}$).
Panels (b)(c) show $\phi$ and $\Delta n$ in the plane of  $D_{01}$ and $U$, respectively. 
The black solid line indicates the boundary of the EI phase ($U_c$), the gray dot-dashed line indicates $U_{\rm BB}$,
the red dot-dashed line indicates $U_{\rm SS}$ (the $B_{\bf 0}^z<0$ regime is below and the $B_{\bf 0}^z>0$ regime is above the line). 
The vertical dashed line is $D_{01}=2.2$.
Here, the system is half filled, and we use $J_{0}({\bf a}_x)=-J_{1}({\bf a}_x)=1.0$, $J_{0}({\bf a}_y)=-J_{1}({\bf a}_y)=0.2$ and $\lambda=0$. 
}
\label{fig:phase}
\end{figure}
\begin{figure*}[t]
     \centering
\includegraphics[width=170mm]{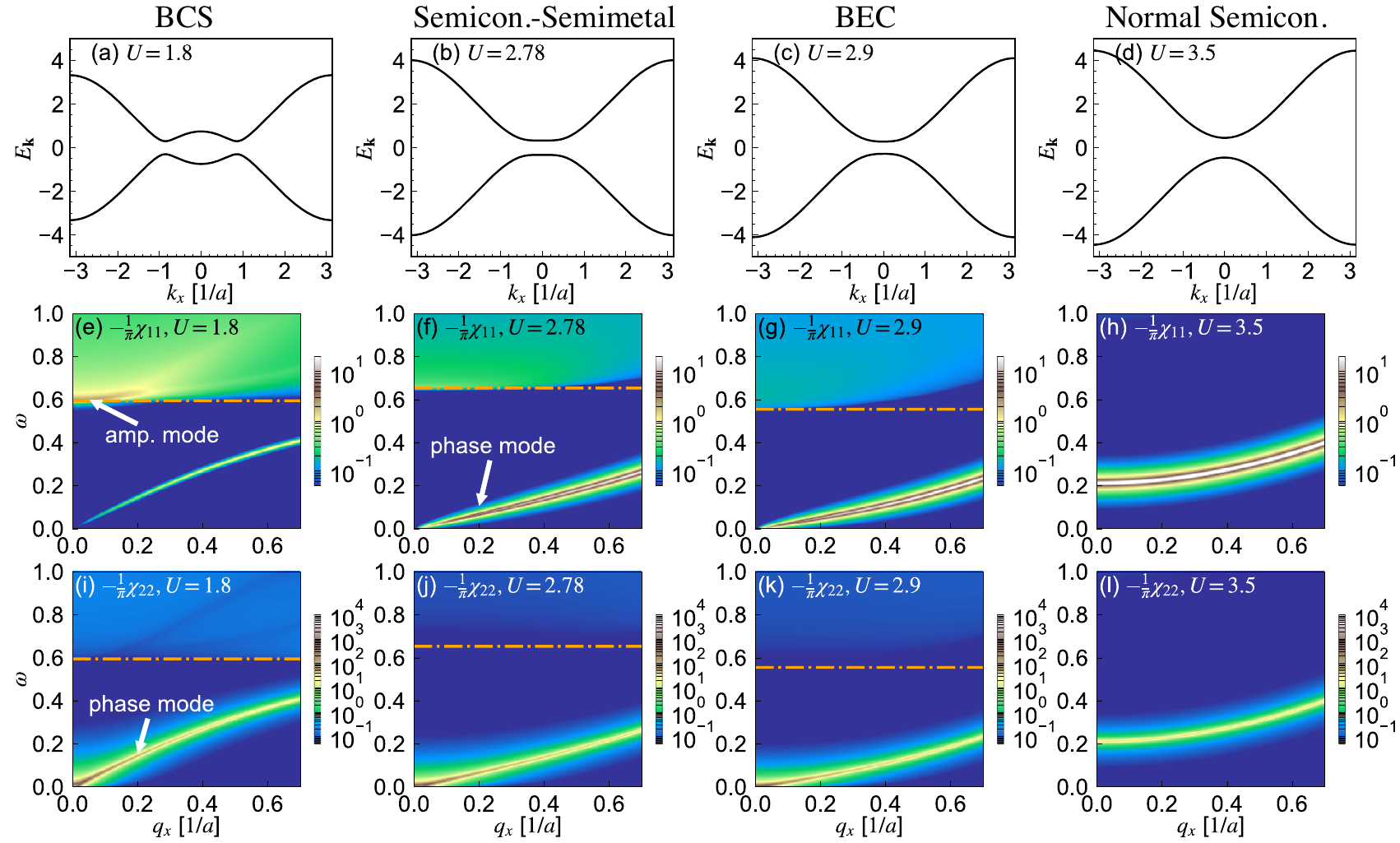}
\caption{(a-d) The MF single-particle spectra along $k_x$ ($k_y=0$). (e-l) Linear response functions for the order parameter in the amplitude direction $-\frac{1}{\pi}{\rm Im}\chi^R_{11}(\omega;q_x,q_y=0)$  (e-h) and in the phase direction  $-\frac{1}{\pi}{\rm Im}\chi^R_{22}(\omega;q_x,q_y=0)$~(i-l) for different values of $U$. The first column corresponds to the BCS regime~($U=1.8)$, the second column to the semiconductor-semimetal crossover~($U=U_{\rm SS}=2.78$), the third column to the BEC regime~($U=2.9$), and the last column to the normal semiconductor~($U=3.5$). 
All figures are for purely electronic systems and the band-level difference is $D_{01}=2.2$. The orange dash-dotted lines indicate the bandgap energies ($E_{\rm gap}$). 
} 
\label{fig:chi_pure_EI}
\end{figure*}

We investigate a two-dimensional system with strongly anisotropic hopping parameters: $J_{0}({\bf a}_x)=-J_{1}({\bf a}_x)=1.0$ and $J_{0}({\bf a}_y)=-J_{1}({\bf a}_y)=0.2$. Here, ${\bf a}_\alpha$ is the lattice vector along the $\alpha$ direction, whose length $a_\alpha$  is set to $1$ for both directions. This model is inspired by TNS, which is a material composed of weakly coupled chains. 
Here, the $x$ direction corresponds to the chain direction and the $y$ direction corresponds to the perpendicular direction of the chain.
We focus on $T=0$ and half-filling $(n_0+n_1=1)$ and fix the phonon frequency to $\omega_0=0.1$.
For the numerical evaluation of $\chi^R_0$ \eqref{eq:chi0}, we use $0^+ = 0.005$ in this work.

\begin{figure}[t]
\centering
\includegraphics[width=60mm]{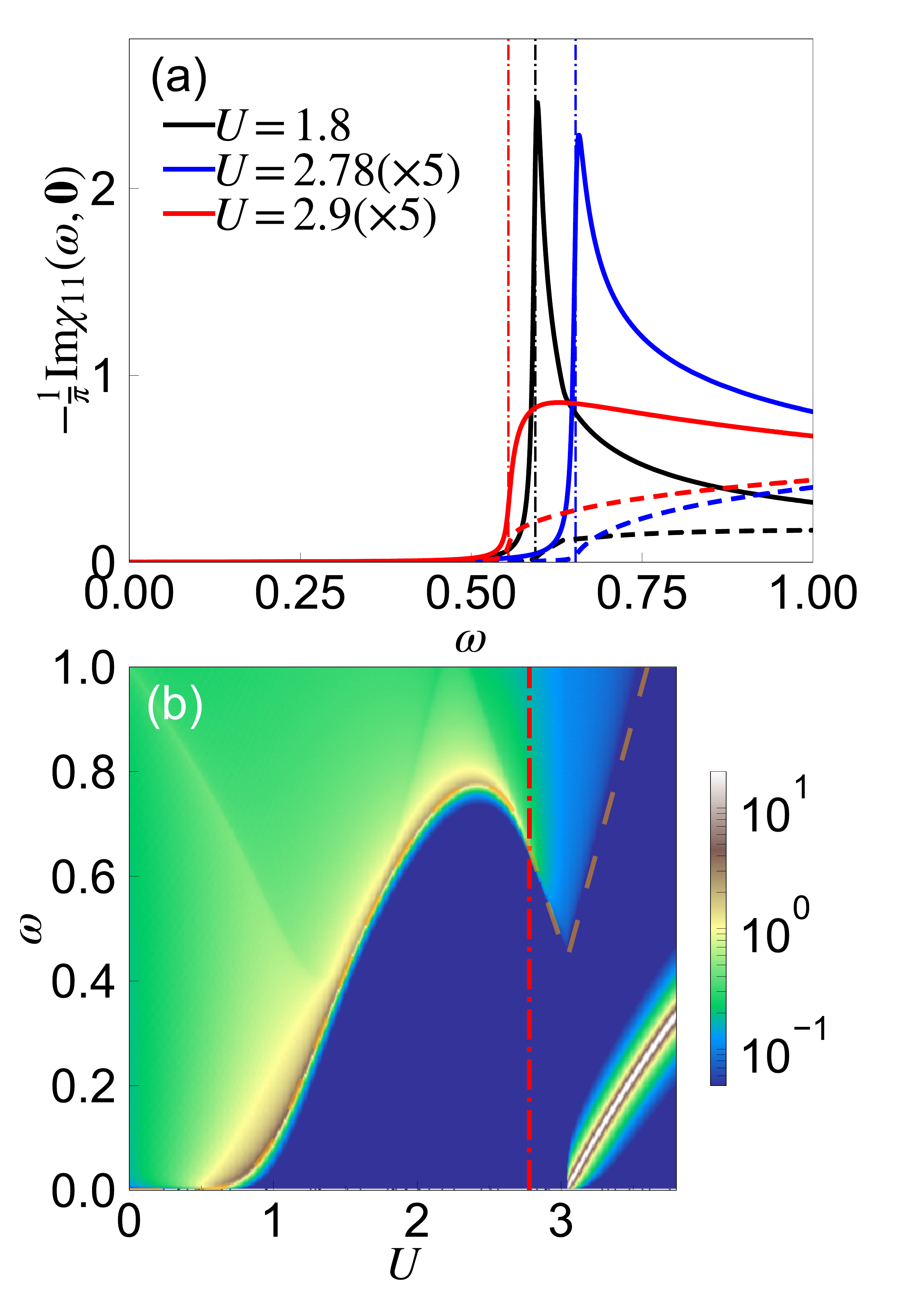}
\caption{Linear response functions for the order parameter in the amplitude direction for the long-wavelength limit. 
(a) Specific examples for the linear response function $-\frac{1}{\pi}{\rm Im}\chi^R_{11}(\omega;{\bf q=0})$ (solid lines) and the corresponding single-particle bubble diagram $-\frac{1}{\pi}{\rm Im}\chi^R_{0,11}(\omega;{\bf q=0})$ (dashed lines) for the specified $U$ values at fixed band-level difference $D_{01}=2.2$. The vertical dot-dashed lines indicate $E_{\rm gap}$ for each $U$. 
(b) $-\frac{1}{\pi}{\rm Im}\chi^R_{11}(\omega;{\bf q=0})$ in the plane of $\omega$ and $U$ at $D_{01}=2.2$. The dot-dashed red line indicates $U=U_{\rm SS}$, and the dashed orange line indicates the bandgap $E_{\rm gap}$.
}
\label{fig:chi_phi}
\end{figure}
\subsection{Pure Excitonic Insulator} \label{sec:pure_EI}
We first revisit the properties of the collective excitations in the pure EI phase ($\lambda=0$).
It is known that a BCS-BEC crossover occurs in the EI phase.\cite{zenker2012} 
In this paper, we define the BCS-BEC crossover point as the maximum of the order parameter for each $D_{01}$, and we denote it by $U_{\rm BB}$.
\footnote{When the system is in the low dimensions and has strong interactions, it can show a gap opening due to the fluctuation above $T_c$\cite{sugimoto2018} even when the systems is in the semi-metallic parameter regime. Thus, strictly speaking, one cannot simply categorize such cases to the BCS type or the BEC type. Therefore, this terminology is meaningful when the interaction is not too large. Indeed, we focus on such cases here.}
In addition, we will demonstrate that the value of $B_{\bf 0}^z$ in Eq.~\eqref{eq:MF_ele} defines the character of the amplitude mode.
For $B_{\bf 0}^z<0$ the disordered system (neglecting the off-diagonal terms in Eq.~\eqref{eq:MF_ele}) is semi-metallic, 
while for $B_{\bf 0}^z>0$ the disordered system is semiconducting. We denote this {\it semiconductor-semimetal} crossover by $U_{\rm SS}$.

\subsubsection{Phase digram and single-particle dispersion}
In this section, we clarify the properties of the ground states obtained within the MF theory to set the stage for the investigation of the collective modes.
In Fig.~\ref{fig:phase}(a), we show the dependence of $\phi$ and $\Delta n$ on the local interaction $U$ for fixed energy levels of the bands $D_{01}=2.2$, as a typical example.
When the interaction is weak, the system is in the ordered phase as in the BCS theory even for infinitesimally small interactions, and the order parameter gradually grows with increasing $U$. 
On the other hand, the interaction also contributes to the Hartree shift of the bands, so that the 
effective band splitting increases with increasing $U$ [see Eq.~\eqref{eq:MF_self_c}]. 
Thus, the order parameter shows a maximum at $U_{\rm BB}(=2.07)$ for fixed $D_{01}$ along $U$ and there is a BCS-BEC crossover.
The number of electrons in the conduction band decreases with increasing $U$ and it becomes close to $0$ in the BEC regime. 
For large enough $U$, the system turns into a normal semiconductor, via a second order phase transition at $U_c(=3.04)$.
By performing similar calculations for different $D_{01}$, we obtain the order parameter ($\phi$) and the difference in the number of electrons between the bands ($\Delta n$) in the ground state in the plane of $D_{01}$ and $U$ [see Fig.~\ref{fig:phase}(b)(c)]. 
The BEC regimes are located on the large $U$ and large $D_{01}$ side, and $B_{\bf 0}^z>0$ is always located in the BEC regime.
For large enough $U$ and $D_{01}$, the ground state becomes a normal semiconductor, which is characterized by $\phi=0$ and $\Delta n=-1$. 
In the following analysis we will focus on $D_{01}=2.2$, where $U_{\rm BB}=2.07$, $U_{\rm SS}=2.78$ and the 
phase boundary with the semiconductor is at $U_c=3.04$.

To clarify the situation further, we show the corresponding MF single-particle spectra for $k_y=0$ in Figs.~\ref{fig:chi_pure_EI}(a-d) for fixed $D_{01}=2.2$.
For $U<U_{\rm BB}$, 
the dispersion shows a characteristic minimum  at the $\Gamma$-point and the band-gap minimum($E_{\rm gap}$) is located at finite momentum [Fig.~\ref{fig:chi_pure_EI}(a)].
For $U_{\rm BB}\leq U<U_{c}$, the dispersion is characterized by the flattening of the band at the $\Gamma$-point [Figs.~\ref{fig:chi_pure_EI}(b,c)], while for $U_c<U$ the cosine dispersion of the free system is recovered [Fig.~\ref{fig:chi_pure_EI}(d)].
These single-particle spectra determine the single-particle contributions in the response functions as will be shown below.

\subsubsection{Linear response functions}
Now we present the linear response functions of the order parameter to see how the collective modes manifest themselves there.
Remember that $\chi^R_{11}(\omega;{\bf q})$ and $\chi^R_{22}(\omega;{\bf q})$ correspond to the dynamics of the order parameter along the amplitude and the phase directions, respectively, 
in the limit of ${\bf q}\rightarrow {\bf 0}$.
In Fig.~\ref{fig:chi_pure_EI}(e-l), we plot $-\frac{1}{\pi} {\rm Im}\chi^R_{11}(\omega;q_x, 0)$ and $-\frac{1}{\pi} {\rm Im}\chi^R_{22}(\omega;q_x, 0)$ for the pure EI in the different regimes 
for fixed 
$D_{01}=2.2$. 
For $U<U_c$ the system is in the EI state and we observe a massless mode
($\omega\rightarrow0$ at ${\bf q}\rightarrow {\bf 0}$) both in $-\frac{1}{\pi} {\rm Im}\chi^R_{11}(\omega;q_x, 0)$ [Fig.~\ref{fig:chi_pure_EI}(e-g)] and in $-\frac{1}{\pi} {\rm Im}\chi^R_{22}(\omega;q_x, 0)$  [Fig.~\ref{fig:chi_pure_EI}(i-k)].
This mode is nothing but the phase mode (the Nambu-Goldstone mode), which is characterized by a much stronger signal in $-\frac{1}{\pi} {\rm Im}\chi^R_{22}$ compared to $-\frac{1}{\pi} {\rm Im}\chi^R_{11}$. Moreover, its weight in $-\frac{1}{\pi} {\rm Im}\chi^R_{11}$ completely disappears at ${\bf q}\rightarrow {\bf 0}$.
We note that the appearance of the phase mode both in $-\frac{1}{\pi} {\rm Im}\chi^R_{11}$ and $-\frac{1}{\pi} {\rm Im}\chi^R_{22}$ means that the dynamics along the phase and amplitude 
direction of the order parameter is not completely separated at nonzero momenta.
The phase mode shows a linear dispersion around ${\bf q}= {\bf 0}$ 
for all $U$,  
but the velocity $v_x =\partial_{q_x}\omega({\bf q})|_{{\bf q}= {\bf 0}}$ gets suppressed as $U$ increases and the system approaches the BEC regime.
When the interaction becomes large ($U>U_c$), the system is in the normal semiconducting phase. There one can observe an in-gap state, which corresponds to an exciton [see Figs.~\ref{fig:chi_pure_EI}(h) and (l)]. 
Now, when we reduce $U$ from the normal semiconductor phase, this mode at ${\bf q=0}$ softens to zero at the boundary to the EI phase [see Fig.~\ref{fig:chi_phi}(b) for details], while at finite momenta, it is 
continuously connected to the phase mode branch.

In addition to the phase mode, there exists the amplitude mode at the bandgap energy ($E_{\rm gap}$), 
which is analogous to the amplitude Higgs mode in the superconducting phase.\cite{littlewood1982,Murakami2016,Tsuji2016,Murakami2016b} 
In the BCS regime, the signature of the amplitude mode is a prominent peak at ${\bf q}= {\bf 0}$ at $\omega=E_{\rm gap}$ in ${\rm Im}\chi^R_{11}(\omega)$ [see Fig.~\ref{fig:chi_pure_EI}(e)].
This mode gets suppressed with increasing $q$ because of the Landau damping of the amplitude mode into particle-hole excitations.\cite{littlewood1982} 
To see the signal of the amplitude mode in detail, we show ${\rm Im}\chi^R_{11}(\omega;{\bf q=0})$ and the single-particle contribution ${\rm Im}\chi^R_{0,11}(\omega;{\bf q=0})$ for $U=1.8$ (BCS), $U=2.78$ $(U_{\rm SS})$ and $U=2.9$ (BEC) in Fig.~\ref{fig:chi_phi}(a) and we show ${\rm Im}\chi^R_{11}(\omega;{\bf q=0})$ as a function of $U$ and $\omega$ in Fig.~\ref{fig:chi_phi}(b).
First, we note that the peak cannot be explained by the single particle excitations ${\rm Im}\chi^R_{0,11}(\omega)$ and it appears only after we take account of the vertex correction (${\boldsymbol{\Theta}}$) and sum up an infinite number of diagrams [see Fig.~\ref{fig:chi_phi}(a)]. This means that this peak originates from the collective motion.
Furthermore, as the system approaches the BEC regime, the peak structure at the gap energy in ${\rm Im}\chi^R_{11}(\omega)$ becomes less clear and it completely disappears for $U\gtrsim U_{\rm SS}$ [see Fig.~\ref{fig:chi_phi}(a)(b)].  This can be associated with the lifetime of the amplitude mode.\cite{Volkov1974,Gurarie2009,Murakami2016b,murakami2017,Behrle2018}  In the previous studies of the amplitude mode dynamics in real time,\cite{Volkov1974,Gurarie2009,Murakami2016b,murakami2017}  it has been shown that $\chi^R_{11}(t; {\bf 0})$ oscillates with a frequency of $\omega_{H}=E_{\rm gap}$ and its amplitude damps with $t^{-\frac{1}{2}}$~($t^{-\frac{3}{2}}$) in the BCS~(BEC) regime.  The corresponding frequency dependence is $(\omega-\omega_{H})^{- \frac{1}{2}}~((\omega-\omega_{H})^{\frac{1}{2}})$ in the BCS~(BEC) regime, which leads to a peak~(dip) structure at  $\omega_{H}=E_{\rm gap}$ [see Fig.~\ref{fig:chi_phi}(a)]. 
 
\begin{figure}[t]
     \centering
\includegraphics[width=60mm]{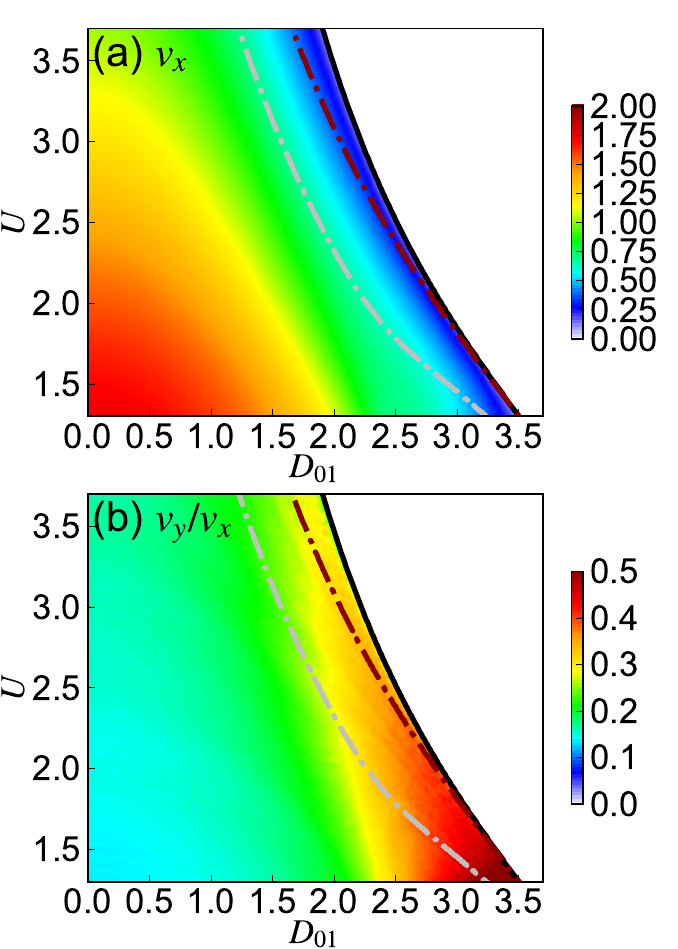}
\caption{(a) Velocity of the phase mode along the $x$ direction ($v_x$) and (b) the ratio between $v_y$ and $v_x$.
The velocities are obtained by a linear fit of the peaks in $-{\rm Im}\chi^R_{22}(\omega;q_x,q_y)$.
 For $v_x$($v_y$), we use $q_x (q_y)\in [0.05,0.15]$ with $q_y=0~(q_x=0)$. The black solid lines indicate $U_c$, the gray dot-dashed lines indicate $U_{\rm BB}$,
the red dot-dashed lines indicate $U_{\rm SS}$ [see the text for detail]. }
\label{fig:velocity} 
\end{figure}
\subsubsection{Anisotropy in velocity of phase mode}
\begin{figure*}[t]
     \centering
\includegraphics[width=170mm]{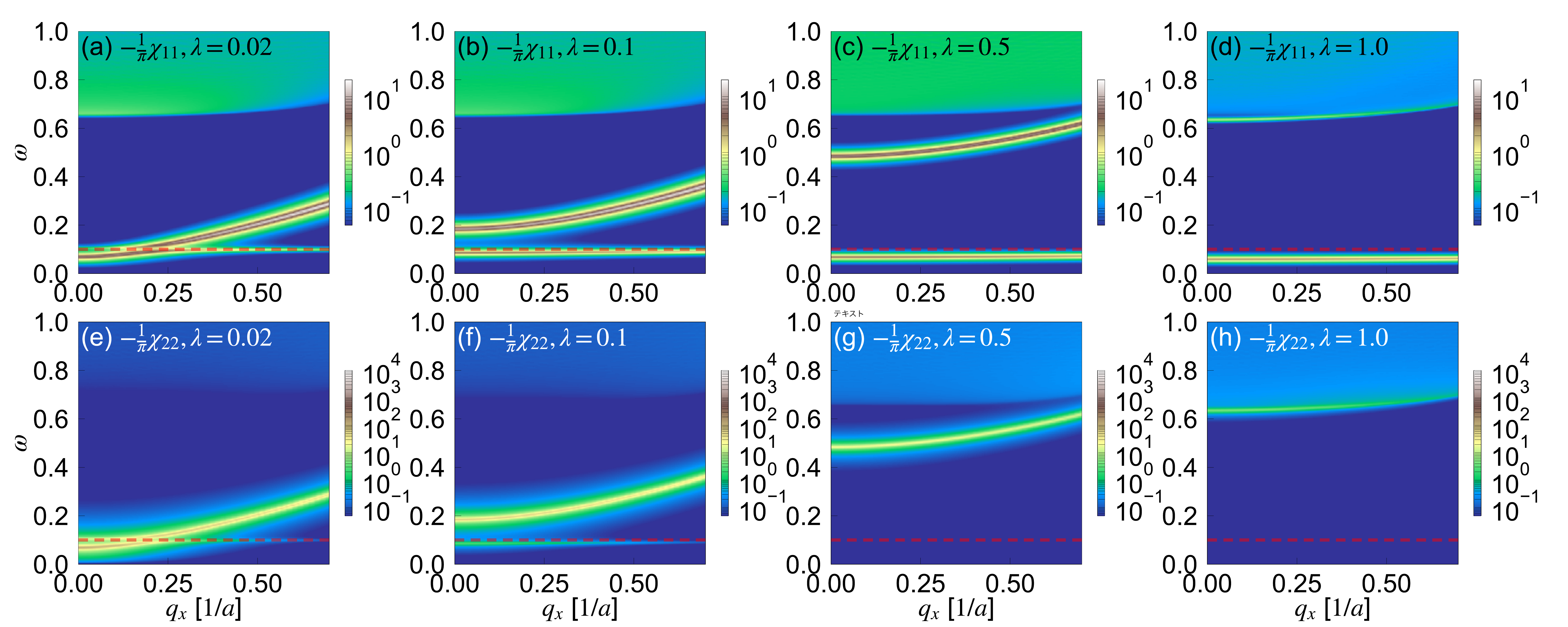}
\caption{(a-d) Linear response functions for the order parameter in the amplitude direction $-\frac{1}{\pi}{\rm Im}\chi^R_{11}(\omega;q_x,q_y=0)$  and~(e-h) in the phase direction $-\frac{1}{\pi}{\rm Im}\chi^R_{22}(\omega;q_x,q_y=0)$ for different values of the electron-phonon coupling $\lambda$.
For all plots we use a bare phonon frequency $\omega_0=0.1$, the reference Coulomb interaction $U_{\rm ref}=U_{\rm SS}=2.78$ and the reference band-level difference $D_{01,{\rm ref}}=2.2$. The red dashed lines indicate $\omega=\omega_0$. }
\label{fig:chi_type1_w}
\end{figure*}
In this section, we study the velocity of the phase mode.
If the hopping parameters are isotropic ($J_a({\bf a}_x)=J_a({\bf a}_y)$), the velocities of the phase mode along the $x$ and $y$ directions are also identical.
On the other hand, in the present setup with  $|J_a({\bf a}_x)|> |J_a({\bf a}_y)|$, these velocities are not equivalent (anisotropy in the phase mode velocity).
We demonstrate that the ratio of the phase mode velocities along the $x$ and $y$ directions depends on whether the system is in the BCS or BEC regime. The velocities of the phase mode ($\omega_{\rm phase}({\bf q})$) along the $\alpha$ direction is defined as 
\begin{align}
v_{\alpha}({\bf q})|_{\bf q =0} = \frac{a_\alpha}{\hbar} \frac{\partial \omega_{\rm phase}({\bf q})}{\partial q_{\alpha} }\Bigl |_{\bf q =0} .
\end{align}
Here, $a_\alpha$ is the lattice constant along the $\alpha$-direction. 
In our study we set $a_\alpha=\hbar=1$ and the velocities are evaluated by linear fitting of the peaks in ${\rm Im} \chi^R_{22}$ in the range $q_x,q_y\in [0.05,0.15]$.
In Fig.~\ref{fig:velocity}, we show the velocity along the chain $v_x $ and the ratio between the velocity along the $x$ direction ($v_x$) and the $y$ direction ($v_y$). 
As mentioned previously, the velocity tends to be larger on the BCS side, i.e. away from the BEC regime [see Fig.~\ref{fig:velocity}(a)].
As for the anisotropy, the difference between $v_x$ and $v_y$ becomes less prominent when $D_{01}$ and/or $U$ increase and the system approaches the BEC regime [see Fig.~\ref{fig:velocity}(b)].
Thus, 
a suppression of the 
anisotropy in the velocity compared to the anisotropy in the hopping parameters is an indication that the system is close
to the BEC regime. 
This result indicates that if one could observe the velocity of the phase mode (if any), one can roughly judge whether the system is close to the BCE regime or to the BEC regime. 
We note that in evaluating the velocity above, we assumed that the lattice constant is the same along the $x$ and $y$ directions.
In reality, an anisotropy in the hopping parameter is usually accompanied by a difference in the lattice constants, which should be taken into account in a realistic estimate  of the velocities.
For instance, in TNS, the distance between the chains is roughly twice longer than the lattice constant along the chain,
when we take into account that two sets of chains are involved in one unit-cell of TNS.

\subsection{Effects of electron-phonon coupling} \label{sec:effect_elph}

In this section, we investigate the effects of the el-ph coupling on the properties of the collective modes. To this end we choose a reference set of parameters $(D_{01},U,\lambda,D_{0}+D_{1}) = (D_{01,{\rm ref}},U_{\rm ref},\lambda_{\rm ref}=0,D_{0,{\rm ref}}+D_{1,{\rm ref}})$.
For nonzero el-ph coupling, we adjust $D_{01}, D_{0}+D_{1}$ and $U$ such that the order parameter and the mean-field single-particle spectrum are the same as for the reference set, i.e., such that condition \eqref{eq:condition} is satisfied.
Since we focus on half-filling ($D_{0}+D_{1}=-U$), the parameter set is specified by $D_{01,{\rm ref}}, U_{\rm ref}, \lambda$.
In the following, we fix $D_{01,{\rm ref}}=2.2$ and consider some values of $U_{\rm ref}$ that represents the BCS regime, semiconductor-semimetal (SS) crossover ($U=U_{\rm SS}$), and BEC regimes [see Fig.~\ref{fig:phase}(a)].
\begin{figure*}[thb]
     \centering
\includegraphics[width=130mm]{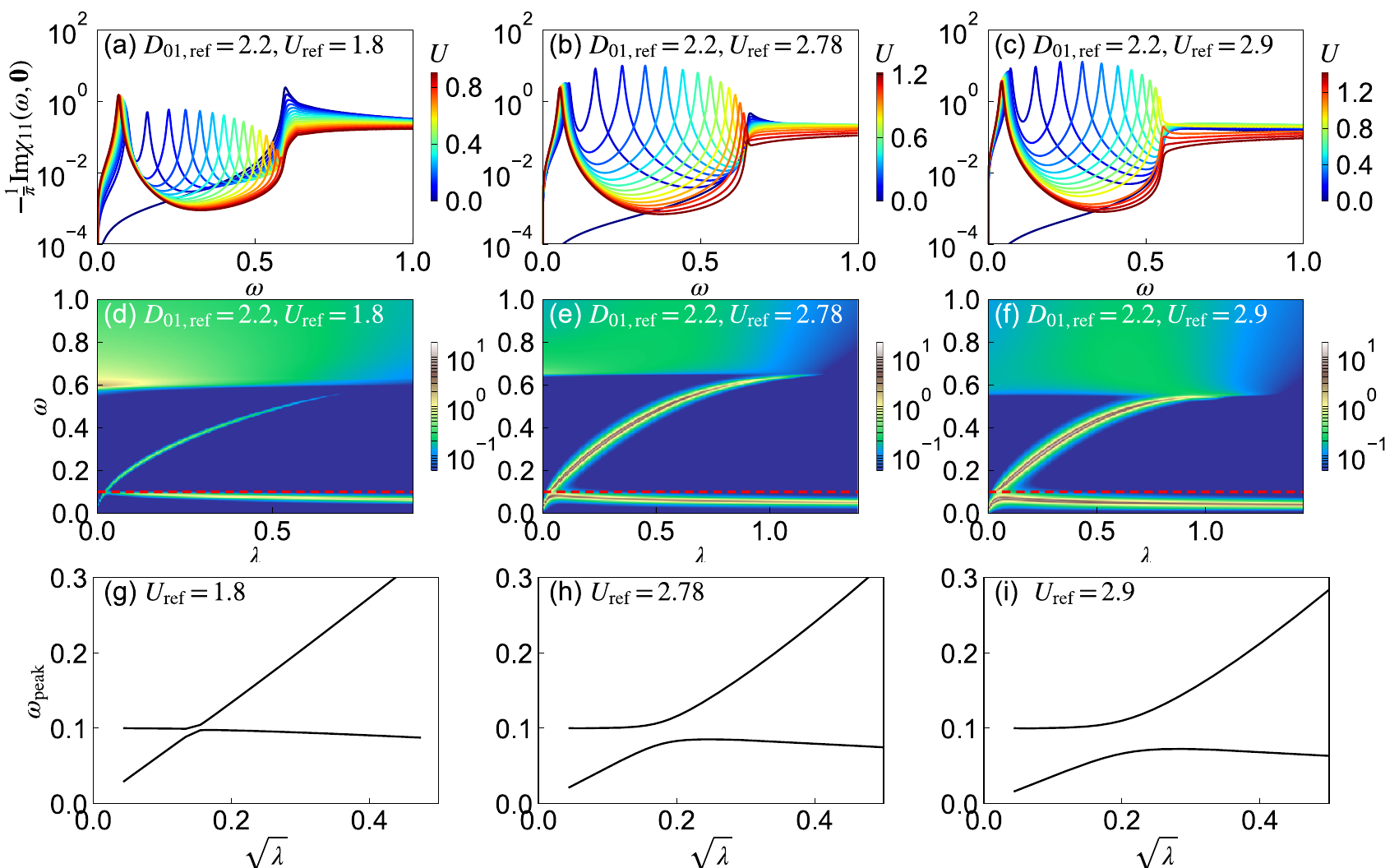}
\caption{(a-f) Linear response functions for the order parameter in the amplitude direction $-\frac{1}{\pi}{\rm Im}\chi^R_{11}(\omega;{\bf 0})$ as a function of $\lambda$ for fixed  $U_{\rm ref}$ and $D_{01,{\rm ref}}$. (g-i) The frequency of the peaks in $-\frac{1}{\pi}{\rm Im}\chi^R_{11}(\omega;{\bf 0})$ against $\sqrt{\lambda}$ for the specified $U_{\rm ref}$. 
The first column corresponds to the BCS regime ($U_{\rm ref}$ = 1.8), the second column to the semiconductor-semimetal crossover ($U_{\rm ref} = U_{\rm SS}=2.78$), and the third column to the BEC regime ($U_{\rm ref}$ = 2.9). 
We use $D_{01,{\rm ref}}=2.2$ and $\omega_0=0.1$. 
}
\label{fig:chi11_type1_lam}
\end{figure*}

In Fig.~\ref{fig:chi_type1_w}, we show  ${\rm Im}\chi^R_{11}$ and ${\rm Im}\chi^R_{22}$ for ${\bf q}= (q_x,0)$ for different values of $\lambda$. Here, we show the results at $U=U_{\rm SS}$, but the main features are the same for the BCS and BEC regimes. 
For nonzero el-ph coupling, the phase mode becomes massive (acquires a finite gap)\cite{zenker2014,murakami2017} and also shows up in ${\rm Im}\chi^R_{11}$ even at ${\bf q} = {\bf 0}$, although its weight 
is still larger in ${\rm Im}\chi^R_{22}$.\cite{murakami2017}
This means that in this case the dynamics of the amplitude and the phase of the order parameter is weakly coupled.
When the el-ph coupling is weak, the gap of the phase mode is small and the phase mode can hybridize with the phonon mode at finite momentum [see Figs.~\ref{fig:chi_type1_w}(a) and \ref{fig:chi_type1_w}(e)]. As we increase the el-ph coupling, the gap is increased and the band corresponding to the phase mode is lifted up.
By further increasing $\lambda$, the massive phase mode gradually merges  into the particle-hole continuum [see Figs.~\ref{fig:chi_type1_w}(d) and \ref{fig:chi_type1_w}(h)]. 

Now we focus on the signatures of the collective modes at ${\bf q}={\bf 0}$ as imprinted in the linear response function of the order parameter in the amplitude direction $\chi^R_{11}(\omega;{\bf 0})$. First, we study how the collective modes change as the relative contribution between the el-ph coupling and the el-el interaction to the order is changed. 
Remember that $U_{\rm ref}=U+2\lambda$ is the effective interaction that drives the system into the ordered phase [see Eq.~\eqref{eq:MF_self}].
In Fig.~\ref{fig:chi11_type1_lam}, we show ${\rm Im}\chi^R_{11}(\omega;{\bf 0})$ as a function of $\lambda$ with $U_{\rm ref}$ fixed to the BCS regime [Figs.~\ref{fig:chi11_type1_lam}(a) and ~\ref{fig:chi11_type1_lam}(d)], the SS crossover regime [Figs.~\ref{fig:chi11_type1_lam}(b) and ~\ref{fig:chi11_type1_lam}(e)]
and the BEC regime [Figs.~\ref{fig:chi11_type1_lam}(c) and ~\ref{fig:chi11_type1_lam}(f)]. 
At $\lambda=0$ ($\lambda=U_{\rm ref}/2$) the order is purely driven by the excitonic scenario (the el-ph coupling).
In all cases, one can see the signature of a massive mode arising from $\omega=0$ at $\lambda=0$, which corresponds to the massive phase mode. At the same time, the phonon frequency is renormalized and reduced from the bare value and the two modes cross at some $\lambda$.
The detailed dependence of the frequencies of the massive phase mode and the phonon mode is shown in Fig.~\ref{fig:chi11_type1_lam}(g-i) as a function of $\sqrt{\lambda}$ $(\propto g)$ for small $\lambda$. 
The gap of the phase mode scales linearly with $\sqrt{\lambda}$. The gap is independent of $\omega_0$ until the phase mode crosses with the phonon mode (not shown). The hybridization of the two modes (the size of the gap caused by the hybridization) is smaller for smaller $U_{\rm ref}$ and for smaller phonon frequencies (the latter is not shown).

Furthermore, the el-ph coupling affects the feature near the gap energy ($E_{\rm gap}$).
In the BCS regime, the well-defined peak at the gap edge in $\frac{1}{\pi}{\rm Im}\chi^R_{11}(\omega;{\bf 0})$, which corresponds to the amplitude (Higgs) mode,
disappears with increasing $\lambda$ [see Figs.~\ref{fig:chi11_type1_lam}(a)(d) and the discussion below]. 
In contrast, in the BEC regime, at some intermediate coupling strengths $\lambda$, the signal above the gap is slightly enhanced due to the 
merging of the massive phase mode with the
 particle-hole continuum [see Figs.~\ref{fig:chi11_type1_lam}(c)(f)].

\begin{figure*}[t]
     \centering
\includegraphics[width=160mm]{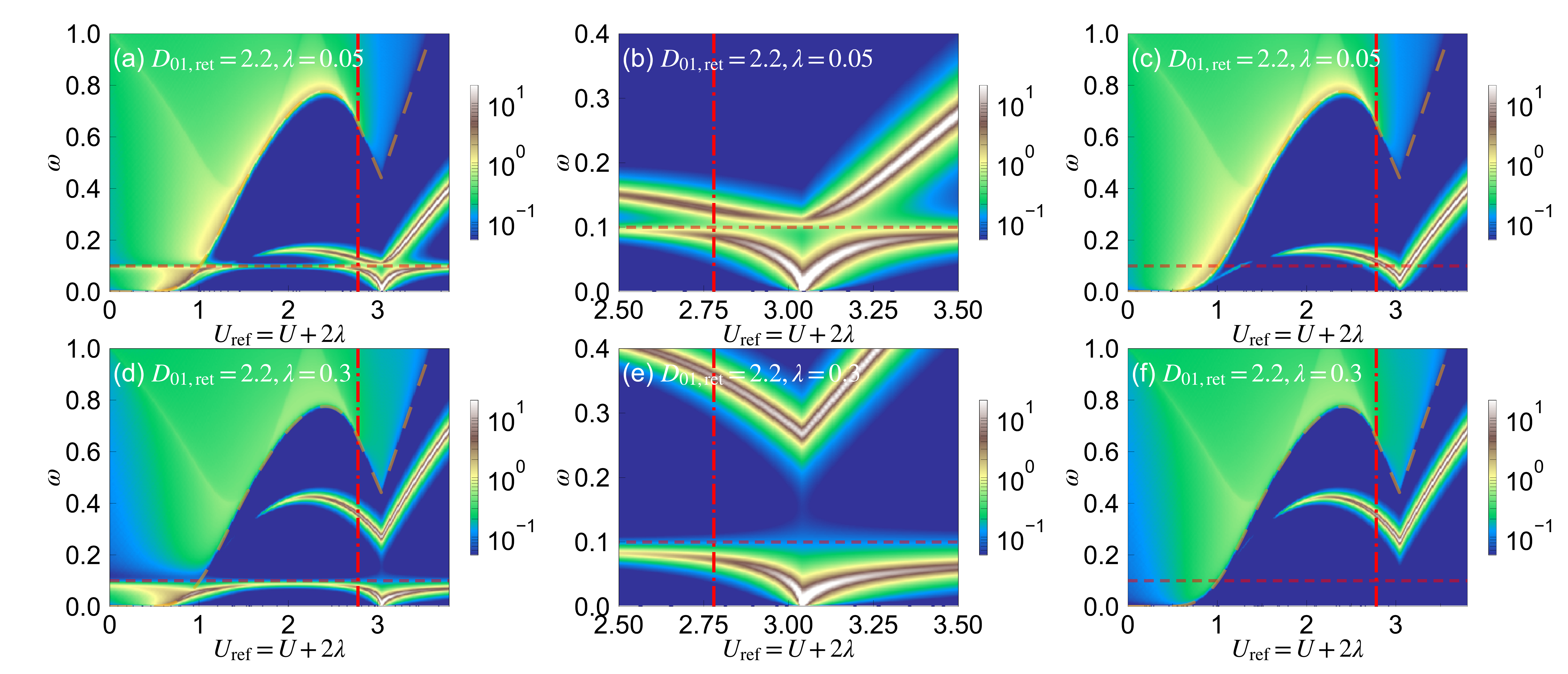}
\caption{(a)(b)(d)(e)~Linear response functions for the order parameter in the amplitude direction $-\frac{1}{\pi}{\rm Im}\chi^R_{11}(\omega;{\bf 0})$ as a function of $U_{\rm ref}$ with fixed electron-phonon interaction $\lambda=0.05~(0.3)$ are shown in the first~(second) row. (b)(e) are a zoom around the phase boundary with the normal semiconductor.~(c)(f) Linear response functions $-\frac{1}{\pi}{\rm Im}\chi_{11,{\rm el}}(\omega;{\bf 0})$, which include only the el-el coupling contribution to the vertex correction $\Theta$ [see Eq.~\eqref{eq:RPA_chi_Uloc} and the corresponding discussion]. The band-level difference is fixed to $D_{01,{\rm ref}}=2.2$, and the bare phonon frequency is fixed to $\omega_0=0.1$. 
The horizontal line indicates the bare phonon frequency, the vertical dashed red line indicates $U=U_{\rm SS}$ (i.e. $B_{\bf 0}^z=0$), and the dashed orange lines indicate the bandgap $E_{\rm gap}$.
}
\label{fig:chi11_type1_Utot}
\end{figure*}

Now, we study the behavior of the collective modes around the phase boundary and the BCS-BEC crossover regime with nonzero el-ph coupling.
In Fig.~\ref{fig:chi11_type1_Utot}, we fix $\lambda$ to a small value ($\lambda=0.05$) and to an intermediate value ($\lambda=0.3$)
and change $U_{\rm ref}=U+2\lambda$. 
For large enough $U_{\rm ref}$, the system is in the normal semiconductor state.
In the semiconductor state, one can see an in-gap mode below the bandgap energy but above the phonon energy, which corresponds to the exciton mode.
As $U_{\rm ref}$ is decreased and the system approaches the ordered phase, the exciton mode as well as the phonon mode soften. 
However, the nature of the mode that softens to zero at the boundary to the ordered phase turns out to be different for weak and  intermediate couplings.
For small $\lambda$, the exciton mode softens, crosses the phonon mode and hybridizes with the phonon mode [see Figs.~\ref{fig:chi11_type1_Utot}(a) and \ref{fig:chi11_type1_Utot}(b)]. The situation becomes clear by looking at the response function defined as 
\begin{align}
\bchi^R_{\rm el}(\omega;{\bf q})&=\bchi^R_{0}(\omega;{\bf q})+\bchi^R_{0}(\omega;{\bf q}) \boldsymbol{\Theta}^{\rm el}(\omega;{\bf q}) \bchi^R_{\rm el}(\omega;{\bf q}),
\end{align}
which only includes the contribution from the el-el interaction to the vertex correction.
The result is shown in Fig.~\ref{fig:chi11_type1_Utot}(c), where one can see the exciton mode ($\omega_{\text{ex},0}$) 
but 
no specific signal at the phonon frequency.
Around the phase boundary, the exciton mode softens below the phonon frequency.
Since the full response function can be expressed as 
\begin{align}
\bchi^R(\omega;{\bf q}) = \bchi^R_{\rm el}(\omega;{\bf q}) + \bchi^R_{\rm el}(\omega;{\bf q})\boldsymbol{\Theta}^{\rm ph}(\omega;{\bf q}) \bchi^R(\omega;{\bf q}),
\end{align}
we can see that in the total response function $\bchi^R(\omega;{\bf 0})$, the exciton mode at $\omega_{\text{ex},0}$ and the bare phonon mode with frequency $\omega_0$ hybridize.
 (Note that $\boldsymbol{\Theta}^{\rm ph}(\omega;{\bf q})$ is almost a phonon propagator.)
When $\omega_{\text{ex},0}<\omega_0$, the hybridization between the two bosonic modes pushes down the energy level of the exciton mode and pushes up the energy of the phonon mode.
Therefore, we can interpret the data for the el-ph coupling $\lambda=0.05$ as an el-ph coupling assisted softening of the exciton mode. However, what is condensing at the EI phase boundary are still excitons.

On the other hand, for intermediate $\lambda$, the exciton energy stays above the phonon mode [see Figs.~\ref{fig:chi11_type1_Utot}(d-f)]. In this case, since $\omega_{\text{ex},0}>\omega_0$, the hybridization pushes the energy of the exciton mode up and pushes the energy of the phonon mode down. Therefore, it is natural to interpret the data as a softening of the phonon mode to zero with the assistance of the coupling to the exciton mode. 
We note that, with this intermediate value of $\lambda$, the Coulomb interaction $U$ is still larger than $2\lambda$ at the boundary between the ordered phase and the normal semiconductor.
Thus, our analysis implies that there also exists a criterion to judge whether the ordered phase is more excitonic or el-ph driven, based on the nature of the mode that softens to zero at the boundary,
and not by just comparing the size of the Coulomb interaction $U$ and the el-ph coupling $2\lambda$.
Although here we focus on $T=0$, we can expect that a similar behavior is found in the phase transition at $T>0$. The behavior at intermediate $\lambda$ is reminiscent of the collective mode behavior observed in $1T-{\rm TiSe_2}$ in Ref.~\onlinecite{kogar2017}, where the softening of the electronic mode occurs above the phonon mode at the phase transition.  

Another important point is the dependence of the amplitude mode signal on the el-ph interaction. On the BCS side, there emerges a peak at the bandgap energy ($E_{\rm gap}$) for weak el-ph coupling, which corresponds to the amplitude mode.  In contrast, at intermediate couplings this peak disappears [see  Figs.~\ref{fig:chi11_type1_Utot}(a)(d) and  Fig.~\ref{fig:chi11_type1_lam}(d)].
The behavior can be understood by looking at the expression for the vertex $\Theta$, Eq.~\eqref{eq:theta}.
As the el-ph coupling $\lambda$ is increased keeping $U_{\rm ref}$ fixed, 
the size of $\Theta(\omega)$ around the bandgap energy $E_{\rm gap}\simeq 0.6$ is decreased as $U_{\rm ref}/2-\lambda$ because of the small phonon frequency $\omega_0\ll E_{\rm gap}$. 
 In other words, the contribution from $\Theta^{\rm ph}(\omega)$ in this energy range is small.
(Around $\omega=0$, $\Theta^{\rm ph}(\omega)$ becomes sizable and $\Theta(\omega)$ becomes approximately independent of the el-ph coupling.) Mathematically, the suppression of the vertex results in the suppression of the divergence around $\omega=E_{\rm gap}$ in $\bchi^R(\omega;{\bf 0})$ [see Eq.~\eqref{eq:RPA_chi_Uloc}]. Physically, this means that the effective el-el interaction mediated by the phonons only applies to the low frequency range and that it does not contribute to the collective motion of the pairs of quasi-particles across the bandgap.
Hence, the amplitude mode, which is a collective motion of such quasiparticles, is weakened. This is why the signature of the amplitude mode is suppressed with increasing $\lambda$ for fixed $U_{\rm ref}$. 

These results can be summarized as follows. When the Coulomb interaction is the dominant mechanism for the ordered phase and the el-ph coupling is very small, the mass of the phase mode is comparable to the phonon energy scale. In this regime, the phase mode and the phonon can cross at finite momentum, and the mode that softens to zero at the phase boundary with the semiconductor is exciton-like. In addition, in the BCS regime, the amplitude mode becomes well defined in the sense that there is a sharp peak in the response function. 
In the intermediate el-ph coupling regime, the mass of the phase mode is of the same order as the single-particle gap energy. 
This can occur even when the el-ph coupling is still smaller than the Coulomb interaction, and the mode that softens to zero at the phase boundary with the semiconductor is now phonon-like.
In addition, for intermediate el-ph coupling, the amplitude mode is not well defined and there is no peak at the corresponding energy. These features of the collective modes can provide a useful guidance to judge the relative importance of the contributions from the el-el interaction and the el-ph interaction for the ordered phase. 

Finally, let us comment on the effects of electronic terms that explicitly break the symmetry, which have been pointed out recently.\cite{mazza2019,Watson2019}
We numerically confirmed that these terms make the phase mode massive as well, and the dependence of the mass depends sensitively on the form of the term.

\subsection{Signature of collective modes in optical response}

\begin{figure}[t]
     \centering
\includegraphics[width=85mm]{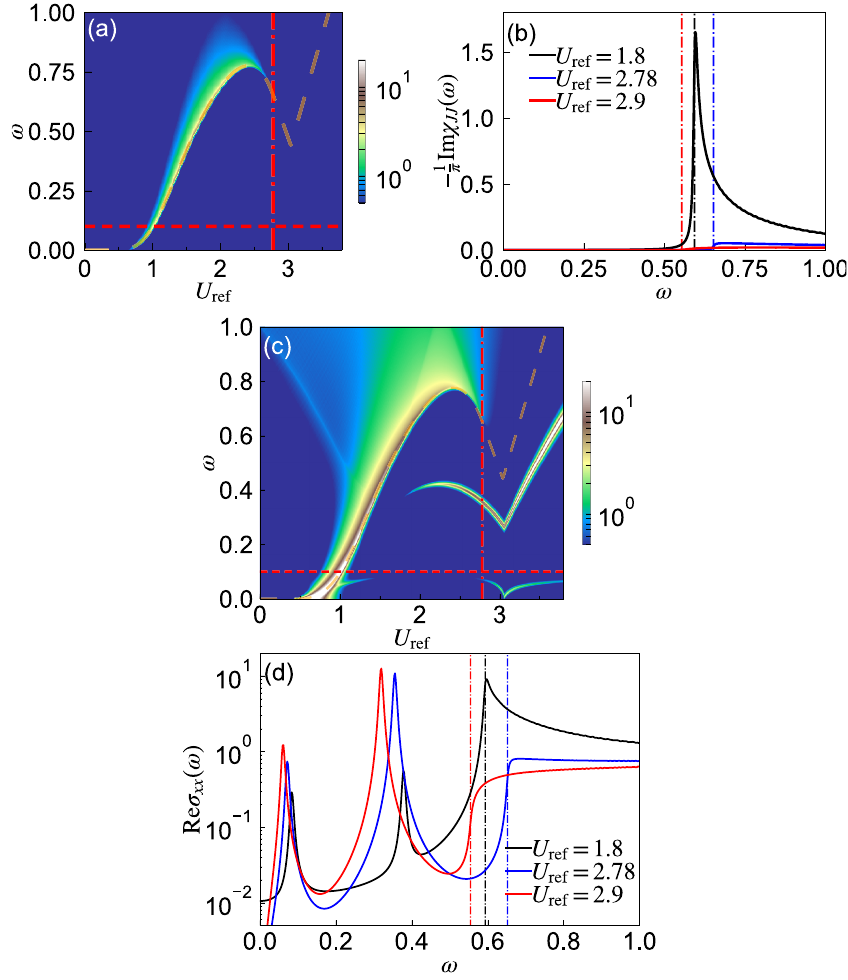}
\caption{(a)(b) The imaginary part of the response function $\chi_{JJ}(\omega)$ for the $x$ direction for $D_{01,{\rm ref}}=2.2$.  (a) $-\frac{1}{\pi}\chi_{JJ}(\omega)$ in the plane of $U_{\rm ref}$ and $\omega$. (b) Specific examples for $\chi_{JJ}(\omega)$ for the specified $U_{\rm ref}$ values.
 We note that the vertex correction is zero for this quantity, so that it only depends on the MF dispersion and it is independent of $\lambda$. 
(c)(d)  The real part of the total optical conductivity $\sigma_{xx}(\omega)$ for $D_{01,{\rm ref}}=2.2$.  
(c) ${\rm Re}\sigma_{xx}(\omega)$  in the plane of $U_{\rm ref}$ and $\omega$.
(d) Specific examples for  ${\rm Re}\sigma_{xx}(\omega)$ for the specified $U_{\rm ref}$ values.
For (c)(d), we use $d_{0,x}=d_{1,x}=1.0$, $\lambda=0.3$, and $\omega_0=0.1$. }
\label{fig:Chi_JJ}
\end{figure}

Here, we show how the collective modes can manifest themselves in the optical response, Eq.~\eqref{eq:opt_cond}.
As was pointed out in Sec.~\ref{sec:opt_response}, ${\boldsymbol \chi}_{JP}^R=0$ and ${\boldsymbol \chi}_{PJ}^R=0$ in the present model,  while ${\boldsymbol \chi}_{PP}^R$ is essentially 
the same as ${\boldsymbol \chi}^R$ discussed above.
Therefore, we first focus on ${\boldsymbol \chi}_{JJ}^R$.
In Fig.~\ref{fig:Chi_JJ}(a)(b), we show the imaginary part of  $[{\boldsymbol \chi}_{JJ}^R(\omega)]_{xx}$ for $D_{01,{\rm ref}}=2.2$, which contributes to the real part of the optical conductivity.
In the normal semiconducting state, it vanishes. 
In the ordered phase, the signal is weak in the BEC regime,
while in the BCS regime it exhibits a peak at the energy of the bandgap.
We note that this peak does not originate from collective excitations since  ${\boldsymbol \chi}_{JJ}^R$ has no contribution from vertex corrections.
This means that when we see a peak structure in the optical conductivity at the gap energy, it can originate from the single-particle excitations and/or the amplitude mode.
A related point has been discussed in the context of the third harmonic generation (THG) in superconductors, where both the amplitude mode and the single-particle excitations can contribute to THG and produce a resonant signal at the gap energy. \cite{Matsunaga2014,Tsuji2015,Cea2016,Tsuji2016,Murotani2019,shimano2020}
In the superconductor, the two different contributions can be distinguished by the polarization dependence,\cite{shimano2020} 
which may be also the case for the EI and is an interesting topic for future studies.
We also note that  ${\boldsymbol \chi}_{JJ}^R$ is the only contribution that was considered in a previous DMRG study\cite{sugimoto2018} motivated by the optical experiments for TNS.\cite{larkin2017}
There, a peak in the optical conductivity shows up at the gap because of the strong fluctuations in the vicinity of
the ordered phase, which reconstructs the band structure and opens a gap.\cite{Monney2012}
How the peak structure in ${\boldsymbol \chi}_{JJ}^R$ is modified by the feedback from collective excitations, which is not taken into account in the present MF formalism, and whether the peak observed in Ref.~\onlinecite{sugimoto2018} originates from the collective motion are interesting open questions. \\

Finally, in Fig.~\ref{fig:Chi_JJ}(c)(d), we show the real part of the total optical conductivity $\sigma_{xx}(\omega)$ for $d_{0,x}=d_{1,x}=1.0$ ( the dipole matrix along the $x$ direction )
and $\lambda=0.3$.
As in the previous section, the sharp peak appears at the bandgap in the BCS regime, which mainly originates from the single-particle excitation.
In addition, the signal of the massive phase mode appears below the band gap. The signals in the lower energy regime becomes less clear because of the factor $\omega$ to $\chi_{PP}^R$ in
the optical conductivity Eq.~\eqref{eq:opt_cond}.

\begin{figure}[t]
     \centering
\includegraphics[width=60mm]{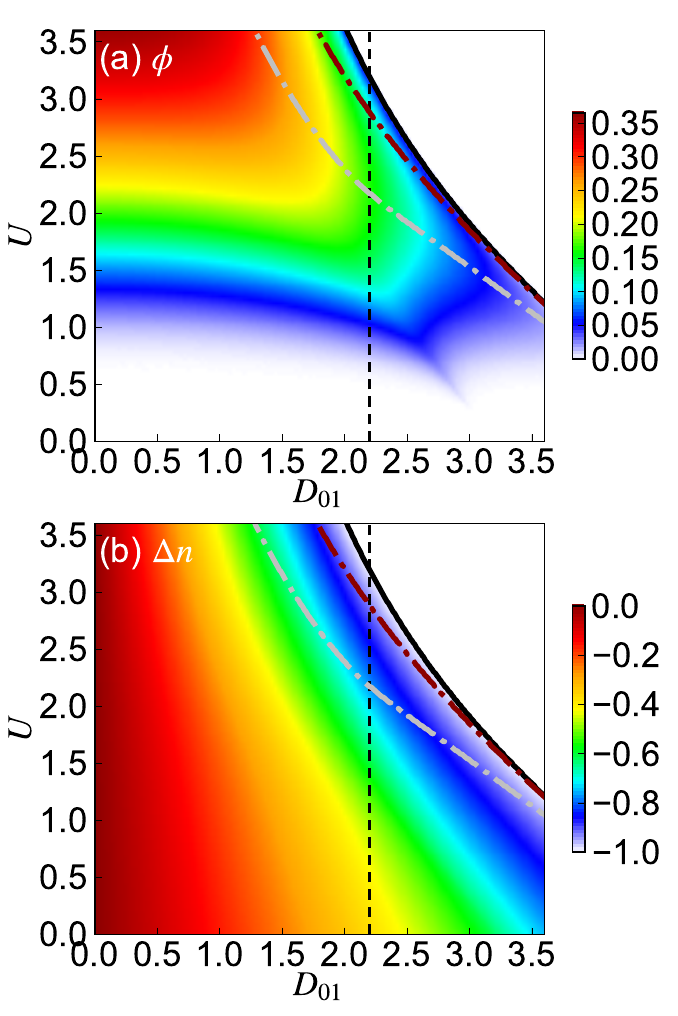}
\caption{(a) Order parameter for a system with nonlocal interactions $\phi = \langle \hc^\dagger_{i,0} \hc_{i,1}\rangle \in \mathbb{R}$ in the plane of $D_{01}$ and $U$. (b) Difference of the occupation in the plane of $D_{01}$ and $U$.
The black solid line indicates the boundary of the EI phase ($U_c$), the gray dot-dashed line indicates $U_{\rm BB}$,
the red dot-dashed line indicates $U_{\rm SS}$ (the $B_{\bf 0}^z<0$ regime is below and the $B_{\bf 0}^z>0$ regime is above the line). 
The vertical dashed line is $D_{01}=2.2$.
Here, the system is half filled, and we use $J_{0}({\bf a}_x)=-J_{1}({\bf a}_x)=1.0$, $J_{0}({\bf a}_y)=-J_{1}({\bf a}_y)=0.2,V_y=0.4U$ and $\lambda=0$.}
\label{fig:phase_nonlocU}
\end{figure}

\subsection{Effects of nonlocal interactions} \label{sec:nonlocU}
\begin{figure}[t]
     \centering
\includegraphics[width=60mm]{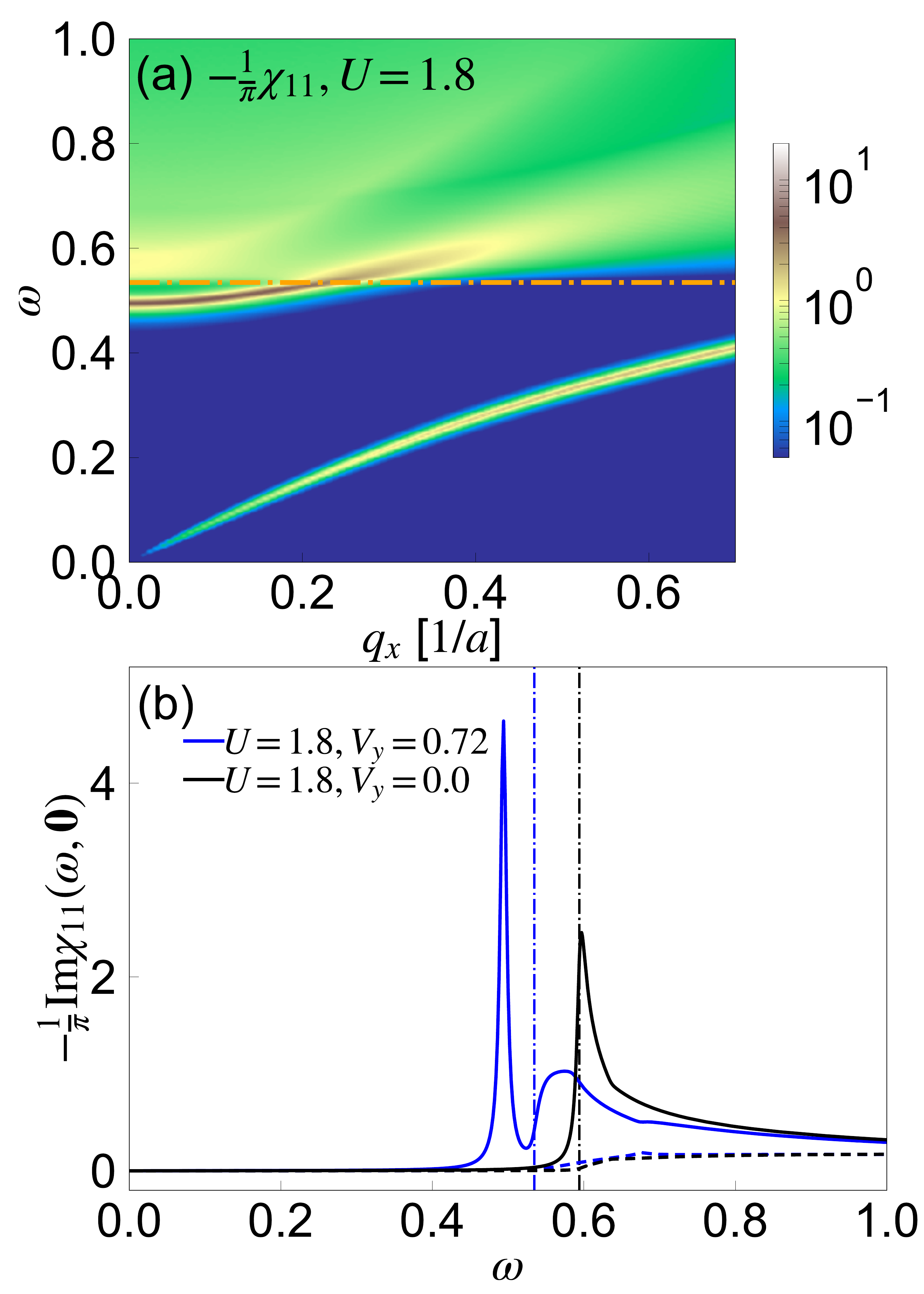}
\caption{ (a) Linear response functions in the amplitude direction at  the momentum cut $q_y=0$ for $-\frac{1}{\pi}{\rm Im}\chi^R_{11}(\omega;q_x,q_y=0)$ for $U=1.8$, $V_y=0.4U ,D_{01}=2.2$. (b) $-\frac{1}{\pi}{\rm Im}\chi^R_{11}(\omega;{\bf q=0})$ and $-\frac{1}{\pi}{\rm Im}\chi^R_{0,11}(\omega;{\bf q=0})$ for specified $(U,V_y)$ at $D_{01}=2.2$. The solid lines correspond to $-\frac{1}{\pi}{\rm Im}\chi^R_{11}(\omega;{\bf q=0})$, while the dashed lines correspond to $-\frac{1}{\pi}{\rm Im}\chi^R_{0,11}(\omega;{\bf q=0})$. The dashed vertical lines indicate the bandgap energies extracted from the MF single-particle spectrum. }
\label{fig:chi_q_w_nonlocU}
\end{figure}

Although we focused on the local Coulomb interaction so far, in realistic materials such as TNS, the interaction can have nonlocal components.
Thus, in this section, we study the effects of nonlocal interactions on the collective modes.  
For simplicity, we choose a specific form of the nonlocal interactions, and point out several potentially interesting effects of it. In the following, we focus on the case without el-ph coupling ($\lambda=0$). 
We leave a systematic analysis of different types of nonlocal interactions for a future study.  
Specifically, we consider a model with nearest-neighbor (NN) interactions along the $y$ direction,
\begin{align}
\hH_{\rm int}=U\sum_i \hat{n}_{i,0}\hat{n}_{i,1} +V_y \sum_{\langle i,j\rangle_y} \hat{n}_{i}\hat{n}_{j},
\end{align}
where $\langle i,j\rangle_y$ denotes the nearest neighbors along the $y$ axis and $\hat{n}_{i}=\hat{n}_{i,0} + \hat{n}_{i,1}$.
The reasons for this choice of interaction are the following.
First, the order parameter and the phase digram turns out to be sensitive to nonlocal interactions along the $x$ direction, which makes it difficult to compare the results with those of the local interaction model.  
On the other hand, the interaction along the $y$ direction has small effects on the order parameter and the phase diagram.
In Fig.~\ref{fig:phase_nonlocU}(a)(b), we show the order parameter and the occupation difference as a function of $U$ and $D_{01}$, which are similar to the previous
results for the local interaction (Fig.~\ref{fig:phase}).
Second, even with the NN interaction along the $x$ direction stronger than $V_y$,
 similar effects 
as discussed here are observed. Therefore, it is sufficient to consider $V_y>0$ for the following issues.
In the following, we evaluate the response function \eqref{eq:chi_munu} as in the case of the local interaction.
Although the expression of the response function is not Eq.~\eqref{eq:RPA_chi_Uloc} anymore, one can 
generalize the strategy based on 
mean-field dynamics to the nonlocal interaction model [see Appendix~\ref{sec:general_rpa} for details].

\begin{figure}[t]
     \centering
\includegraphics[width=70mm]{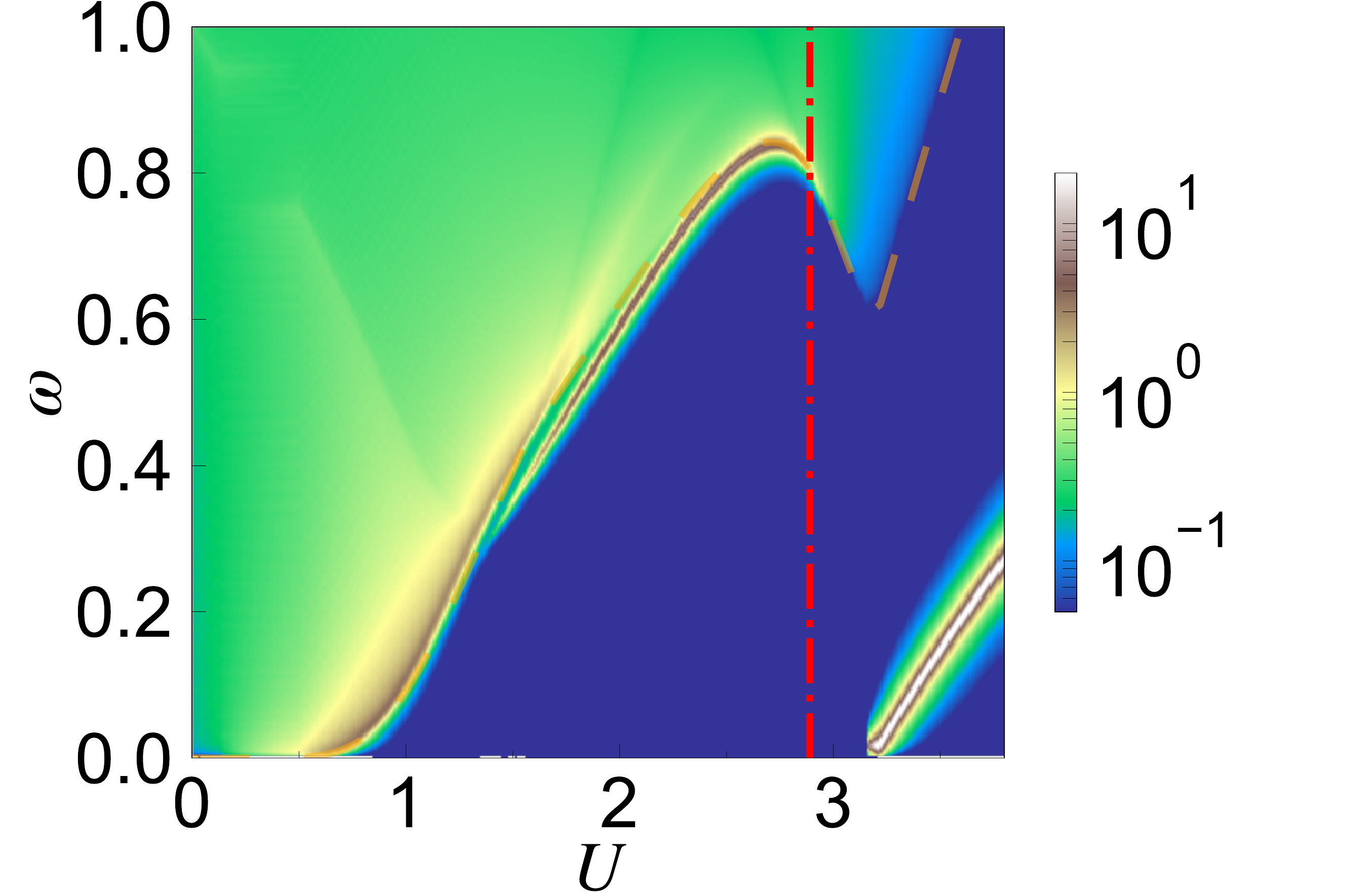}
\caption{Linear response functions for the order parameter in the amplitude direction $-\frac{1}{\pi}{\rm Im}\chi^R_{11}(\omega;{\bf q=0})$ in the plane of $\omega$ and $U$  for $D_{01}=2.2$ and $V_y=0.4U$. Here, $\lambda=0.0$. The dot-dashed red line indicates $U=U_{\rm SS}$, and the dashed orange line indicates the bandgap $E_{\rm gap}$. }
\label{fig:chi_nonlocU}
\end{figure}

One expected effect of non-local interactions is the appearance of multiple uncondensed exciton states (in-gap states) in the EI phase, as was originally discussed for semiconductors with long-range Coulomb interactions.\cite{halperin1968RMP} 
We will demonstrate that reminiscent phenomena can be observed with the NN interaction, and, in the intermediate coupling regime of $U$, 
the amplitude mode is split into two modes, one of which becomes an in-gap mode.
We show the response function along the amplitude direction (${\rm Im}\chi^R_{11}(\omega;{\bf q})$)  for $V_y=0.4U$, $U=1.8$ and $D_{01}=2.2$ (the BCS regime) in Fig.~\ref{fig:chi_q_w_nonlocU}(a) as a function of $\omega$ and $q_x$ and in Fig.~\ref{fig:chi_q_w_nonlocU}(b) at ${\bf q=0}$, respectively. 
There, one can observe two peaks around the bandgap energy, one of which is below the gap [ see Fig.~\ref{fig:chi_q_w_nonlocU}(a)(b) and compare them
with Fig.~\ref{fig:chi_pure_EI}(e)].
The new in-gap mode has a strong signal since the decay into particle-hole excitations is now completely forbidden.
As for the phase mode, it is massless due to the $U(1)$ symmetry.
${\rm Im}\chi^R_{11}(\omega;{\bf q=0})$ is shown in Fig.~\ref{fig:chi_nonlocU} as a function of $U$ for $V_y=0.4U$ and $D_{01}=2.2$.
The two modes can only be clearly identified around the intermediate interaction $U=1.8$.
By decreasing the interaction, the in-gap mode gradually disappears and only the above-gap mode survives, 
where the signal of the latter becomes sharp.
On the other hand, by increasing the interaction from $U=1.8$, the above-gap mode disappears first.
By further increasing $U$, the in-gap mode also disappears around $U_{\rm BB}$ as in the case of local interactions only. 
By comparing Fig.~\ref{fig:chi_nonlocU} and Fig.~\ref{fig:chi_phi}(b), one can see that these two modes are continuously connected to the amplitude mode 
observed in the case of local interactions ($V_y=0.0$).
Thus, our results suggest that the amplitude mode can become more prominent in the presence of nonlocal interactions since the mode is pushed below the particle-hole continuum.

\begin{figure}[t]
     \centering
\includegraphics[width=60mm]{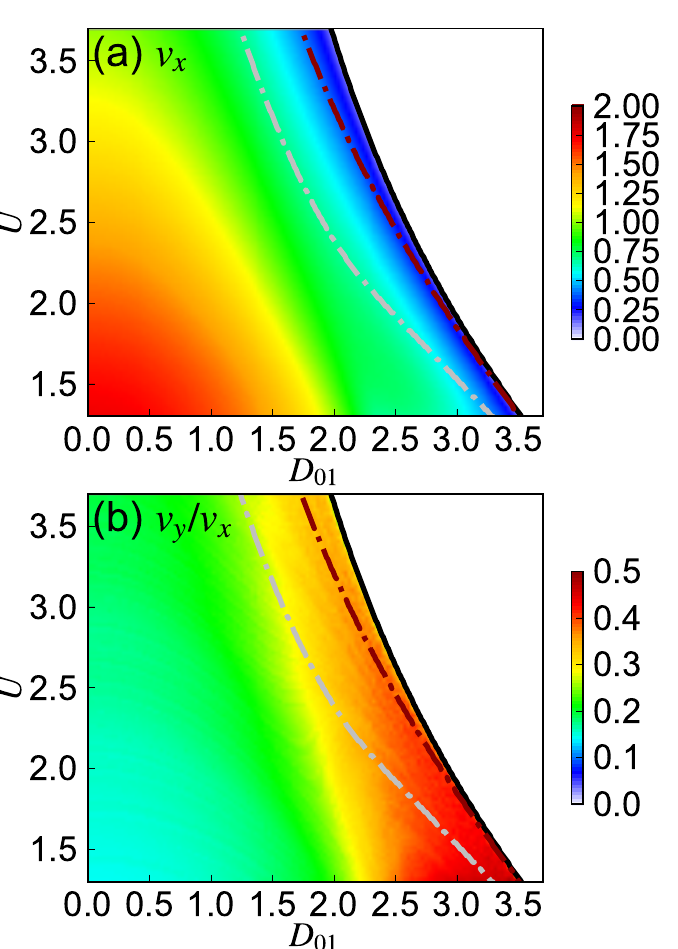}
\caption{(a) Velocity of the phase mode along the $x$ direction ($v_x$) and (b) ratio between $v_x$ and $v_y$ for the system with non-local interaction.
The velocity $v_x$($v_y$) is obtained from a linear fit of the peaks in $-{\rm Im}\chi^R_{22}(\omega;q_x,q_y)$ for $q_x (q_y)\in [0.05,0.15]$ with $q_y=0~(q_x=0)$. 
 The black solid lines indicate $U_c$, the gray dot-dashed lines indicate $U_{\rm BB}$,
the red dot-dashed lines indicate $U_{\rm SS}$.
Here, $V_y=0.4U$ and $\lambda=0.0$.
 }
\label{fig:velocity_nonolocU}
\end{figure}

Another potential effect of the non-local interaction is a change of the velocity of the phase mode.
Since we only consider the interaction along the $y$ direction, the Fock term can strongly affect the dispersion of the single-particle spectrum along the $y$ direction, which may change the velocity of the phase mode as well.
The results of the estimated velocity and the ratio of the velocities in the $x$ and $y$ directions are shown in Fig.~\ref{fig:velocity_nonolocU}. 
One can find that the velocity along the $x$ direction is not sensitive to the NN interaction [compare Fig.~\ref{fig:velocity_nonolocU}(a) and Fig.~\ref{fig:velocity}(a)].
As for the ratio, again the anisotropy in the phase-mode velocity is reduced when the system approaches the BEC regime. One can see that the anisotropy is slightly relaxed with the NN interaction compared to the case with the local interaction only [Fig.~\ref{fig:velocity_nonolocU}(b) and Fig.~\ref{fig:velocity}(b)] , as is expected above. 

\section{Conclusions}
\label{sec:conclusion}
In this paper, we systematically studied the collective excitations in the ordered phase of a spinless two band model,
which is driven by the local interband Coulomb interaction (the excitonic scenario) and the electron-phonon interaction.
The TNS inspired model is defined on the two-dimensional square lattice with anisotropic hopping parameters.  The linear response functions were 
evaluated on the RPA level, which is consistent with the nonequilibrium MF formalism. 

For the pure EI, we showed that the massless phase mode and the amplitude mode appear.
The signal of the amplitude mode is prominent in the BCS regime, while in the BEC regime the signal is suppressed, as in the superconducting phase.
In addition, the anisotropy in the phase-mode velocity originating from the anisotropic hopping parameters is relaxed in the BEC regime.
For nonzero el-ph coupling, the phase mode acquires a finite mass and the signal of the amplitude mode becomes less clear even in the BCS regime.
We argued that the latter originates from the fact that the el-ph coupling only affects the collective motion in the low energy regime.
Furthermore, we pointed out that, even for moderate el-ph coupling smaller than the Coulomb interaction, the mode that softens to zero at the boundary between the  semiconductor phase and the ordered phase can be more phonon-like and less exciton-like.
These behaviors of the collective modes discussed here can provide a useful guidance to judge the relative importance of the contributions from the el-el interaction and the el-ph interaction to the ordered phase. We also discussed how these modes can be observed in the optical conductivity within the present model.

Another interesting point is the effect of the nonlocal interactions, which can be important in transition metal chalcogenides.
We revealed that in the presence of nonlocal interactions, the amplitude mode can be split into two parts. One of them becomes an in-gap mode and acquires a long life-time.
In addition, we showed that the NN interactions can further relax the anisotropy in the phase mode velocity.

The collective modes can be experimentally observed through optics or nonequilibrium setups.
Indeed, strange behaviors of the phonons in the pump-probe experiments for TNS are already observed and it will be important to 
theoretically clarify the origin of these behaviors and their relation with the collective excitations.\cite{werdehausen2018} 
Furthermore, time- and space- resolved pump-probe experiments should be a promising tool to observe the phase mode of the ordered phase. 
Indeed, a recent experiment reports a fast spatial propagation of the phonon oscillation in the optical response, 
which can originate from the mixing between the phonon and the phase mode in the ordered phase.\cite{Paolo2020}
Another important problem is the effect of the electronic terms that break the continuous symmetry.\cite{mazza2019,Watson2019} 
While we have explicitly checked that such terms also lead to the massive phase mode, the mass sensitively depends on the functional form of the hybridization. This calls for an ab-initio description of TNS in order to describe the intriguing interplay of different microscopic terms and to quantitatively  capture the properties of collective modes.

\acknowledgements
We would like to thank P. Andrich, H. Bretscher and B. Remez for fruitful discussion.
We appreciate the CECAM workshop "Excitonic insulator: New perspectives in long-range interacting systems" at EPFL Lausanne in 2018 
for providing us motivation to initiate this study.
This work was supported by Grant-in-Aid for Scientific Research from JSPS, KAKENHI Grant Nos. JP19K23425 (Y.M.), JP18K13509 (T.K.), JP19H05821, JP18K04678, JP17K05536 (A.K.),
JST CREST Grant No. JPMJCR1901 (Y.M.), ERC Consolidator Grant No. 724103 (P.W.) and the Swiss National Science Foundation via NCCR Marvel (P.W.). 
T.K. acknowledges support from the JSPS Overseas Research Fellowship.
The calculations were run on the Beo05 cluster at the University of Fribourg.  

\appendix 
\section{General expression for the RPA-type susceptibility}\label{sec:general_rpa}
In this Appendix, we derive the linear response functions consistent with the tdMF 
theory for general models, which can include long-range interactions.
 The Hamiltonian we consider is 
\begin{align}
\hH(t) = \hH_{\rm kin} + \hH_{\rm int} +\hH_{\rm el-ph} + \hH_{\rm ph} + \hH_{\rm ex}(t),
\end{align}
where 
\begin{subequations}
\begin{align}
 \hH_{\rm kin} &= -\sum_{\alpha\neq \beta} J_{\alpha,\beta} \hc_\alpha^\dagger \hc_\beta
+ \sum_\alpha D_\alpha \hn_\alpha, \\
 \hH_{\rm int} &= \frac{1}{2}\sum_{\alpha,\beta} V_{\alpha\beta} \hn_\alpha \hn_\beta,
\end{align}
\end{subequations}
and for the phonon part
\begin{equation}
\begin{split}
\hH_{\rm el-ph} &= \sum_{Y,\alpha,\beta} g_{Y,\alpha\beta} (\ha^\dagger_Y + \ha_Y ) \hc^\dagger_\alpha \hc_\beta, \\
 \hH_{\rm ph} &= \sum_{Y} \omega_{Y} \ha^\dagger_Y\ha_Y.
\end{split}
\end{equation}
Here, we assume $V_{\alpha\beta}=V_{\beta\alpha}$  and $V_{\alpha\alpha}=0$.
$\alpha$ and $\beta$ are composite indices consisting of the site-index, orbital-index and spin-index, and 
$Y$ is a composite index consisting of the site-index and the phonon-band index.
The single-particle density matrix is defined as $\hrho_{\al\be}=\hc^\dagger_\be \hc_\al$ and the phonon displacement is $\hat{X}_Y = \ha_Y^\dagger + \ha_Y$.

The MF Hamiltonians for the electrons and phonons become
\begin{align}
\hH^{\rm MF}_{\rm el}(t) &= \hH_{\rm kin} + \hH^{\rm H}(t) + \hH^{\rm F}(t)  + \hH^{\rm MF,el}_{\rm el-ph}(t)+ \hH_{\rm ex}(t),\nonumber\\
\hH^{\rm MF}_{\rm ph}(t) &=  \hH_{\rm ph} + \hH^{\rm MF,ph}_{\rm el-ph}(t),
\end{align}
with 
\begin{equation}
\begin{split}
\hH^{\rm H}(t)  &= \sum_{\alpha,\beta} V_{\alpha\beta}\rho_{\alpha\alpha}(t)\hrho_{\be\be},\\
\hH^{\rm F}(t) &= -\sum_{\al,\be} V_{\al\be} \rho_{\be\al}(t)\hrho_{\al\be},\\ 
\hH^{\rm MF,el}_{\rm el-ph}(t) &= \sum_{Y,\alpha,\beta} g_{Y,\alpha\beta} X_{Y}(t)\hrho_{\beta\alpha}, \\
\hH^{\rm MF,ph}_{\rm el-ph}(t) &=\sum_{Y,\alpha,\beta} g_{Y,\alpha\beta} \hat{X}_Y \rho_{\beta\alpha}(t).
\end{split}
\end{equation}

Now, we consider an excitation from a weak pulse $\hH_{\rm ex}(t) = \delta F^{\rm ex}_{\al_0\be_0}(t) \hc^\dagger_{\al_0} \hc_{\be_0}$.
Then, the response of $\hc^\dagger_{\al} \hc_{\be}$ to $\hc^\dagger_{\al_0} \hc_{\be_0}$ is expressed as 
\begin{align} \label{eq:chi_def}
\chi^R_{\be\al;\be_0\al_0}(t) \equiv  -i \theta(t) \langle [\hrho_{\be\al}(t),\hrho_{\be_0\al_0}(0)]\rangle.
\end{align}
With this, the response to $\hH_{\rm ex}(t)$ becomes 
\begin{align} \label{eq:rho_chi}
\delta \rho_{\be\al}(t) = \int d0 \;\chi^R_{\be\al;\be_0\al_0}(t-t_0) \delta  F^{\rm ex}_{\al_0\be_0}(t_0).
\end{align}
Here, we use the notation $\int d0 = \sum_{\al_0\be_0}\int dt_0$. 
On the other hand, in the linear response regime, we can write 
\begin{align}
\hH^{\rm MF,el}(t) &= \hH^{\rm MF,el}_{\rm eq} + \sum_{\al_0\be_0} \bigl[ \delta F^{\rm MF,el}_{\al_0\be_0}(t) + \delta F^{\rm ex}_{\al_0\be_0}(t) \bigl] \hrho_{\be_0\al_0}, \nonumber\\
\hH^{\rm MF,ph}(t) &= \hH^{\rm MF,ph}_{\rm eq} + \sum_{Y} \delta F_{Y}^{\rm MF,ph} (t) \hat{X}_Y, 
\end{align}
with 
\begin{align}
 \delta F^{\rm MF,el}_{\al_0\be_0}(t) &= \sum_{\al_2\be_2} \Theta^{\rm el}_{\be_0\al_0;\be_2\al_2} \delta \rho_{\be_2\al_2}(t) + \sum_Y g_{Y,\al_0\be_0} \delta X_Y(t),\nonumber\\
  \delta F^{\rm MF,ph}_{Y}(t) & =  \sum_{\al_0\be_0} g_{Y,\al_0\be_0} \delta\rho_{\be_0\al_0}(t) \label{eq:del_meanfield}.
\end{align}
Here, $\hH^{\rm MF,el}_{\rm eq}$ and $\hH^{\rm MF,ph}_{\rm eq}$ indicate the equilibrium MF Hamiltonians and $\delta \mathcal{O}$ indicates the deviation from the equilibrium value of $\mathcal{O}$.
We furthermore introduced  
\begin{align}
 \Theta^{\rm el}_{\be_0\al_0;\be_2\al_2}  = \delta_{\al_0\be_0} \delta_{\al_2\be_2}V_{\be_2\be_0} -\delta_{\al_0\be_2}\delta_{\be_0\al_2}V_{\be_0\al_0}.
\end{align}

Regarding Eq.~\eqref{eq:del_meanfield} as an extra external field to a free system described by $\hH^{\rm MF,el}_{\rm eq}$ and $\hH^{\rm MF,ph}_{\rm eq}$,
we obtain
\begin{subequations} 
\begin{align}
\delta \rho_{\be\al}(t) &= \int d0 \; \chi^R_{0,\be\al;\be_0\al_0}(t-t_0) [\delta  F^{\rm ex}_{\al_0\be_0}(t_0)\label{eq:del_rho2}
+\delta  F^{\rm MF,el}_{\al_0\be_0}(t_0)], \\
\delta X_Y(t) &=  \int dt_0 D^R_{Y,0}(t-t_0)  \delta F^{\rm MF,ph}_{Y}(t), 
\end{align}
\end{subequations}
where $\chi^R_0$ is the susceptibility without updating the mean fields  and 
$D^R_{Y,0}(t) \equiv -i\theta(t)\langle [\hat{X}_Y(t),\hat{X}_Y(0)]\rangle_{\hH_{\rm ph}}$. 
Here, $\langle \cdots \rangle_{\hH}$ denotes the thermal ensemble with respect to the Hamiltonian ${\hH}$.

Using Eq.~\eqref{eq:del_meanfield}, we express Eq.~\eqref{eq:del_rho2} in terms of  $\delta  F^{\rm ex}_{\al_0\be_0}(t_0)$, and compare the results with Eq.~\eqref{eq:rho_chi}.
Thus, we obtain 
\small
\begin{align}
& \chi^R_{\be\al;\be_0\al_0}(t-t_0) = \chi^R_{0,\be\al;\be_0\al_0}(t-t_0)\nonumber\\
& \quad+  \int d2\;d3  \; 
\chi^R_{0,\be\al;\be_3\al_3}(t-t_3)  \Theta_{\be_3\al_3;\be_2\al_2}(t_3-t_2)  \nonumber\\
&\quad \hspace{15mm} \times \chi^R_{\be_2\al_2;\be_0\al_0}(t_2-t_0) \label{eq:RPA_general}
\end{align}
\normalsize
with  $\Theta_{\be_3\al_3;\be_2\al_2}(t_3-t_2) =\Theta_{\be_3\al_3;\be_2\al_2}^{\rm el} \delta(t_3-t_2) + \Theta^{\rm ph}_{\be_3\al_3;\be_2\al_2}(t_3-t_2)$
and 
\begin{align}
\Theta^{\rm ph}_{\be_3\al_3;\be_2\al_2}(t_3-t_2) = \sum_Y g_{Y,\al_3\be_3} D^R_{Y,0}(t_3-t_2) g_{Y,\al_2\be_2}.
\end{align}

We can simplify the expression by  i) rewriting  the response function, ii) assuming translational invariance of the Hamiltonian and the equilibrium state and iii) considering Holstein-type phonons.
In the following, we assume that  $\alpha$ ($\beta$) consist of a site index ``$i$" (``$j$") and the rest ``$a$" (``$b$"), which represents orbitals and/or spins,
and that $Y$ consists of a site index ``$m$" and the band index ``$\nu$". 
We also assume a Holstein-type coupling, $g_{(m,\nu),(i,a)(j,b)} = g_{m,ab}\delta_{m,i}\delta_{i,j}$ and $\omega_{(m,\nu)} = \omega_\nu$.
First, we rewrite the response function as 
\begin{align}
\chi^R_{b_1a_1l_1;b_0a_0l_0}(t;i_1,i_0) \equiv \chi^R_{(i_1+l_1,b_1),(i_1,a_1);(i_0+l_0,b_0)(i_0,a_0)}(t).
\end{align}
We also do the same rewriting for $\hat{\Theta}$.
From ii) and iii) above,  $\chi^R_{b_1a_1l_1;b_0a_0l_0}(t;i_1,i_0) = \chi^R_{b_1a_1l_1;b_0a_0l_0}(t;i_1-i_0)$ and $\Theta_{b_1a_1l_1;b_0a_0l_0}(t;i_1,i_0)=\Theta_{b_1a_1l_1;b_0a_0l_0}(t;i_1-i_0)$.
We then introduce the Fourier transformation as $\chi^R_{b_1 a_1 l_1; b_0a_0l_0} (\omega;{\bf q})\equiv \sum_i \int dt\; \chi^R_{b_1 a_1 l_1; b_0a_0l_0} ( t ;i) e^{i\omega t-i{\bf q}\cdot{\bf r}_i}$. 
Regarding $\chi$ and $\Theta$ as a matrix whose index is $(b,a,l)$, we can write Eq.~\eqref{eq:RPA_general} as 
\small
\begin{align}
\bchi^R(\omega ; {\bf q}) = \bchi^R_0(\omega ; {\bf q}) + \bchi^R_0(\omega ; {\bf q}) \boldsymbol{\Theta}(\omega; {\bf q}) \bchi^R(\omega ; {\bf q}), \label{eq:RPA_general2}
\end{align}
\normalsize
where $ \boldsymbol{\Theta} =  \boldsymbol{\Theta}^{\rm el} +  \boldsymbol{\Theta}^{\rm ph}$ and 
\begin{align}
 \Theta^{\rm el}_{b_3a_3l_3;b_2a_2l_2}(\omega;{\bf q}) = & \delta_{a_3,b_3} \delta_{l_3,0} \delta_{a_2,b_2}\delta_{l_2,0}V_{a_3,a_2}({\bf q})  \nonumber \\
& - \delta_{a_3,b_2}\delta_{b_3,a_2} \delta_{l_3,-l_2}  e^{i{\bf q}\cdot {\bf r}_{l_3}}V_{b_3,a_3}(l_3). \nonumber\\
 \Theta^{\rm ph}_{b_3a_3l_3;b_2a_2l_2}(\omega;{\bf q}) = & \delta_{l_3,0}\delta_{l_2,0}\sum_\nu g_{\nu,a_3b_3} D^R_\nu(\omega) g_{\nu,a_2b_2}\label{eq:theta_general}.
\end{align}

Now the remaining question is the expression of $\chi^R_0$.
Using the Wick theorem we obtain 
\begin{align}
\chi_{0,b_1a_1l_1;b_0a_0l_0}^R&(t;{\bf q}) = -i \theta(t) 
 \frac{1}{N} \sum_{\bf k} e^{i({\bf k-q})\cdot {\bf r}_{l_0}} e^{i{\bf k}\cdot{\bf r}_{l_1}}
\nonumber\\
&\times\Bigl\{ g^<_{0,b_0a_1}(-t;{\bf k-q})g^>_{0,b_1a_0}(t;{\bf k}) \nonumber\\
& -  g^>_{0,b_0a_1}(-t;{\bf k-q})g^<_{0,b_1a_0}(t;{\bf k}) \Bigl\}.
 \end{align}
  Here, $g^<$ and $g^>$ are the lesser and greater Green's function of the MF theory in equilibrium.
They can be written in the following form using the eigenenergies $E_c({\bf k})$ of $\hH^{\rm MF,el}_{\rm eq}$ with $c$ an index of the eigenstates
\begin{align}
g^<_{0,ba}(t;{\bf k}) &=i\langle \hc^\dagger_{{\bf k},a}(0) \hc_{{\bf k},b}(t) \rangle_{\hat{H}^{\rm MF}_{\rm el,eq}} \nonumber\\
&= i\sum_c W^c_{ba}({\bf k}) e^{-iE_c({\bf k})t} f(E_c({\bf k}),T),\nonumber \\
g^>_{0,ba}(t;{\bf k}) &=-i\langle  \hc_{{\bf k},b}(t) \hc^\dagger_{{\bf k},a}(0)\rangle_{\hat{H}^{\rm MF}_{\rm el,eq}} \\
&= -i\sum_c W^c_{ba}({\bf k}) e^{-iE_c({\bf k})t} (1-f(E_c({\bf k}),T)).\nonumber
\end{align}
Here, $W^c_{ba}({\bf k})$ are some coefficients that satisfy the above equations and they can be evaluated by expressing $g_{0,ba}$ in terms of 
the Fermion operators that diagonalize $\hat{H}^{\rm MF}_{\rm el,eq}$.
Using this we finally obtain 
 \begin{align}
& \chi_{0,b_1a_1l_1;b_0a_0l_0}^R(\omega;{\bf q}) = \frac{1}{N} \sum_{\bf k} \sum_{c_1,c_2} e^{i({\bf k-q})\cdot {\bf r}_{l_0}} e^{i{\bf k}\cdot{\bf r}_{l_1}} \nonumber \\
& \times W^{c_1}_{b_0a_1}({\bf k-q}) W_{b_1a_0}^{c_2}({\bf k})
\frac{f(E_{c_1}({\bf k-q}),T) -f(E_{c_2}({\bf k}),T)}{\omega + 0^+ - ( E_{c_2}({\bf k})-E_{c_1}({\bf k-q}))}.
 \end{align} 

Let us comment on a technically important point to solve Eq.~\eqref{eq:RPA_general2}.
In Eq.~\eqref{eq:RPA_general2}, the summation over the site index $l$ runs over the whole system. 
However, in Eq.~\eqref{eq:theta_general}, $\Theta$ is zero if $V({\bf r}_{l_3})$ is zero and $(l_3,l_2)\neq (0,0)$.
We can use this fact to further simplify Eq.~\eqref{eq:RPA_general2}.
To see this, we first introduce a T-matrix as 
\begin{align}
\bT &\equiv \boldsymbol{\Theta} +\boldsymbol{\Theta}\bchi_0\boldsymbol{\Theta}+\boldsymbol{\Theta}\bchi_0\boldsymbol{\Theta}\bchi_0\boldsymbol{\Theta}  + \cdots \nonumber \\
       &=   \boldsymbol{\Theta} +\boldsymbol{\Theta}\bchi\boldsymbol{\Theta}.
\end{align}
With this matrix,
\begin{align}
\bchi^R = \bchi^R_0 + \bchi^R_0 \bT\bchi^R_0. \label{eq:RPA_tmatrix}
\end{align}
Then, we also  introduce a set of "$l$" as $\Lambda \equiv \{l\;  |\text{ at least one component of } V({\bf r}_{l}) \text{ is not zero, or }  {\bf r}_l={\bf 0}\} $.
Then $ \Theta_{b_3a_3l_3;b_2a_2l_2}({\bf q})$ can be potentially finite only when $l_3\in \Lambda $ and $l_2\in \Lambda$.
The T-matrix $\bT$ also possesses the same structure as $ \boldsymbol{\Theta}$.
Now, we introduce new matrices, whose indices are $(b,a,\xi)$ with $\xi \in \Lambda$, for $\bchi_0$,$\bchi$, $\boldsymbol{\Theta}$ and $\bT$ and 
express them as $\underline{\bchi_0}$,$\underline{\bchi}$, $\underline{\boldsymbol{\Theta}}$ and $\underline{\bT}$, respectively. 
($\underline{\bf A}$ is a sub-matrix of ${\bf A}$.) Then, we obtain
\begin{subequations}
\begin{align}
\underline{\bT} &\equiv \underline{\boldsymbol{\Theta}} +\underline{\boldsymbol{\Theta}}\underline{\bchi}_0\underline{\boldsymbol{\Theta}}+\underline{\boldsymbol{\Theta}}\underline{\bchi}_0\underline{\boldsymbol{\Theta}}\underline{\bchi}_0\underline{\boldsymbol{\Theta}}  + \cdots \nonumber \\
       &=   \underline{\boldsymbol{\Theta}} +\underline{\boldsymbol{\Theta}}\underline{\bchi}\underline{\boldsymbol{\Theta}},\\
 \underline{\bchi}^R &= \underline{\bchi}^R_0 + \underline{\bchi}^R_0 \underline{\boldsymbol{T}}\underline{\bchi}^R_0 \nonumber \\
  & = \underline{\bchi}^R_0 + \underline{\bchi}^R_0 \underline{\boldsymbol{\Theta}}\underline{\bchi}^R .
\end{align}
\end{subequations}
Note that for the product of the matrix, the space index $\xi$ runs over $\Lambda$ and not over the whole range, hence the computational cost is reduced.
This is particularly relevant for the case of a local interaction, where $\Lambda = \{{\bf 0}\}$.
In this case, we can focus on $\bchi_{0,b_1a_1 0;b_0a_0 0}^R$ 
and the corresponding RPA expression becomes Eq.~\eqref{eq:RPA_chi_Uloc} (after properly redefining the susceptibility as Eq.~\eqref{eq:chi_munu}).
In order to obtain the full $\bchi^R$, we use Eq.~\eqref{eq:RPA_tmatrix} since the components of $\bT$ other than $\underline{\bT}$ are zero,
\begin{align}
&\chi^R_{b_1a_1l_1;b_0a_0l_0} = \chi^R_{0,b_1a_1l_1;b_0a_0l_0}  \nonumber\\
&+ \sum_{b_2,a_2,b_3,a_3}\sum_{l_2,l_3\in\Lambda}\chi^R_{0,b_1a_1l_1;b_2a_2l_2} T_{b_2a_2l_2;b_3a_3l_3}\chi^R_{0,b_3a_3l_3;b_0a_0l_0}. \label{eq:RPA_tmatrix2}
\end{align}

Finally, we note that the same strategy can be used to obtain the expression for the renormalized Green's functions of phonons.
To evaluate them, we consider $\hH_{\rm ex}$ proportional to the phonon displacement $\hX$ and calculate the evolution of the phonon displacement within the tdMF.
In addition, to go beyond the MF theory, the response functions obtained here can be used to evaluate the feedbacks from the collective motions to the single-particle properties~\cite{Monney2012}.

\section{Optical Conductivity} \label{sec:opt_cond2}
Here, we explain the derivation of Eq.~\eqref{eq:opt_cond} and show the explicit expression of Eq.~\eqref{eq:Chi_opt} in terms of $\chi_{0,b_1a_1l_1;b_0a_0l_0}^R(t;{\bf q})$.
In the linear response regime, the linear components of the external field in Eq.~\eqref{eq:ham_optic} can be written as 
\begin{align}
\hH_{\rm lin}(t) = -{\bf A}(t)\cdot \hat{{\bf J}} - {\bf E}(t)\cdot\hat{{\bf P}}.
\end{align}
In addition, in this regime, the intraband current is expressed as 
\begin{align}
{\bf J}_{\rm intra}(t) =\langle  \hat{\bf J}(t)\rangle  - q^2\sum_{\langle i,j\rangle,a} {\bf r}_{ij}  J_a({\bf r}_{ij})({\bf r}_{ij}\cdot {\bf A}(t)) \rho_{ja,ia,{\rm eq}}.
\end{align}
With these equations and the fact that ${\bf E}(\omega) = i\omega {\bf A}(\omega)$ and ${\bf J}_{\rm inter}(\omega) = -i\omega {\bf P}(\omega)$ 
we obtain Eq.~\eqref{eq:opt_cond}.
The explicit expression for Eq.~\eqref{eq:Chi_opt} becomes
\begin{equation}
\begin{split}
[{\boldsymbol \chi}_{JJ}^R(t)]_{\alpha\beta}  &= -Nq^2\sum_{a_1,a_2}\sum_{l_1,l_2} ({\bf r}_{l_1})_\alpha  ({\bf r}_{l_2})_\beta\\
& \times J_{a_1}(-{\bf r}_{l_1})  J_{a_2}(-{\bf r}_{l_2})  \chi_{a_1 a_1 l_1;a_2 a_2 l_2 }(t;{\bf 0}), \\
[{\boldsymbol \chi}_{JP}^R(t)]_{\alpha\beta}  &= -iNq^2\sum_{a_1,a_2}\sum_{l_1} ({\bf r}_{l_1})_\alpha  ({\bf d}_{a_2})_\beta\\
& \times J_{a_1}(-{\bf r}_{l_1})   \chi_{a_1 a_1 l_1;\bar{a}_2 a_2 0}(t;{\bf 0}),  \\
[{\boldsymbol \chi}_{PJ}^R(t)]_{\alpha\beta}  &= -iNq^2\sum_{a_1,a_2}\sum_{l_2} ({\bf d}_{a_1})_\alpha  ({\bf r}_{l_2})_\beta\\
& \times J_{a_2}(-{\bf r}_{l_2})  \chi_{\bar{a}_1 a_1 0;a_2 a_2 l_2 }(t;{\bf 0}),  \\
[{\boldsymbol \chi}_{PP}^R(t)]_{\alpha\beta}  &= Nq^2\sum_{a_1,a_2} ({\bf d}_{a_1})_\alpha  ({\bf d}_{a_2})_\beta  
 \chi_{\bar{a}_1 a_1 0;\bar{a}_2 a_2 0 }(t;{\bf 0})
\end{split}
\end{equation}

\bibliography{../ref}

\begin{thebibliography}{70}%
\makeatletter
\providecommand \@ifxundefined [1]{%
 \@ifx{#1\undefined}
}%
\providecommand \@ifnum [1]{%
 \ifnum #1\expandafter \@firstoftwo
 \else \expandafter \@secondoftwo
 \fi
}%
\providecommand \@ifx [1]{%
 \ifx #1\expandafter \@firstoftwo
 \else \expandafter \@secondoftwo
 \fi
}%
\providecommand \natexlab [1]{#1}%
\providecommand \enquote  [1]{``#1''}%
\providecommand \bibnamefont  [1]{#1}%
\providecommand \bibfnamefont [1]{#1}%
\providecommand \citenamefont [1]{#1}%
\providecommand \href@noop [0]{\@secondoftwo}%
\providecommand \href [0]{\begingroup \@sanitize@url \@href}%
\providecommand \@href[1]{\@@startlink{#1}\@@href}%
\providecommand \@@href[1]{\endgroup#1\@@endlink}%
\providecommand \@sanitize@url [0]{\catcode `\\12\catcode `\$12\catcode
  `\&12\catcode `\#12\catcode `\^12\catcode `\_12\catcode `\%12\relax}%
\providecommand \@@startlink[1]{}%
\providecommand \@@endlink[0]{}%
\providecommand \url  [0]{\begingroup\@sanitize@url \@url }%
\providecommand \@url [1]{\endgroup\@href {#1}{\urlprefix }}%
\providecommand \urlprefix  [0]{URL }%
\providecommand \Eprint [0]{\href }%
\providecommand \doibase [0]{http://dx.doi.org/}%
\providecommand \selectlanguage [0]{\@gobble}%
\providecommand \bibinfo  [0]{\@secondoftwo}%
\providecommand \bibfield  [0]{\@secondoftwo}%
\providecommand \translation [1]{[#1]}%
\providecommand \BibitemOpen [0]{}%
\providecommand \bibitemStop [0]{}%
\providecommand \bibitemNoStop [0]{.\EOS\space}%
\providecommand \EOS [0]{\spacefactor3000\relax}%
\providecommand \BibitemShut  [1]{\csname bibitem#1\endcsname}%
\let\auto@bib@innerbib\@empty
\bibitem [{\citenamefont {Anderson}(1958)}]{Anderson1958}%
  \BibitemOpen
  \bibfield  {author} {\bibinfo {author} {\bibfnamefont {P.~W.}\ \bibnamefont
  {Anderson}},\ }\href {\doibase 10.1103/PhysRev.112.1900} {\bibfield
  {journal} {\bibinfo  {journal} {Phys. Rev.}\ }\textbf {\bibinfo {volume}
  {112}},\ \bibinfo {pages} {1900} (\bibinfo {year} {1958})}\BibitemShut
  {NoStop}%
\bibitem [{\citenamefont {Anderson}(1963)}]{Anderson1963}%
  \BibitemOpen
  \bibfield  {author} {\bibinfo {author} {\bibfnamefont {P.~W.}\ \bibnamefont
  {Anderson}},\ }\href {\doibase 10.1103/PhysRev.130.439} {\bibfield  {journal}
  {\bibinfo  {journal} {Phys. Rev.}\ }\textbf {\bibinfo {volume} {130}},\
  \bibinfo {pages} {439} (\bibinfo {year} {1963})}\BibitemShut {NoStop}%
\bibitem [{\citenamefont {Littlewood}\ and\ \citenamefont
  {Varma}(1982)}]{littlewood1982}%
  \BibitemOpen
  \bibfield  {author} {\bibinfo {author} {\bibfnamefont {P.~B.}\ \bibnamefont
  {Littlewood}}\ and\ \bibinfo {author} {\bibfnamefont {C.~M.}\ \bibnamefont
  {Varma}},\ }\href {\doibase 10.1103/PhysRevB.26.4883} {\bibfield  {journal}
  {\bibinfo  {journal} {Phys. Rev. B}\ }\textbf {\bibinfo {volume} {26}},\
  \bibinfo {pages} {4883} (\bibinfo {year} {1982})}\BibitemShut {NoStop}%
\bibitem [{\citenamefont {Matsunaga}\ \emph {et~al.}(2013)\citenamefont
  {Matsunaga}, \citenamefont {Hamada}, \citenamefont {Makise}, \citenamefont
  {Uzawa}, \citenamefont {Terai}, \citenamefont {Wang},\ and\ \citenamefont
  {Shimano}}]{Matsunaga2013}%
  \BibitemOpen
  \bibfield  {author} {\bibinfo {author} {\bibfnamefont {R.}~\bibnamefont
  {Matsunaga}}, \bibinfo {author} {\bibfnamefont {Y.~I.}\ \bibnamefont
  {Hamada}}, \bibinfo {author} {\bibfnamefont {K.}~\bibnamefont {Makise}},
  \bibinfo {author} {\bibfnamefont {Y.}~\bibnamefont {Uzawa}}, \bibinfo
  {author} {\bibfnamefont {H.}~\bibnamefont {Terai}}, \bibinfo {author}
  {\bibfnamefont {Z.}~\bibnamefont {Wang}}, \ and\ \bibinfo {author}
  {\bibfnamefont {R.}~\bibnamefont {Shimano}},\ }\href {\doibase
  10.1103/PhysRevLett.111.057002} {\bibfield  {journal} {\bibinfo  {journal}
  {Phys. Rev. Lett.}\ }\textbf {\bibinfo {volume} {111}},\ \bibinfo {pages}
  {057002} (\bibinfo {year} {2013})}\BibitemShut {NoStop}%
\bibitem [{\citenamefont {Matsunaga}\ \emph {et~al.}(2014)\citenamefont
  {Matsunaga}, \citenamefont {Tsuji}, \citenamefont {Fujita}, \citenamefont
  {Sugioka}, \citenamefont {Makise}, \citenamefont {Uzawa}, \citenamefont
  {Terai}, \citenamefont {Wang}, \citenamefont {Aoki},\ and\ \citenamefont
  {Shimano}}]{Matsunaga2014}%
  \BibitemOpen
  \bibfield  {author} {\bibinfo {author} {\bibfnamefont {R.}~\bibnamefont
  {Matsunaga}}, \bibinfo {author} {\bibfnamefont {N.}~\bibnamefont {Tsuji}},
  \bibinfo {author} {\bibfnamefont {H.}~\bibnamefont {Fujita}}, \bibinfo
  {author} {\bibfnamefont {A.}~\bibnamefont {Sugioka}}, \bibinfo {author}
  {\bibfnamefont {K.}~\bibnamefont {Makise}}, \bibinfo {author} {\bibfnamefont
  {Y.}~\bibnamefont {Uzawa}}, \bibinfo {author} {\bibfnamefont
  {H.}~\bibnamefont {Terai}}, \bibinfo {author} {\bibfnamefont
  {Z.}~\bibnamefont {Wang}}, \bibinfo {author} {\bibfnamefont {H.}~\bibnamefont
  {Aoki}}, \ and\ \bibinfo {author} {\bibfnamefont {R.}~\bibnamefont
  {Shimano}},\ }\href {\doibase 10.1126/science.1254697} {\bibfield  {journal}
  {\bibinfo  {journal} {Science}\ }\textbf {\bibinfo {volume} {345}},\ \bibinfo
  {pages} {1145} (\bibinfo {year} {2014})}\BibitemShut {NoStop}%
\bibitem [{\citenamefont {M\'easson}\ \emph {et~al.}(2014)\citenamefont
  {M\'easson}, \citenamefont {Gallais}, \citenamefont {Cazayous}, \citenamefont
  {Clair}, \citenamefont {Rodi\`ere}, \citenamefont {Cario},\ and\
  \citenamefont {Sacuto}}]{Measson2014}%
  \BibitemOpen
  \bibfield  {author} {\bibinfo {author} {\bibfnamefont {M.-A.}\ \bibnamefont
  {M\'easson}}, \bibinfo {author} {\bibfnamefont {Y.}~\bibnamefont {Gallais}},
  \bibinfo {author} {\bibfnamefont {M.}~\bibnamefont {Cazayous}}, \bibinfo
  {author} {\bibfnamefont {B.}~\bibnamefont {Clair}}, \bibinfo {author}
  {\bibfnamefont {P.}~\bibnamefont {Rodi\`ere}}, \bibinfo {author}
  {\bibfnamefont {L.}~\bibnamefont {Cario}}, \ and\ \bibinfo {author}
  {\bibfnamefont {A.}~\bibnamefont {Sacuto}},\ }\href {\doibase
  10.1103/PhysRevB.89.060503} {\bibfield  {journal} {\bibinfo  {journal} {Phys.
  Rev. B}\ }\textbf {\bibinfo {volume} {89}},\ \bibinfo {pages} {060503}
  (\bibinfo {year} {2014})}\BibitemShut {NoStop}%
\bibitem [{\citenamefont {Cea}\ and\ \citenamefont
  {Benfatto}(2014)}]{Benfatto2014}%
  \BibitemOpen
  \bibfield  {author} {\bibinfo {author} {\bibfnamefont {T.}~\bibnamefont
  {Cea}}\ and\ \bibinfo {author} {\bibfnamefont {L.}~\bibnamefont {Benfatto}},\
  }\href {\doibase 10.1103/PhysRevB.90.224515} {\bibfield  {journal} {\bibinfo
  {journal} {Phys. Rev. B}\ }\textbf {\bibinfo {volume} {90}},\ \bibinfo
  {pages} {224515} (\bibinfo {year} {2014})}\BibitemShut {NoStop}%
\bibitem [{\citenamefont {Pekker}\ and\ \citenamefont
  {Varma}(2015)}]{pekker2015}%
  \BibitemOpen
  \bibfield  {author} {\bibinfo {author} {\bibfnamefont {D.}~\bibnamefont
  {Pekker}}\ and\ \bibinfo {author} {\bibfnamefont {C.}~\bibnamefont {Varma}},\
  }\href {\doibase 10.1146/annurev-conmatphys-031214-014350} {\bibfield
  {journal} {\bibinfo  {journal} {Annu. Rev. Condens. Matter Phys.}\ }\textbf
  {\bibinfo {volume} {6}},\ \bibinfo {pages} {269} (\bibinfo {year}
  {2015})}\BibitemShut {NoStop}%
\bibitem [{\citenamefont {Murakami}\ \emph
  {et~al.}(2016{\natexlab{a}})\citenamefont {Murakami}, \citenamefont {Werner},
  \citenamefont {Tsuji},\ and\ \citenamefont {Aoki}}]{Murakami2016}%
  \BibitemOpen
  \bibfield  {author} {\bibinfo {author} {\bibfnamefont {Y.}~\bibnamefont
  {Murakami}}, \bibinfo {author} {\bibfnamefont {P.}~\bibnamefont {Werner}},
  \bibinfo {author} {\bibfnamefont {N.}~\bibnamefont {Tsuji}}, \ and\ \bibinfo
  {author} {\bibfnamefont {H.}~\bibnamefont {Aoki}},\ }\href {\doibase
  10.1103/PhysRevB.93.094509} {\bibfield  {journal} {\bibinfo  {journal} {Phys.
  Rev. B}\ }\textbf {\bibinfo {volume} {93}},\ \bibinfo {pages} {094509}
  (\bibinfo {year} {2016}{\natexlab{a}})}\BibitemShut {NoStop}%
\bibitem [{\citenamefont {Katsumi}\ \emph {et~al.}(2018)\citenamefont
  {Katsumi}, \citenamefont {Tsuji}, \citenamefont {Hamada}, \citenamefont
  {Matsunaga}, \citenamefont {Schneeloch}, \citenamefont {Zhong}, \citenamefont
  {Gu}, \citenamefont {Aoki}, \citenamefont {Gallais},\ and\ \citenamefont
  {Shimano}}]{Katsumi2018}%
  \BibitemOpen
  \bibfield  {author} {\bibinfo {author} {\bibfnamefont {K.}~\bibnamefont
  {Katsumi}}, \bibinfo {author} {\bibfnamefont {N.}~\bibnamefont {Tsuji}},
  \bibinfo {author} {\bibfnamefont {Y.~I.}\ \bibnamefont {Hamada}}, \bibinfo
  {author} {\bibfnamefont {R.}~\bibnamefont {Matsunaga}}, \bibinfo {author}
  {\bibfnamefont {J.}~\bibnamefont {Schneeloch}}, \bibinfo {author}
  {\bibfnamefont {R.~D.}\ \bibnamefont {Zhong}}, \bibinfo {author}
  {\bibfnamefont {G.~D.}\ \bibnamefont {Gu}}, \bibinfo {author} {\bibfnamefont
  {H.}~\bibnamefont {Aoki}}, \bibinfo {author} {\bibfnamefont {Y.}~\bibnamefont
  {Gallais}}, \ and\ \bibinfo {author} {\bibfnamefont {R.}~\bibnamefont
  {Shimano}},\ }\href {\doibase 10.1103/PhysRevLett.120.117001} {\bibfield
  {journal} {\bibinfo  {journal} {Phys. Rev. Lett.}\ }\textbf {\bibinfo
  {volume} {120}},\ \bibinfo {pages} {117001} (\bibinfo {year}
  {2018})}\BibitemShut {NoStop}%
\bibitem [{\citenamefont {Shimano}\ and\ \citenamefont
  {Tsuji}(2020)}]{shimano2020}%
  \BibitemOpen
  \bibfield  {author} {\bibinfo {author} {\bibfnamefont {R.}~\bibnamefont
  {Shimano}}\ and\ \bibinfo {author} {\bibfnamefont {N.}~\bibnamefont
  {Tsuji}},\ }\href {\doibase 10.1146/annurev-conmatphys-031119-050813}
  {\bibfield  {journal} {\bibinfo  {journal} {Annu. Rev. Condens. Matter
  Phys.}\ }\textbf {\bibinfo {volume} {11}},\ \bibinfo {pages} {null} (\bibinfo
  {year} {2020})}\BibitemShut {NoStop}%
\bibitem [{\citenamefont {Schwarz}\ \emph {et~al.}(2020)\citenamefont
  {Schwarz}, \citenamefont {Fauseweh}, \citenamefont {Tsuji}, \citenamefont
  {Cheng}, \citenamefont {Bittner}, \citenamefont {Krull}, \citenamefont
  {Berciu}, \citenamefont {Uhrig}, \citenamefont {Schnyder}, \citenamefont
  {Kaiser},\ and\ \citenamefont {Manske}}]{Schwarz2020}%
  \BibitemOpen
  \bibfield  {author} {\bibinfo {author} {\bibfnamefont {L.}~\bibnamefont
  {Schwarz}}, \bibinfo {author} {\bibfnamefont {B.}~\bibnamefont {Fauseweh}},
  \bibinfo {author} {\bibfnamefont {N.}~\bibnamefont {Tsuji}}, \bibinfo
  {author} {\bibfnamefont {N.}~\bibnamefont {Cheng}}, \bibinfo {author}
  {\bibfnamefont {N.}~\bibnamefont {Bittner}}, \bibinfo {author} {\bibfnamefont
  {H.}~\bibnamefont {Krull}}, \bibinfo {author} {\bibfnamefont
  {M.}~\bibnamefont {Berciu}}, \bibinfo {author} {\bibfnamefont {G.~S.}\
  \bibnamefont {Uhrig}}, \bibinfo {author} {\bibfnamefont {A.~P.}\ \bibnamefont
  {Schnyder}}, \bibinfo {author} {\bibfnamefont {S.}~\bibnamefont {Kaiser}}, \
  and\ \bibinfo {author} {\bibfnamefont {D.}~\bibnamefont {Manske}},\ }\href
  {\doibase 10.1038/s41467-019-13763-5} {\bibfield  {journal} {\bibinfo
  {journal} {Nature Communications}\ }\textbf {\bibinfo {volume} {11}},\
  \bibinfo {pages} {287} (\bibinfo {year} {2020})}\BibitemShut {NoStop}%
\bibitem [{\citenamefont {J\'erome}\ \emph {et~al.}(1967)\citenamefont
  {J\'erome}, \citenamefont {Rice},\ and\ \citenamefont {Kohn}}]{jerome1967}%
  \BibitemOpen
  \bibfield  {author} {\bibinfo {author} {\bibfnamefont {D.}~\bibnamefont
  {J\'erome}}, \bibinfo {author} {\bibfnamefont {T.~M.}\ \bibnamefont {Rice}},
  \ and\ \bibinfo {author} {\bibfnamefont {W.}~\bibnamefont {Kohn}},\ }\href
  {\doibase 10.1103/PhysRev.158.462} {\bibfield  {journal} {\bibinfo  {journal}
  {Phys. Rev.}\ }\textbf {\bibinfo {volume} {158}},\ \bibinfo {pages} {462}
  (\bibinfo {year} {1967})}\BibitemShut {NoStop}%
\bibitem [{\citenamefont {Kohn}(1967)}]{kohn1967}%
  \BibitemOpen
  \bibfield  {author} {\bibinfo {author} {\bibfnamefont {W.}~\bibnamefont
  {Kohn}},\ }\href {\doibase 10.1103/PhysRevLett.19.439} {\bibfield  {journal}
  {\bibinfo  {journal} {Phys. Rev. Lett.}\ }\textbf {\bibinfo {volume} {19}},\
  \bibinfo {pages} {439} (\bibinfo {year} {1967})}\BibitemShut {NoStop}%
\bibitem [{\citenamefont {{Keldysh}}\ and\ \citenamefont
  {{Kozlov}}(1968)}]{keldysh1968}%
  \BibitemOpen
  \bibfield  {author} {\bibinfo {author} {\bibfnamefont {L.~V.}\ \bibnamefont
  {{Keldysh}}}\ and\ \bibinfo {author} {\bibfnamefont {A.~N.}\ \bibnamefont
  {{Kozlov}}},\ }\href@noop {} {\bibfield  {journal} {\bibinfo  {journal} {Sov.
  Phys. JETP}\ }\textbf {\bibinfo {volume} {27}},\ \bibinfo {pages} {521}
  (\bibinfo {year} {1968})}\BibitemShut {NoStop}%
\bibitem [{\citenamefont {Halperin}\ and\ \citenamefont
  {Rice}(1968)}]{halperin1968}%
  \BibitemOpen
  \bibfield  {author} {\bibinfo {author} {\bibfnamefont {B.}~\bibnamefont
  {Halperin}}\ and\ \bibinfo {author} {\bibfnamefont {T.}~\bibnamefont
  {Rice}},\ }\href {\doibase http://dx.doi.org/10.1016/S0081-1947(08)60740-7}
  {\bibfield  {journal} {\bibinfo  {journal} {Solid State Physics}\ }\textbf
  {\bibinfo {volume} {21}},\ \bibinfo {pages} {115 } (\bibinfo {year}
  {1968})}\BibitemShut {NoStop}%
\bibitem [{\citenamefont {HALPERIN}\ and\ \citenamefont
  {RICE}(1968)}]{halperin1968RMP}%
  \BibitemOpen
  \bibfield  {author} {\bibinfo {author} {\bibfnamefont {B.~I.}\ \bibnamefont
  {HALPERIN}}\ and\ \bibinfo {author} {\bibfnamefont {T.~M.}\ \bibnamefont
  {RICE}},\ }\href {\doibase 10.1103/RevModPhys.40.755} {\bibfield  {journal}
  {\bibinfo  {journal} {Rev. Mod. Phys.}\ }\textbf {\bibinfo {volume} {40}},\
  \bibinfo {pages} {755} (\bibinfo {year} {1968})}\BibitemShut {NoStop}%
\bibitem [{\citenamefont {Butov}\ \emph {et~al.}(2002)\citenamefont {Butov},
  \citenamefont {Gossard},\ and\ \citenamefont {Chemla}}]{butov2002}%
  \BibitemOpen
  \bibfield  {author} {\bibinfo {author} {\bibfnamefont {L.}~\bibnamefont
  {Butov}}, \bibinfo {author} {\bibfnamefont {A.}~\bibnamefont {Gossard}}, \
  and\ \bibinfo {author} {\bibfnamefont {D.}~\bibnamefont {Chemla}},\ }\href
  {https://www.nature.com/articles/nature00943} {\bibfield  {journal} {\bibinfo
   {journal} {Nature (London)}\ }\textbf {\bibinfo {volume} {418}},\ \bibinfo
  {pages} {751} (\bibinfo {year} {2002})}\BibitemShut {NoStop}%
\bibitem [{\citenamefont {Eisenstein}\ and\ \citenamefont
  {MacDonald}(2004)}]{eisenstein2004}%
  \BibitemOpen
  \bibfield  {author} {\bibinfo {author} {\bibfnamefont {J.}~\bibnamefont
  {Eisenstein}}\ and\ \bibinfo {author} {\bibfnamefont {A.~H.}\ \bibnamefont
  {MacDonald}},\ }\href {https://www.nature.com/articles/nature03081}
  {\bibfield  {journal} {\bibinfo  {journal} {Nature (London)}\ }\textbf
  {\bibinfo {volume} {432}},\ \bibinfo {pages} {691} (\bibinfo {year}
  {2004})}\BibitemShut {NoStop}%
\bibitem [{\citenamefont {Wang}\ \emph {et~al.}(2019)\citenamefont {Wang},
  \citenamefont {Rhodes}, \citenamefont {Watanabe}, \citenamefont {Taniguchi},
  \citenamefont {Hone}, \citenamefont {Shan},\ and\ \citenamefont
  {Mak}}]{wang2019}%
  \BibitemOpen
  \bibfield  {author} {\bibinfo {author} {\bibfnamefont {Z.}~\bibnamefont
  {Wang}}, \bibinfo {author} {\bibfnamefont {D.~A.}\ \bibnamefont {Rhodes}},
  \bibinfo {author} {\bibfnamefont {K.}~\bibnamefont {Watanabe}}, \bibinfo
  {author} {\bibfnamefont {T.}~\bibnamefont {Taniguchi}}, \bibinfo {author}
  {\bibfnamefont {J.~C.}\ \bibnamefont {Hone}}, \bibinfo {author}
  {\bibfnamefont {J.}~\bibnamefont {Shan}}, \ and\ \bibinfo {author}
  {\bibfnamefont {K.~F.}\ \bibnamefont {Mak}},\ }\href {\doibase
  10.1038/s41586-019-1591-7} {\bibfield  {journal} {\bibinfo  {journal} {Nature
  (London)}\ }\textbf {\bibinfo {volume} {574}},\ \bibinfo {pages} {76}
  (\bibinfo {year} {2019})}\BibitemShut {NoStop}%
\bibitem [{\citenamefont {Nandi}\ \emph {et~al.}(2012)\citenamefont {Nandi},
  \citenamefont {Finck}, \citenamefont {Eisenstein}, \citenamefont {Pfeiffer},\
  and\ \citenamefont {West}}]{nandi2012}%
  \BibitemOpen
  \bibfield  {author} {\bibinfo {author} {\bibfnamefont {D.}~\bibnamefont
  {Nandi}}, \bibinfo {author} {\bibfnamefont {A.}~\bibnamefont {Finck}},
  \bibinfo {author} {\bibfnamefont {J.}~\bibnamefont {Eisenstein}}, \bibinfo
  {author} {\bibfnamefont {L.}~\bibnamefont {Pfeiffer}}, \ and\ \bibinfo
  {author} {\bibfnamefont {K.}~\bibnamefont {West}},\ }\href
  {https://www.nature.com/articles/nature11302} {\bibfield  {journal} {\bibinfo
   {journal} {Nature (London)}\ }\textbf {\bibinfo {volume} {488}},\ \bibinfo
  {pages} {481} (\bibinfo {year} {2012})}\BibitemShut {NoStop}%
\bibitem [{\citenamefont {Cercellier}\ \emph {et~al.}(2007)\citenamefont
  {Cercellier}, \citenamefont {Monney}, \citenamefont {Clerc}, \citenamefont
  {Battaglia}, \citenamefont {Despont}, \citenamefont {Garnier}, \citenamefont
  {Beck}, \citenamefont {Aebi}, \citenamefont {Patthey}, \citenamefont
  {Berger},\ and\ \citenamefont {Forr\'o}}]{cercellier2007}%
  \BibitemOpen
  \bibfield  {author} {\bibinfo {author} {\bibfnamefont {H.}~\bibnamefont
  {Cercellier}}, \bibinfo {author} {\bibfnamefont {C.}~\bibnamefont {Monney}},
  \bibinfo {author} {\bibfnamefont {F.}~\bibnamefont {Clerc}}, \bibinfo
  {author} {\bibfnamefont {C.}~\bibnamefont {Battaglia}}, \bibinfo {author}
  {\bibfnamefont {L.}~\bibnamefont {Despont}}, \bibinfo {author} {\bibfnamefont
  {M.~G.}\ \bibnamefont {Garnier}}, \bibinfo {author} {\bibfnamefont
  {H.}~\bibnamefont {Beck}}, \bibinfo {author} {\bibfnamefont {P.}~\bibnamefont
  {Aebi}}, \bibinfo {author} {\bibfnamefont {L.}~\bibnamefont {Patthey}},
  \bibinfo {author} {\bibfnamefont {H.}~\bibnamefont {Berger}}, \ and\ \bibinfo
  {author} {\bibfnamefont {L.}~\bibnamefont {Forr\'o}},\ }\href {\doibase
  10.1103/PhysRevLett.99.146403} {\bibfield  {journal} {\bibinfo  {journal}
  {Phys. Rev. Lett.}\ }\textbf {\bibinfo {volume} {99}},\ \bibinfo {pages}
  {146403} (\bibinfo {year} {2007})}\BibitemShut {NoStop}%
\bibitem [{\citenamefont {Monney}\ \emph {et~al.}(2011)\citenamefont {Monney},
  \citenamefont {Battaglia}, \citenamefont {Cercellier}, \citenamefont {Aebi},\
  and\ \citenamefont {Beck}}]{Monney2011PRL}%
  \BibitemOpen
  \bibfield  {author} {\bibinfo {author} {\bibfnamefont {C.}~\bibnamefont
  {Monney}}, \bibinfo {author} {\bibfnamefont {C.}~\bibnamefont {Battaglia}},
  \bibinfo {author} {\bibfnamefont {H.}~\bibnamefont {Cercellier}}, \bibinfo
  {author} {\bibfnamefont {P.}~\bibnamefont {Aebi}}, \ and\ \bibinfo {author}
  {\bibfnamefont {H.}~\bibnamefont {Beck}},\ }\href {\doibase
  10.1103/PhysRevLett.106.106404} {\bibfield  {journal} {\bibinfo  {journal}
  {Phys. Rev. Lett.}\ }\textbf {\bibinfo {volume} {106}},\ \bibinfo {pages}
  {106404} (\bibinfo {year} {2011})}\BibitemShut {NoStop}%
\bibitem [{\citenamefont {Monney}\ \emph {et~al.}(2012)\citenamefont {Monney},
  \citenamefont {Monney}, \citenamefont {Aebi},\ and\ \citenamefont
  {Beck}}]{Monney2012}%
  \BibitemOpen
  \bibfield  {author} {\bibinfo {author} {\bibfnamefont {C.}~\bibnamefont
  {Monney}}, \bibinfo {author} {\bibfnamefont {G.}~\bibnamefont {Monney}},
  \bibinfo {author} {\bibfnamefont {P.}~\bibnamefont {Aebi}}, \ and\ \bibinfo
  {author} {\bibfnamefont {H.}~\bibnamefont {Beck}},\ }\href {\doibase
  10.1088/1367-2630/14/7/075026} {\bibfield  {journal} {\bibinfo  {journal}
  {New Journal of Physics}\ }\textbf {\bibinfo {volume} {14}},\ \bibinfo
  {pages} {075026} (\bibinfo {year} {2012})}\BibitemShut {NoStop}%
\bibitem [{\citenamefont {Monney}\ \emph {et~al.}(2015)\citenamefont {Monney},
  \citenamefont {Monney}, \citenamefont {Hildebrand}, \citenamefont {Aebi},\
  and\ \citenamefont {Beck}}]{Monney2015PRL}%
  \BibitemOpen
  \bibfield  {author} {\bibinfo {author} {\bibfnamefont {G.}~\bibnamefont
  {Monney}}, \bibinfo {author} {\bibfnamefont {C.}~\bibnamefont {Monney}},
  \bibinfo {author} {\bibfnamefont {B.}~\bibnamefont {Hildebrand}}, \bibinfo
  {author} {\bibfnamefont {P.}~\bibnamefont {Aebi}}, \ and\ \bibinfo {author}
  {\bibfnamefont {H.}~\bibnamefont {Beck}},\ }\href {\doibase
  10.1103/PhysRevLett.114.086402} {\bibfield  {journal} {\bibinfo  {journal}
  {Phys. Rev. Lett.}\ }\textbf {\bibinfo {volume} {114}},\ \bibinfo {pages}
  {086402} (\bibinfo {year} {2015})}\BibitemShut {NoStop}%
\bibitem [{\citenamefont {Kogar}\ \emph {et~al.}(2017)\citenamefont {Kogar},
  \citenamefont {Rak}, \citenamefont {Vig}, \citenamefont {Husain},
  \citenamefont {Flicker}, \citenamefont {Joe}, \citenamefont {Venema},
  \citenamefont {MacDougall}, \citenamefont {Chiang}, \citenamefont {Fradkin},
  \citenamefont {van Wezel},\ and\ \citenamefont {Abbamonte}}]{kogar2017}%
  \BibitemOpen
  \bibfield  {author} {\bibinfo {author} {\bibfnamefont {A.}~\bibnamefont
  {Kogar}}, \bibinfo {author} {\bibfnamefont {M.~S.}\ \bibnamefont {Rak}},
  \bibinfo {author} {\bibfnamefont {S.}~\bibnamefont {Vig}}, \bibinfo {author}
  {\bibfnamefont {A.~A.}\ \bibnamefont {Husain}}, \bibinfo {author}
  {\bibfnamefont {F.}~\bibnamefont {Flicker}}, \bibinfo {author} {\bibfnamefont
  {Y.~I.}\ \bibnamefont {Joe}}, \bibinfo {author} {\bibfnamefont
  {L.}~\bibnamefont {Venema}}, \bibinfo {author} {\bibfnamefont {G.~J.}\
  \bibnamefont {MacDougall}}, \bibinfo {author} {\bibfnamefont {T.~C.}\
  \bibnamefont {Chiang}}, \bibinfo {author} {\bibfnamefont {E.}~\bibnamefont
  {Fradkin}}, \bibinfo {author} {\bibfnamefont {J.}~\bibnamefont {van Wezel}},
  \ and\ \bibinfo {author} {\bibfnamefont {P.}~\bibnamefont {Abbamonte}},\
  }\href {\doibase 10.1126/science.aam6432} {\bibfield  {journal} {\bibinfo
  {journal} {Science}\ }\textbf {\bibinfo {volume} {358}},\ \bibinfo {pages}
  {1314} (\bibinfo {year} {2017})}\BibitemShut {NoStop}%
\bibitem [{\citenamefont {Kaneko}\ \emph {et~al.}(2018)\citenamefont {Kaneko},
  \citenamefont {Ohta},\ and\ \citenamefont {Yunoki}}]{kaneko2018}%
  \BibitemOpen
  \bibfield  {author} {\bibinfo {author} {\bibfnamefont {T.}~\bibnamefont
  {Kaneko}}, \bibinfo {author} {\bibfnamefont {Y.}~\bibnamefont {Ohta}}, \ and\
  \bibinfo {author} {\bibfnamefont {S.}~\bibnamefont {Yunoki}},\ }\href
  {\doibase 10.1103/PhysRevB.97.155131} {\bibfield  {journal} {\bibinfo
  {journal} {Phys. Rev. B}\ }\textbf {\bibinfo {volume} {97}},\ \bibinfo
  {pages} {155131} (\bibinfo {year} {2018})}\BibitemShut {NoStop}%
\bibitem [{\citenamefont {Chen}\ \emph {et~al.}(2018)\citenamefont {Chen},
  \citenamefont {Singh}, \citenamefont {Lin},\ and\ \citenamefont
  {Pereira}}]{chen2018}%
  \BibitemOpen
  \bibfield  {author} {\bibinfo {author} {\bibfnamefont {C.}~\bibnamefont
  {Chen}}, \bibinfo {author} {\bibfnamefont {B.}~\bibnamefont {Singh}},
  \bibinfo {author} {\bibfnamefont {H.}~\bibnamefont {Lin}}, \ and\ \bibinfo
  {author} {\bibfnamefont {V.~M.}\ \bibnamefont {Pereira}},\ }\href {\doibase
  10.1103/PhysRevLett.121.226602} {\bibfield  {journal} {\bibinfo  {journal}
  {Phys. Rev. Lett.}\ }\textbf {\bibinfo {volume} {121}},\ \bibinfo {pages}
  {226602} (\bibinfo {year} {2018})}\BibitemShut {NoStop}%
\bibitem [{\citenamefont {Chernikov}\ \emph {et~al.}(2014)\citenamefont
  {Chernikov}, \citenamefont {Berkelbach}, \citenamefont {Hill}, \citenamefont
  {Rigosi}, \citenamefont {Li}, \citenamefont {Aslan}, \citenamefont
  {Reichman}, \citenamefont {Hybertsen},\ and\ \citenamefont
  {Heinz}}]{chernikov14}%
  \BibitemOpen
  \bibfield  {author} {\bibinfo {author} {\bibfnamefont {A.}~\bibnamefont
  {Chernikov}}, \bibinfo {author} {\bibfnamefont {T.~C.}\ \bibnamefont
  {Berkelbach}}, \bibinfo {author} {\bibfnamefont {H.~M.}\ \bibnamefont
  {Hill}}, \bibinfo {author} {\bibfnamefont {A.}~\bibnamefont {Rigosi}},
  \bibinfo {author} {\bibfnamefont {Y.}~\bibnamefont {Li}}, \bibinfo {author}
  {\bibfnamefont {O.~B.}\ \bibnamefont {Aslan}}, \bibinfo {author}
  {\bibfnamefont {D.~R.}\ \bibnamefont {Reichman}}, \bibinfo {author}
  {\bibfnamefont {M.~S.}\ \bibnamefont {Hybertsen}}, \ and\ \bibinfo {author}
  {\bibfnamefont {T.~F.}\ \bibnamefont {Heinz}},\ }\href {\doibase
  10.1103/PhysRevLett.113.076802} {\bibfield  {journal} {\bibinfo  {journal}
  {Phys. Rev. Lett.}\ }\textbf {\bibinfo {volume} {113}},\ \bibinfo {pages}
  {076802} (\bibinfo {year} {2014})}\BibitemShut {NoStop}%
\bibitem [{\citenamefont {He}\ \emph {et~al.}(2014)\citenamefont {He},
  \citenamefont {Kumar}, \citenamefont {Zhao}, \citenamefont {Wang},
  \citenamefont {Mak}, \citenamefont {Zhao},\ and\ \citenamefont
  {Shan}}]{he2014}%
  \BibitemOpen
  \bibfield  {author} {\bibinfo {author} {\bibfnamefont {K.}~\bibnamefont
  {He}}, \bibinfo {author} {\bibfnamefont {N.}~\bibnamefont {Kumar}}, \bibinfo
  {author} {\bibfnamefont {L.}~\bibnamefont {Zhao}}, \bibinfo {author}
  {\bibfnamefont {Z.}~\bibnamefont {Wang}}, \bibinfo {author} {\bibfnamefont
  {K.~F.}\ \bibnamefont {Mak}}, \bibinfo {author} {\bibfnamefont
  {H.}~\bibnamefont {Zhao}}, \ and\ \bibinfo {author} {\bibfnamefont
  {J.}~\bibnamefont {Shan}},\ }\href {\doibase 10.1103/PhysRevLett.113.026803}
  {\bibfield  {journal} {\bibinfo  {journal} {Phys. Rev. Lett.}\ }\textbf
  {\bibinfo {volume} {113}},\ \bibinfo {pages} {026803} (\bibinfo {year}
  {2014})}\BibitemShut {NoStop}%
\bibitem [{\citenamefont {Ugeda}\ \emph {et~al.}(2014)\citenamefont {Ugeda},
  \citenamefont {Bradley}, \citenamefont {Shi}, \citenamefont {da~Jornada},
  \citenamefont {Zhang}, \citenamefont {Qiu}, \citenamefont {Ruan},
  \citenamefont {Mo}, \citenamefont {Hussain}, \citenamefont {Shen},
  \citenamefont {Wang}, \citenamefont {Louie},\ and\ \citenamefont
  {Crommie}}]{ugeda2014}%
  \BibitemOpen
  \bibfield  {author} {\bibinfo {author} {\bibfnamefont {M.~M.}\ \bibnamefont
  {Ugeda}}, \bibinfo {author} {\bibfnamefont {A.~J.}\ \bibnamefont {Bradley}},
  \bibinfo {author} {\bibfnamefont {S.-F.}\ \bibnamefont {Shi}}, \bibinfo
  {author} {\bibfnamefont {F.~H.}\ \bibnamefont {da~Jornada}}, \bibinfo
  {author} {\bibfnamefont {Y.}~\bibnamefont {Zhang}}, \bibinfo {author}
  {\bibfnamefont {D.~Y.}\ \bibnamefont {Qiu}}, \bibinfo {author} {\bibfnamefont
  {W.}~\bibnamefont {Ruan}}, \bibinfo {author} {\bibfnamefont {S.-K.}\
  \bibnamefont {Mo}}, \bibinfo {author} {\bibfnamefont {Z.}~\bibnamefont
  {Hussain}}, \bibinfo {author} {\bibfnamefont {Z.-X.}\ \bibnamefont {Shen}},
  \bibinfo {author} {\bibfnamefont {F.}~\bibnamefont {Wang}}, \bibinfo {author}
  {\bibfnamefont {S.~G.}\ \bibnamefont {Louie}}, \ and\ \bibinfo {author}
  {\bibfnamefont {M.~F.}\ \bibnamefont {Crommie}},\ }\href {\doibase
  10.1038/nmat4061} {\bibfield  {journal} {\bibinfo  {journal} {Nat. Mater.}\
  }\textbf {\bibinfo {volume} {13}},\ \bibinfo {pages} {1091} (\bibinfo {year}
  {2014})}\BibitemShut {NoStop}%
\bibitem [{\citenamefont {Mueller}\ and\ \citenamefont
  {Malic}(2018)}]{mueller2018}%
  \BibitemOpen
  \bibfield  {author} {\bibinfo {author} {\bibfnamefont {T.}~\bibnamefont
  {Mueller}}\ and\ \bibinfo {author} {\bibfnamefont {E.}~\bibnamefont
  {Malic}},\ }\href {\doibase 10.1038/s41699-018-0074-2} {\bibfield  {journal}
  {\bibinfo  {journal} {npj 2D Mater. Appl.}\ }\textbf {\bibinfo {volume}
  {2}},\ \bibinfo {pages} {29} (\bibinfo {year} {2018})}\BibitemShut {NoStop}%
\bibitem [{\citenamefont {Sunshine}\ and\ \citenamefont
  {Ibers}(1985)}]{sunshine1985}%
  \BibitemOpen
  \bibfield  {author} {\bibinfo {author} {\bibfnamefont {S.~A.}\ \bibnamefont
  {Sunshine}}\ and\ \bibinfo {author} {\bibfnamefont {J.~A.}\ \bibnamefont
  {Ibers}},\ }\href {\doibase 10.1021/ic00216a027} {\bibfield  {journal}
  {\bibinfo  {journal} {Inorg. Chem.}\ }\textbf {\bibinfo {volume} {24}},\
  \bibinfo {pages} {3611} (\bibinfo {year} {1985})}\BibitemShut {NoStop}%
\bibitem [{\citenamefont {Di~Salvo}\ \emph {et~al.}(1986)\citenamefont
  {Di~Salvo}, \citenamefont {Chen}, \citenamefont {Fleming}, \citenamefont
  {Waszczak}, \citenamefont {Dunn}, \citenamefont {Sunshine},\ and\
  \citenamefont {Ibers}}]{disalvo1986}%
  \BibitemOpen
  \bibfield  {author} {\bibinfo {author} {\bibfnamefont {F.}~\bibnamefont
  {Di~Salvo}}, \bibinfo {author} {\bibfnamefont {C.}~\bibnamefont {Chen}},
  \bibinfo {author} {\bibfnamefont {R.}~\bibnamefont {Fleming}}, \bibinfo
  {author} {\bibfnamefont {J.}~\bibnamefont {Waszczak}}, \bibinfo {author}
  {\bibfnamefont {R.}~\bibnamefont {Dunn}}, \bibinfo {author} {\bibfnamefont
  {S.}~\bibnamefont {Sunshine}}, \ and\ \bibinfo {author} {\bibfnamefont
  {J.~A.}\ \bibnamefont {Ibers}},\ }\href
  {https://www.sciencedirect.com/science/article/abs/pii/002250888690216X}
  {\bibfield  {journal} {\bibinfo  {journal} {J. Less-Common Met.}\ }\textbf
  {\bibinfo {volume} {116}},\ \bibinfo {pages} {51} (\bibinfo {year}
  {1986})}\BibitemShut {NoStop}%
\bibitem [{\citenamefont {Wakisaka}\ \emph {et~al.}(2009)\citenamefont
  {Wakisaka}, \citenamefont {Sudayama}, \citenamefont {Takubo}, \citenamefont
  {Mizokawa}, \citenamefont {Arita}, \citenamefont {Namatame}, \citenamefont
  {Taniguchi}, \citenamefont {Katayama}, \citenamefont {Nohara},\ and\
  \citenamefont {Takagi}}]{wakisaka2009}%
  \BibitemOpen
  \bibfield  {author} {\bibinfo {author} {\bibfnamefont {Y.}~\bibnamefont
  {Wakisaka}}, \bibinfo {author} {\bibfnamefont {T.}~\bibnamefont {Sudayama}},
  \bibinfo {author} {\bibfnamefont {K.}~\bibnamefont {Takubo}}, \bibinfo
  {author} {\bibfnamefont {T.}~\bibnamefont {Mizokawa}}, \bibinfo {author}
  {\bibfnamefont {M.}~\bibnamefont {Arita}}, \bibinfo {author} {\bibfnamefont
  {H.}~\bibnamefont {Namatame}}, \bibinfo {author} {\bibfnamefont
  {M.}~\bibnamefont {Taniguchi}}, \bibinfo {author} {\bibfnamefont
  {N.}~\bibnamefont {Katayama}}, \bibinfo {author} {\bibfnamefont
  {M.}~\bibnamefont {Nohara}}, \ and\ \bibinfo {author} {\bibfnamefont
  {H.}~\bibnamefont {Takagi}},\ }\href {\doibase
  10.1103/PhysRevLett.103.026402} {\bibfield  {journal} {\bibinfo  {journal}
  {Phys. Rev. Lett.}\ }\textbf {\bibinfo {volume} {103}},\ \bibinfo {pages}
  {026402} (\bibinfo {year} {2009})}\BibitemShut {NoStop}%
\bibitem [{\citenamefont {Kaneko}\ \emph
  {et~al.}(2013{\natexlab{a}})\citenamefont {Kaneko}, \citenamefont {Toriyama},
  \citenamefont {Konishi},\ and\ \citenamefont {Ohta}}]{kaneko2013}%
  \BibitemOpen
  \bibfield  {author} {\bibinfo {author} {\bibfnamefont {T.}~\bibnamefont
  {Kaneko}}, \bibinfo {author} {\bibfnamefont {T.}~\bibnamefont {Toriyama}},
  \bibinfo {author} {\bibfnamefont {T.}~\bibnamefont {Konishi}}, \ and\
  \bibinfo {author} {\bibfnamefont {Y.}~\bibnamefont {Ohta}},\ }\href {\doibase
  10.1103/PhysRevB.87.035121} {\bibfield  {journal} {\bibinfo  {journal} {Phys.
  Rev. B}\ }\textbf {\bibinfo {volume} {87}},\ \bibinfo {pages} {035121}
  (\bibinfo {year} {2013}{\natexlab{a}})}\BibitemShut {NoStop}%
\bibitem [{\citenamefont {Kaneko}\ \emph
  {et~al.}(2013{\natexlab{b}})\citenamefont {Kaneko}, \citenamefont {Toriyama},
  \citenamefont {Konishi},\ and\ \citenamefont {Ohta}}]{kaneko2013e}%
  \BibitemOpen
  \bibfield  {author} {\bibinfo {author} {\bibfnamefont {T.}~\bibnamefont
  {Kaneko}}, \bibinfo {author} {\bibfnamefont {T.}~\bibnamefont {Toriyama}},
  \bibinfo {author} {\bibfnamefont {T.}~\bibnamefont {Konishi}}, \ and\
  \bibinfo {author} {\bibfnamefont {Y.}~\bibnamefont {Ohta}},\ }\href {\doibase
  10.1103/PhysRevB.87.199902} {\bibfield  {journal} {\bibinfo  {journal} {Phys.
  Rev. B}\ }\textbf {\bibinfo {volume} {87}},\ \bibinfo {pages} {199902}
  (\bibinfo {year} {2013}{\natexlab{b}})}\BibitemShut {NoStop}%
\bibitem [{\citenamefont {Seki}\ \emph {et~al.}(2014)\citenamefont {Seki},
  \citenamefont {Wakisaka}, \citenamefont {Kaneko}, \citenamefont {Toriyama},
  \citenamefont {Konishi}, \citenamefont {Sudayama}, \citenamefont {Saini},
  \citenamefont {Arita}, \citenamefont {Namatame}, \citenamefont {Taniguchi},
  \citenamefont {Katayama}, \citenamefont {Nohara}, \citenamefont {Takagi},
  \citenamefont {Mizokawa},\ and\ \citenamefont {Ohta}}]{seki2014}%
  \BibitemOpen
  \bibfield  {author} {\bibinfo {author} {\bibfnamefont {K.}~\bibnamefont
  {Seki}}, \bibinfo {author} {\bibfnamefont {Y.}~\bibnamefont {Wakisaka}},
  \bibinfo {author} {\bibfnamefont {T.}~\bibnamefont {Kaneko}}, \bibinfo
  {author} {\bibfnamefont {T.}~\bibnamefont {Toriyama}}, \bibinfo {author}
  {\bibfnamefont {T.}~\bibnamefont {Konishi}}, \bibinfo {author} {\bibfnamefont
  {T.}~\bibnamefont {Sudayama}}, \bibinfo {author} {\bibfnamefont {N.~L.}\
  \bibnamefont {Saini}}, \bibinfo {author} {\bibfnamefont {M.}~\bibnamefont
  {Arita}}, \bibinfo {author} {\bibfnamefont {H.}~\bibnamefont {Namatame}},
  \bibinfo {author} {\bibfnamefont {M.}~\bibnamefont {Taniguchi}}, \bibinfo
  {author} {\bibfnamefont {N.}~\bibnamefont {Katayama}}, \bibinfo {author}
  {\bibfnamefont {M.}~\bibnamefont {Nohara}}, \bibinfo {author} {\bibfnamefont
  {H.}~\bibnamefont {Takagi}}, \bibinfo {author} {\bibfnamefont
  {T.}~\bibnamefont {Mizokawa}}, \ and\ \bibinfo {author} {\bibfnamefont
  {Y.}~\bibnamefont {Ohta}},\ }\href {\doibase 10.1103/PhysRevB.90.155116}
  {\bibfield  {journal} {\bibinfo  {journal} {Phys. Rev. B}\ }\textbf {\bibinfo
  {volume} {90}},\ \bibinfo {pages} {155116} (\bibinfo {year}
  {2014})}\BibitemShut {NoStop}%
\bibitem [{\citenamefont {Mor}\ \emph {et~al.}(2017)\citenamefont {Mor},
  \citenamefont {Herzog}, \citenamefont {Gole\v{z}}, \citenamefont {Werner},
  \citenamefont {Eckstein}, \citenamefont {Katayama}, \citenamefont {Nohara},
  \citenamefont {Takagi}, \citenamefont {Mizokawa}, \citenamefont {Monney},\
  and\ \citenamefont {St\"ahler}}]{mor2017}%
  \BibitemOpen
  \bibfield  {author} {\bibinfo {author} {\bibfnamefont {S.}~\bibnamefont
  {Mor}}, \bibinfo {author} {\bibfnamefont {M.}~\bibnamefont {Herzog}},
  \bibinfo {author} {\bibfnamefont {D.}~\bibnamefont {Gole\v{z}}}, \bibinfo
  {author} {\bibfnamefont {P.}~\bibnamefont {Werner}}, \bibinfo {author}
  {\bibfnamefont {M.}~\bibnamefont {Eckstein}}, \bibinfo {author}
  {\bibfnamefont {N.}~\bibnamefont {Katayama}}, \bibinfo {author}
  {\bibfnamefont {M.}~\bibnamefont {Nohara}}, \bibinfo {author} {\bibfnamefont
  {H.}~\bibnamefont {Takagi}}, \bibinfo {author} {\bibfnamefont
  {T.}~\bibnamefont {Mizokawa}}, \bibinfo {author} {\bibfnamefont
  {C.}~\bibnamefont {Monney}}, \ and\ \bibinfo {author} {\bibfnamefont
  {J.}~\bibnamefont {St\"ahler}},\ }\href {\doibase
  10.1103/PhysRevLett.119.086401} {\bibfield  {journal} {\bibinfo  {journal}
  {Phys. Rev. Lett.}\ }\textbf {\bibinfo {volume} {119}},\ \bibinfo {pages}
  {086401} (\bibinfo {year} {2017})}\BibitemShut {NoStop}%
\bibitem [{\citenamefont {Okazaki}\ \emph {et~al.}(2018)\citenamefont
  {Okazaki}, \citenamefont {Ogawa}, \citenamefont {Suzuki}, \citenamefont
  {Yamamoto}, \citenamefont {Someya}, \citenamefont {Michimae}, \citenamefont
  {Watanabe}, \citenamefont {Lu}, \citenamefont {Nohara}, \citenamefont
  {Takagi}, \citenamefont {Katayama}, \citenamefont {Sawa}, \citenamefont
  {Fujisawa}, \citenamefont {Kanai}, \citenamefont {Ishii}, \citenamefont
  {Itatani}, \citenamefont {Mizokawa},\ and\ \citenamefont
  {Shin}}]{okazaki2018}%
  \BibitemOpen
  \bibfield  {author} {\bibinfo {author} {\bibfnamefont {K.}~\bibnamefont
  {Okazaki}}, \bibinfo {author} {\bibfnamefont {Y.}~\bibnamefont {Ogawa}},
  \bibinfo {author} {\bibfnamefont {T.}~\bibnamefont {Suzuki}}, \bibinfo
  {author} {\bibfnamefont {T.}~\bibnamefont {Yamamoto}}, \bibinfo {author}
  {\bibfnamefont {T.}~\bibnamefont {Someya}}, \bibinfo {author} {\bibfnamefont
  {S.}~\bibnamefont {Michimae}}, \bibinfo {author} {\bibfnamefont
  {M.}~\bibnamefont {Watanabe}}, \bibinfo {author} {\bibfnamefont
  {Y.}~\bibnamefont {Lu}}, \bibinfo {author} {\bibfnamefont {M.}~\bibnamefont
  {Nohara}}, \bibinfo {author} {\bibfnamefont {H.}~\bibnamefont {Takagi}},
  \bibinfo {author} {\bibfnamefont {N.}~\bibnamefont {Katayama}}, \bibinfo
  {author} {\bibfnamefont {H.}~\bibnamefont {Sawa}}, \bibinfo {author}
  {\bibfnamefont {M.}~\bibnamefont {Fujisawa}}, \bibinfo {author}
  {\bibfnamefont {T.}~\bibnamefont {Kanai}}, \bibinfo {author} {\bibfnamefont
  {N.}~\bibnamefont {Ishii}}, \bibinfo {author} {\bibfnamefont
  {J.}~\bibnamefont {Itatani}}, \bibinfo {author} {\bibfnamefont
  {T.}~\bibnamefont {Mizokawa}}, \ and\ \bibinfo {author} {\bibfnamefont
  {S.}~\bibnamefont {Shin}},\ }\href {\doibase 10.1038/s41467-018-06801-1}
  {\bibfield  {journal} {\bibinfo  {journal} {Nat. Commun.}\ }\textbf {\bibinfo
  {volume} {9}},\ \bibinfo {pages} {4322} (\bibinfo {year} {2018})}\BibitemShut
  {NoStop}%
\bibitem [{\citenamefont {Lu}\ \emph {et~al.}(2017)\citenamefont {Lu},
  \citenamefont {Kono}, \citenamefont {Larkin}, \citenamefont {Rost},
  \citenamefont {Takayama}, \citenamefont {Boris}, \citenamefont {Keimer},\
  and\ \citenamefont {Takagi}}]{lu2017}%
  \BibitemOpen
  \bibfield  {author} {\bibinfo {author} {\bibfnamefont {Y.~F.}\ \bibnamefont
  {Lu}}, \bibinfo {author} {\bibfnamefont {H.}~\bibnamefont {Kono}}, \bibinfo
  {author} {\bibfnamefont {T.~I.}\ \bibnamefont {Larkin}}, \bibinfo {author}
  {\bibfnamefont {A.~W.}\ \bibnamefont {Rost}}, \bibinfo {author}
  {\bibfnamefont {T.}~\bibnamefont {Takayama}}, \bibinfo {author}
  {\bibfnamefont {A.~V.}\ \bibnamefont {Boris}}, \bibinfo {author}
  {\bibfnamefont {B.}~\bibnamefont {Keimer}}, \ and\ \bibinfo {author}
  {\bibfnamefont {H.}~\bibnamefont {Takagi}},\ }\href
  {http://dx.doi.org/10.1038/ncomms14408} {\bibfield  {journal} {\bibinfo
  {journal} {Nat. Commun.}\ }\textbf {\bibinfo {volume} {8}},\ \bibinfo {pages}
  {14408} (\bibinfo {year} {2017})}\BibitemShut {NoStop}%
\bibitem [{\citenamefont {Nakano}\ \emph {et~al.}(2019)\citenamefont {Nakano},
  \citenamefont {Nagai}, \citenamefont {Katayama}, \citenamefont {Sawa},
  \citenamefont {Taniguchi},\ and\ \citenamefont {Terasaki}}]{nakano2019}%
  \BibitemOpen
  \bibfield  {author} {\bibinfo {author} {\bibfnamefont {A.}~\bibnamefont
  {Nakano}}, \bibinfo {author} {\bibfnamefont {T.}~\bibnamefont {Nagai}},
  \bibinfo {author} {\bibfnamefont {N.}~\bibnamefont {Katayama}}, \bibinfo
  {author} {\bibfnamefont {H.}~\bibnamefont {Sawa}}, \bibinfo {author}
  {\bibfnamefont {H.}~\bibnamefont {Taniguchi}}, \ and\ \bibinfo {author}
  {\bibfnamefont {I.}~\bibnamefont {Terasaki}},\ }\href {\doibase
  10.7566/JPSJ.88.113706} {\bibfield  {journal} {\bibinfo  {journal} {J. Phys.
  Soc. Jpn.}\ }\textbf {\bibinfo {volume} {88}},\ \bibinfo {pages} {113706}
  (\bibinfo {year} {2019})}\BibitemShut {NoStop}%
\bibitem [{\citenamefont {Werdehausen}\ \emph {et~al.}(2018)\citenamefont
  {Werdehausen}, \citenamefont {Takayama}, \citenamefont {H{\"o}ppner},
  \citenamefont {Albrecht}, \citenamefont {Rost}, \citenamefont {Lu},
  \citenamefont {Manske}, \citenamefont {Takagi},\ and\ \citenamefont
  {Kaiser}}]{werdehausen2018}%
  \BibitemOpen
  \bibfield  {author} {\bibinfo {author} {\bibfnamefont {D.}~\bibnamefont
  {Werdehausen}}, \bibinfo {author} {\bibfnamefont {T.}~\bibnamefont
  {Takayama}}, \bibinfo {author} {\bibfnamefont {M.}~\bibnamefont
  {H{\"o}ppner}}, \bibinfo {author} {\bibfnamefont {G.}~\bibnamefont
  {Albrecht}}, \bibinfo {author} {\bibfnamefont {A.~W.}\ \bibnamefont {Rost}},
  \bibinfo {author} {\bibfnamefont {Y.}~\bibnamefont {Lu}}, \bibinfo {author}
  {\bibfnamefont {D.}~\bibnamefont {Manske}}, \bibinfo {author} {\bibfnamefont
  {H.}~\bibnamefont {Takagi}}, \ and\ \bibinfo {author} {\bibfnamefont
  {S.}~\bibnamefont {Kaiser}},\ }\href
  {http://advances.sciencemag.org/content/4/3/eaap8652} {\bibfield  {journal}
  {\bibinfo  {journal} {Sci. Adv.}\ }\textbf {\bibinfo {volume} {4}},\ \bibinfo
  {pages} {eaap8652} (\bibinfo {year} {2018})}\BibitemShut {NoStop}%
\bibitem [{\citenamefont {Nakano}\ \emph {et~al.}(2018)\citenamefont {Nakano},
  \citenamefont {Hasegawa}, \citenamefont {Tamura}, \citenamefont {Katayama},
  \citenamefont {Tsutsui},\ and\ \citenamefont {Sawa}}]{nakano2018}%
  \BibitemOpen
  \bibfield  {author} {\bibinfo {author} {\bibfnamefont {A.}~\bibnamefont
  {Nakano}}, \bibinfo {author} {\bibfnamefont {T.}~\bibnamefont {Hasegawa}},
  \bibinfo {author} {\bibfnamefont {S.}~\bibnamefont {Tamura}}, \bibinfo
  {author} {\bibfnamefont {N.}~\bibnamefont {Katayama}}, \bibinfo {author}
  {\bibfnamefont {S.}~\bibnamefont {Tsutsui}}, \ and\ \bibinfo {author}
  {\bibfnamefont {H.}~\bibnamefont {Sawa}},\ }\href {\doibase
  10.1103/PhysRevB.98.045139} {\bibfield  {journal} {\bibinfo  {journal} {Phys.
  Rev. B}\ }\textbf {\bibinfo {volume} {98}},\ \bibinfo {pages} {045139}
  (\bibinfo {year} {2018})}\BibitemShut {NoStop}%
\bibitem [{\citenamefont {Larkin}\ \emph {et~al.}(2018)\citenamefont {Larkin},
  \citenamefont {Dawson}, \citenamefont {H\"oppner}, \citenamefont {Takayama},
  \citenamefont {Isobe}, \citenamefont {Mathis}, \citenamefont {Takagi},
  \citenamefont {Keimer},\ and\ \citenamefont {Boris}}]{larkin2018}%
  \BibitemOpen
  \bibfield  {author} {\bibinfo {author} {\bibfnamefont {T.~I.}\ \bibnamefont
  {Larkin}}, \bibinfo {author} {\bibfnamefont {R.~D.}\ \bibnamefont {Dawson}},
  \bibinfo {author} {\bibfnamefont {M.}~\bibnamefont {H\"oppner}}, \bibinfo
  {author} {\bibfnamefont {T.}~\bibnamefont {Takayama}}, \bibinfo {author}
  {\bibfnamefont {M.}~\bibnamefont {Isobe}}, \bibinfo {author} {\bibfnamefont
  {Y.-L.}\ \bibnamefont {Mathis}}, \bibinfo {author} {\bibfnamefont
  {H.}~\bibnamefont {Takagi}}, \bibinfo {author} {\bibfnamefont
  {B.}~\bibnamefont {Keimer}}, \ and\ \bibinfo {author} {\bibfnamefont {A.~V.}\
  \bibnamefont {Boris}},\ }\href {\doibase 10.1103/PhysRevB.98.125113}
  {\bibfield  {journal} {\bibinfo  {journal} {Phys. Rev. B}\ }\textbf {\bibinfo
  {volume} {98}},\ \bibinfo {pages} {125113} (\bibinfo {year}
  {2018})}\BibitemShut {NoStop}%
\bibitem [{\citenamefont {Zenker}\ \emph {et~al.}(2014)\citenamefont {Zenker},
  \citenamefont {Fehske},\ and\ \citenamefont {Beck}}]{zenker2014}%
  \BibitemOpen
  \bibfield  {author} {\bibinfo {author} {\bibfnamefont {B.}~\bibnamefont
  {Zenker}}, \bibinfo {author} {\bibfnamefont {H.}~\bibnamefont {Fehske}}, \
  and\ \bibinfo {author} {\bibfnamefont {H.}~\bibnamefont {Beck}},\ }\href
  {\doibase 10.1103/PhysRevB.90.195118} {\bibfield  {journal} {\bibinfo
  {journal} {Phys. Rev. B}\ }\textbf {\bibinfo {volume} {90}},\ \bibinfo
  {pages} {195118} (\bibinfo {year} {2014})}\BibitemShut {NoStop}%
\bibitem [{\citenamefont {Murakami}\ \emph {et~al.}(2017)\citenamefont
  {Murakami}, \citenamefont {Gole\ifmmode~\check{z}\else \v{z}\fi{}},
  \citenamefont {Eckstein},\ and\ \citenamefont {Werner}}]{murakami2017}%
  \BibitemOpen
  \bibfield  {author} {\bibinfo {author} {\bibfnamefont {Y.}~\bibnamefont
  {Murakami}}, \bibinfo {author} {\bibfnamefont {D.}~\bibnamefont
  {Gole\ifmmode~\check{z}\else \v{z}\fi{}}}, \bibinfo {author} {\bibfnamefont
  {M.}~\bibnamefont {Eckstein}}, \ and\ \bibinfo {author} {\bibfnamefont
  {P.}~\bibnamefont {Werner}},\ }\href {\doibase
  10.1103/PhysRevLett.119.247601} {\bibfield  {journal} {\bibinfo  {journal}
  {Phys. Rev. Lett.}\ }\textbf {\bibinfo {volume} {119}},\ \bibinfo {pages}
  {247601} (\bibinfo {year} {2017})}\BibitemShut {NoStop}%
\bibitem [{\citenamefont {{Mazza}}\ \emph {et~al.}()\citenamefont {{Mazza}},
  \citenamefont {{R{\"o}sner}}, \citenamefont {{Windg{\"a}tter}}, \citenamefont
  {{Latini}}, \citenamefont {{H{\"u}bener}}, \citenamefont {{Millis}},
  \citenamefont {{Rubio}},\ and\ \citenamefont {{Geroges}}}]{mazza2019}%
  \BibitemOpen
  \bibfield  {author} {\bibinfo {author} {\bibfnamefont {G.}~\bibnamefont
  {{Mazza}}}, \bibinfo {author} {\bibfnamefont {M.}~\bibnamefont
  {{R{\"o}sner}}}, \bibinfo {author} {\bibfnamefont {L.}~\bibnamefont
  {{Windg{\"a}tter}}}, \bibinfo {author} {\bibfnamefont {S.}~\bibnamefont
  {{Latini}}}, \bibinfo {author} {\bibfnamefont {H.}~\bibnamefont
  {{H{\"u}bener}}}, \bibinfo {author} {\bibfnamefont {A.~J.}\ \bibnamefont
  {{Millis}}}, \bibinfo {author} {\bibfnamefont {A.}~\bibnamefont {{Rubio}}}, \
  and\ \bibinfo {author} {\bibfnamefont {A.}~\bibnamefont {{Geroges}}},\
  }\href@noop {} {}\Eprint {http://arxiv.org/abs/1911.11835} {arXiv:1911.11835}
  \BibitemShut {NoStop}%
\bibitem [{\citenamefont {Watson}\ \emph {et~al.}(2019)\citenamefont {Watson},
  \citenamefont {Markovi\'c}, \citenamefont {Morales}, \citenamefont
  {Le~F\`evre}, \citenamefont {Merz},\ and\ \citenamefont {King}}]{Watson2019}%
  \BibitemOpen
  \bibfield  {author} {\bibinfo {author} {\bibfnamefont {M.~D.}\ \bibnamefont
  {Watson}}, \bibinfo {author} {\bibfnamefont {I.}~\bibnamefont {Markovi\'c}},
  \bibinfo {author} {\bibfnamefont {E.~A.}\ \bibnamefont {Morales}}, \bibinfo
  {author} {\bibfnamefont {P.}~\bibnamefont {Le~F\`evre}}, \bibinfo {author}
  {\bibfnamefont {A.~A.}\ \bibnamefont {Merz}, \bibfnamefont {M.~Haghighirad}},
  \ and\ \bibinfo {author} {\bibfnamefont {P.~D.~C.}\ \bibnamefont {King}},\
  }\href@noop {} {\bibfield  {journal} {\bibinfo  {journal} {arXiv:1912.01591}\
  } (\bibinfo {year} {2019})}\BibitemShut {NoStop}%
\bibitem [{\citenamefont {Kaneko}\ \emph {et~al.}(2015)\citenamefont {Kaneko},
  \citenamefont {Zenker}, \citenamefont {Fehske},\ and\ \citenamefont
  {Ohta}}]{kaneko2015}%
  \BibitemOpen
  \bibfield  {author} {\bibinfo {author} {\bibfnamefont {T.}~\bibnamefont
  {Kaneko}}, \bibinfo {author} {\bibfnamefont {B.}~\bibnamefont {Zenker}},
  \bibinfo {author} {\bibfnamefont {H.}~\bibnamefont {Fehske}}, \ and\ \bibinfo
  {author} {\bibfnamefont {Y.}~\bibnamefont {Ohta}},\ }\href {\doibase
  10.1103/PhysRevB.92.115106} {\bibfield  {journal} {\bibinfo  {journal} {Phys.
  Rev. B}\ }\textbf {\bibinfo {volume} {92}},\ \bibinfo {pages} {115106}
  (\bibinfo {year} {2015})}\BibitemShut {NoStop}%
\bibitem [{\citenamefont {Ihle}\ \emph {et~al.}(2008)\citenamefont {Ihle},
  \citenamefont {Pfafferott}, \citenamefont {Burovski}, \citenamefont
  {Bronold},\ and\ \citenamefont {Fehske}}]{ihle2008}%
  \BibitemOpen
  \bibfield  {author} {\bibinfo {author} {\bibfnamefont {D.}~\bibnamefont
  {Ihle}}, \bibinfo {author} {\bibfnamefont {M.}~\bibnamefont {Pfafferott}},
  \bibinfo {author} {\bibfnamefont {E.}~\bibnamefont {Burovski}}, \bibinfo
  {author} {\bibfnamefont {F.~X.}\ \bibnamefont {Bronold}}, \ and\ \bibinfo
  {author} {\bibfnamefont {H.}~\bibnamefont {Fehske}},\ }\href {\doibase
  10.1103/PhysRevB.78.193103} {\bibfield  {journal} {\bibinfo  {journal} {Phys.
  Rev. B}\ }\textbf {\bibinfo {volume} {78}},\ \bibinfo {pages} {193103}
  (\bibinfo {year} {2008})}\BibitemShut {NoStop}%
\bibitem [{\citenamefont {Seki}\ \emph {et~al.}(2011)\citenamefont {Seki},
  \citenamefont {Eder},\ and\ \citenamefont {Ohta}}]{seki2011}%
  \BibitemOpen
  \bibfield  {author} {\bibinfo {author} {\bibfnamefont {K.}~\bibnamefont
  {Seki}}, \bibinfo {author} {\bibfnamefont {R.}~\bibnamefont {Eder}}, \ and\
  \bibinfo {author} {\bibfnamefont {Y.}~\bibnamefont {Ohta}},\ }\href {\doibase
  10.1103/PhysRevB.84.245106} {\bibfield  {journal} {\bibinfo  {journal} {Phys.
  Rev. B}\ }\textbf {\bibinfo {volume} {84}},\ \bibinfo {pages} {245106}
  (\bibinfo {year} {2011})}\BibitemShut {NoStop}%
\bibitem [{\citenamefont {Zenker}\ \emph {et~al.}(2012)\citenamefont {Zenker},
  \citenamefont {Ihle}, \citenamefont {Bronold},\ and\ \citenamefont
  {Fehske}}]{zenker2012}%
  \BibitemOpen
  \bibfield  {author} {\bibinfo {author} {\bibfnamefont {B.}~\bibnamefont
  {Zenker}}, \bibinfo {author} {\bibfnamefont {D.}~\bibnamefont {Ihle}},
  \bibinfo {author} {\bibfnamefont {F.~X.}\ \bibnamefont {Bronold}}, \ and\
  \bibinfo {author} {\bibfnamefont {H.}~\bibnamefont {Fehske}},\ }\href
  {\doibase 10.1103/PhysRevB.85.121102} {\bibfield  {journal} {\bibinfo
  {journal} {Phys. Rev. B}\ }\textbf {\bibinfo {volume} {85}},\ \bibinfo
  {pages} {121102} (\bibinfo {year} {2012})}\BibitemShut {NoStop}%
\bibitem [{\citenamefont {Murakami}\ \emph {et~al.}(2020)\citenamefont
  {Murakami}, \citenamefont {Sch\"{u}ler}, \citenamefont {Takayoshi},\ and\
  \citenamefont {Werner}}]{Murakami2020PRB}%
  \BibitemOpen
  \bibfield  {author} {\bibinfo {author} {\bibfnamefont {Y.}~\bibnamefont
  {Murakami}}, \bibinfo {author} {\bibfnamefont {M.}~\bibnamefont
  {Sch\"{u}ler}}, \bibinfo {author} {\bibfnamefont {S.}~\bibnamefont
  {Takayoshi}}, \ and\ \bibinfo {author} {\bibfnamefont {P.}~\bibnamefont
  {Werner}},\ }\href {\doibase 10.1103/PhysRevB.101.035203} {\bibfield
  {journal} {\bibinfo  {journal} {Phys. Rev. B}\ }\textbf {\bibinfo {volume}
  {101}},\ \bibinfo {pages} {035203} (\bibinfo {year} {2020})}\BibitemShut
  {NoStop}%
\bibitem [{\citenamefont {Basov}\ \emph {et~al.}(2011)\citenamefont {Basov},
  \citenamefont {Averitt}, \citenamefont {van~der Marel}, \citenamefont
  {Dressel},\ and\ \citenamefont {Haule}}]{basov2011}%
  \BibitemOpen
  \bibfield  {author} {\bibinfo {author} {\bibfnamefont {D.~N.}\ \bibnamefont
  {Basov}}, \bibinfo {author} {\bibfnamefont {R.~D.}\ \bibnamefont {Averitt}},
  \bibinfo {author} {\bibfnamefont {D.}~\bibnamefont {van~der Marel}}, \bibinfo
  {author} {\bibfnamefont {M.}~\bibnamefont {Dressel}}, \ and\ \bibinfo
  {author} {\bibfnamefont {K.}~\bibnamefont {Haule}},\ }\href {\doibase
  10.1103/RevModPhys.83.471} {\bibfield  {journal} {\bibinfo  {journal} {Rev.
  Mod. Phys.}\ }\textbf {\bibinfo {volume} {83}},\ \bibinfo {pages} {471}
  (\bibinfo {year} {2011})}\BibitemShut {NoStop}%
\bibitem [{\citenamefont {Giannetti}\ \emph {et~al.}(2016)\citenamefont
  {Giannetti}, \citenamefont {Capone}, \citenamefont {Fausti}, \citenamefont
  {Fabrizio}, \citenamefont {Parmigiani},\ and\ \citenamefont
  {Mihailovic}}]{giannetti2016}%
  \BibitemOpen
  \bibfield  {author} {\bibinfo {author} {\bibfnamefont {C.}~\bibnamefont
  {Giannetti}}, \bibinfo {author} {\bibfnamefont {M.}~\bibnamefont {Capone}},
  \bibinfo {author} {\bibfnamefont {D.}~\bibnamefont {Fausti}}, \bibinfo
  {author} {\bibfnamefont {M.}~\bibnamefont {Fabrizio}}, \bibinfo {author}
  {\bibfnamefont {F.}~\bibnamefont {Parmigiani}}, \ and\ \bibinfo {author}
  {\bibfnamefont {D.}~\bibnamefont {Mihailovic}},\ }\href {\doibase
  10.1080/00018732.2016.1194044} {\bibfield  {journal} {\bibinfo  {journal}
  {Adv. Phys.}\ }\textbf {\bibinfo {volume} {65}},\ \bibinfo {pages} {58}
  (\bibinfo {year} {2016})}\BibitemShut {NoStop}%
\bibitem [{\citenamefont {Tsuji}\ and\ \citenamefont {Aoki}(2015)}]{Tsuji2015}%
  \BibitemOpen
  \bibfield  {author} {\bibinfo {author} {\bibfnamefont {N.}~\bibnamefont
  {Tsuji}}\ and\ \bibinfo {author} {\bibfnamefont {H.}~\bibnamefont {Aoki}},\
  }\href {\doibase 10.1103/PhysRevB.92.064508} {\bibfield  {journal} {\bibinfo
  {journal} {Phys. Rev. B}\ }\textbf {\bibinfo {volume} {92}},\ \bibinfo
  {pages} {064508} (\bibinfo {year} {2015})}\BibitemShut {NoStop}%
\bibitem [{\citenamefont {Cea}\ \emph {et~al.}(2016)\citenamefont {Cea},
  \citenamefont {Castellani},\ and\ \citenamefont {Benfatto}}]{Cea2016}%
  \BibitemOpen
  \bibfield  {author} {\bibinfo {author} {\bibfnamefont {T.}~\bibnamefont
  {Cea}}, \bibinfo {author} {\bibfnamefont {C.}~\bibnamefont {Castellani}}, \
  and\ \bibinfo {author} {\bibfnamefont {L.}~\bibnamefont {Benfatto}},\ }\href
  {\doibase 10.1103/PhysRevB.93.180507} {\bibfield  {journal} {\bibinfo
  {journal} {Phys. Rev. B}\ }\textbf {\bibinfo {volume} {93}},\ \bibinfo
  {pages} {180507} (\bibinfo {year} {2016})}\BibitemShut {NoStop}%
\bibitem [{\citenamefont {Tsuji}\ \emph {et~al.}(2016)\citenamefont {Tsuji},
  \citenamefont {Murakami},\ and\ \citenamefont {Aoki}}]{Tsuji2016}%
  \BibitemOpen
  \bibfield  {author} {\bibinfo {author} {\bibfnamefont {N.}~\bibnamefont
  {Tsuji}}, \bibinfo {author} {\bibfnamefont {Y.}~\bibnamefont {Murakami}}, \
  and\ \bibinfo {author} {\bibfnamefont {H.}~\bibnamefont {Aoki}},\ }\href
  {\doibase 10.1103/PhysRevB.94.224519} {\bibfield  {journal} {\bibinfo
  {journal} {Phys. Rev. B}\ }\textbf {\bibinfo {volume} {94}},\ \bibinfo
  {pages} {224519} (\bibinfo {year} {2016})}\BibitemShut {NoStop}%
\bibitem [{\citenamefont {Murotani}\ and\ \citenamefont
  {Shimano}(2019)}]{Murotani2019}%
  \BibitemOpen
  \bibfield  {author} {\bibinfo {author} {\bibfnamefont {Y.}~\bibnamefont
  {Murotani}}\ and\ \bibinfo {author} {\bibfnamefont {R.}~\bibnamefont
  {Shimano}},\ }\href {\doibase 10.1103/PhysRevB.99.224510} {\bibfield
  {journal} {\bibinfo  {journal} {Phys. Rev. B}\ }\textbf {\bibinfo {volume}
  {99}},\ \bibinfo {pages} {224510} (\bibinfo {year} {2019})}\BibitemShut
  {NoStop}%
\bibitem [{\citenamefont {Gole\ifmmode~\check{z}\else \v{z}\fi{}}\ \emph
  {et~al.}(2019)\citenamefont {Gole\ifmmode~\check{z}\else \v{z}\fi{}},
  \citenamefont {Eckstein},\ and\ \citenamefont {Werner}}]{golez2019multiband}%
  \BibitemOpen
  \bibfield  {author} {\bibinfo {author} {\bibfnamefont {D.}~\bibnamefont
  {Gole\ifmmode~\check{z}\else \v{z}\fi{}}}, \bibinfo {author} {\bibfnamefont
  {M.}~\bibnamefont {Eckstein}}, \ and\ \bibinfo {author} {\bibfnamefont
  {P.}~\bibnamefont {Werner}},\ }\href {\doibase 10.1103/PhysRevB.100.235117}
  {\bibfield  {journal} {\bibinfo  {journal} {Phys. Rev. B}\ }\textbf {\bibinfo
  {volume} {100}},\ \bibinfo {pages} {235117} (\bibinfo {year}
  {2019})}\BibitemShut {NoStop}%
\bibitem [{\citenamefont {Georges}\ \emph {et~al.}(1996)\citenamefont
  {Georges}, \citenamefont {Kotliar}, \citenamefont {Krauth},\ and\
  \citenamefont {Rozenberg}}]{georges1996}%
  \BibitemOpen
  \bibfield  {author} {\bibinfo {author} {\bibfnamefont {A.}~\bibnamefont
  {Georges}}, \bibinfo {author} {\bibfnamefont {G.}~\bibnamefont {Kotliar}},
  \bibinfo {author} {\bibfnamefont {W.}~\bibnamefont {Krauth}}, \ and\ \bibinfo
  {author} {\bibfnamefont {M.~J.}\ \bibnamefont {Rozenberg}},\ }\href {\doibase
  10.1103/RevModPhys.68.13} {\bibfield  {journal} {\bibinfo  {journal} {Rev.
  Mod. Phys.}\ }\textbf {\bibinfo {volume} {68}},\ \bibinfo {pages} {13}
  (\bibinfo {year} {1996})}\BibitemShut {NoStop}%
\bibitem [{Note1()}]{Note1}%
  \BibitemOpen
  \bibinfo {note} {When the system is in the low dimensions and has strong
  interactions, it can show a gap opening due to the fluctuation above
  $T_c$\cite {sugimoto2018} even when the systems is in the semi-metallic
  parameter regime. Thus, strictly speaking, one cannot simply categorize such
  cases to the BCS type or the BEC type. Therefore, this terminology is
  meaningful when the interaction is not too large. Indeed, we focus on such
  cases here.}\BibitemShut {Stop}%
\bibitem [{\citenamefont {Murakami}\ \emph
  {et~al.}(2016{\natexlab{b}})\citenamefont {Murakami}, \citenamefont {Werner},
  \citenamefont {Tsuji},\ and\ \citenamefont {Aoki}}]{Murakami2016b}%
  \BibitemOpen
  \bibfield  {author} {\bibinfo {author} {\bibfnamefont {Y.}~\bibnamefont
  {Murakami}}, \bibinfo {author} {\bibfnamefont {P.}~\bibnamefont {Werner}},
  \bibinfo {author} {\bibfnamefont {N.}~\bibnamefont {Tsuji}}, \ and\ \bibinfo
  {author} {\bibfnamefont {H.}~\bibnamefont {Aoki}},\ }\href {\doibase
  10.1103/PhysRevB.94.115126} {\bibfield  {journal} {\bibinfo  {journal} {Phys.
  Rev. B}\ }\textbf {\bibinfo {volume} {94}},\ \bibinfo {pages} {115126}
  (\bibinfo {year} {2016}{\natexlab{b}})}\BibitemShut {NoStop}%
\bibitem [{\citenamefont {Volkov}\ and\ \citenamefont
  {Kogan}(1974)}]{Volkov1974}%
  \BibitemOpen
  \bibfield  {author} {\bibinfo {author} {\bibfnamefont {A.}~\bibnamefont
  {Volkov}}\ and\ \bibinfo {author} {\bibfnamefont {S.}~\bibnamefont {Kogan}},\
  }\href {http://adsabs.harvard.edu/abs/1974JETP...38.1018V} {\bibfield
  {journal} {\bibinfo  {journal} {Sov. Phys. JETP}\ }\textbf {\bibinfo {volume}
  {38}},\ \bibinfo {pages} {1018} (\bibinfo {year} {1974})}\BibitemShut
  {NoStop}%
\bibitem [{\citenamefont {Gurarie}(2009)}]{Gurarie2009}%
  \BibitemOpen
  \bibfield  {author} {\bibinfo {author} {\bibfnamefont {V.}~\bibnamefont
  {Gurarie}},\ }\href {\doibase 10.1103/PhysRevLett.103.075301} {\bibfield
  {journal} {\bibinfo  {journal} {Phys. Rev. Lett.}\ }\textbf {\bibinfo
  {volume} {103}},\ \bibinfo {pages} {075301} (\bibinfo {year}
  {2009})}\BibitemShut {NoStop}%
\bibitem [{\citenamefont {Behrle}\ \emph {et~al.}(2018)\citenamefont {Behrle},
  \citenamefont {Harrison}, \citenamefont {Kombe}, \citenamefont {Gao},
  \citenamefont {Link}, \citenamefont {Bernier}, \citenamefont {Kollath},\ and\
  \citenamefont {K{\"o}hl}}]{Behrle2018}%
  \BibitemOpen
  \bibfield  {author} {\bibinfo {author} {\bibfnamefont {A.}~\bibnamefont
  {Behrle}}, \bibinfo {author} {\bibfnamefont {T.}~\bibnamefont {Harrison}},
  \bibinfo {author} {\bibfnamefont {J.}~\bibnamefont {Kombe}}, \bibinfo
  {author} {\bibfnamefont {K.}~\bibnamefont {Gao}}, \bibinfo {author}
  {\bibfnamefont {M.}~\bibnamefont {Link}}, \bibinfo {author} {\bibfnamefont
  {J.-S.}\ \bibnamefont {Bernier}}, \bibinfo {author} {\bibfnamefont
  {C.}~\bibnamefont {Kollath}}, \ and\ \bibinfo {author} {\bibfnamefont
  {M.}~\bibnamefont {K{\"o}hl}},\ }\href {\doibase 10.1038/s41567-018-0128-6}
  {\bibfield  {journal} {\bibinfo  {journal} {Nature Physics}\ }\textbf
  {\bibinfo {volume} {14}},\ \bibinfo {pages} {781} (\bibinfo {year}
  {2018})}\BibitemShut {NoStop}%
\bibitem [{\citenamefont {Sugimoto}\ \emph {et~al.}(2018)\citenamefont
  {Sugimoto}, \citenamefont {Nishimoto}, \citenamefont {Kaneko},\ and\
  \citenamefont {Ohta}}]{sugimoto2018}%
  \BibitemOpen
  \bibfield  {author} {\bibinfo {author} {\bibfnamefont {K.}~\bibnamefont
  {Sugimoto}}, \bibinfo {author} {\bibfnamefont {S.}~\bibnamefont {Nishimoto}},
  \bibinfo {author} {\bibfnamefont {T.}~\bibnamefont {Kaneko}}, \ and\ \bibinfo
  {author} {\bibfnamefont {Y.}~\bibnamefont {Ohta}},\ }\href {\doibase
  10.1103/PhysRevLett.120.247602} {\bibfield  {journal} {\bibinfo  {journal}
  {Phys. Rev. Lett.}\ }\textbf {\bibinfo {volume} {120}},\ \bibinfo {pages}
  {247602} (\bibinfo {year} {2018})}\BibitemShut {NoStop}%
\bibitem [{\citenamefont {Larkin}\ \emph {et~al.}(2017)\citenamefont {Larkin},
  \citenamefont {Yaresko}, \citenamefont {Pr\"opper}, \citenamefont {Kikoin},
  \citenamefont {Lu}, \citenamefont {Takayama}, \citenamefont {Mathis},
  \citenamefont {Rost}, \citenamefont {Takagi}, \citenamefont {Keimer},\ and\
  \citenamefont {Boris}}]{larkin2017}%
  \BibitemOpen
  \bibfield  {author} {\bibinfo {author} {\bibfnamefont {T.~I.}\ \bibnamefont
  {Larkin}}, \bibinfo {author} {\bibfnamefont {A.~N.}\ \bibnamefont {Yaresko}},
  \bibinfo {author} {\bibfnamefont {D.}~\bibnamefont {Pr\"opper}}, \bibinfo
  {author} {\bibfnamefont {K.~A.}\ \bibnamefont {Kikoin}}, \bibinfo {author}
  {\bibfnamefont {Y.~F.}\ \bibnamefont {Lu}}, \bibinfo {author} {\bibfnamefont
  {T.}~\bibnamefont {Takayama}}, \bibinfo {author} {\bibfnamefont {Y.-L.}\
  \bibnamefont {Mathis}}, \bibinfo {author} {\bibfnamefont {A.~W.}\
  \bibnamefont {Rost}}, \bibinfo {author} {\bibfnamefont {H.}~\bibnamefont
  {Takagi}}, \bibinfo {author} {\bibfnamefont {B.}~\bibnamefont {Keimer}}, \
  and\ \bibinfo {author} {\bibfnamefont {A.~V.}\ \bibnamefont {Boris}},\ }\href
  {\doibase 10.1103/PhysRevB.95.195144} {\bibfield  {journal} {\bibinfo
  {journal} {Phys. Rev. B}\ }\textbf {\bibinfo {volume} {95}},\ \bibinfo
  {pages} {195144} (\bibinfo {year} {2017})}\BibitemShut {NoStop}%
\bibitem [{\citenamefont {Andrich}\ \emph {et~al.}(2020)\citenamefont
  {Andrich}, \citenamefont {Bretscher}, \citenamefont {Murakami}, \citenamefont
  {Gole\v{z}}, \citenamefont {Remez}, \citenamefont {Telang}, \citenamefont
  {A}, \citenamefont {Harnagea}, \citenamefont {Cooper}, \citenamefont
  {Millis}, \citenamefont {Werner}, \citenamefont {Sood},\ and\ \citenamefont
  {Rao}}]{Paolo2020}%
  \BibitemOpen
  \bibfield  {author} {\bibinfo {author} {\bibfnamefont {P.}~\bibnamefont
  {Andrich}}, \bibinfo {author} {\bibfnamefont {H.~M.}\ \bibnamefont
  {Bretscher}}, \bibinfo {author} {\bibfnamefont {Y.}~\bibnamefont {Murakami}},
  \bibinfo {author} {\bibfnamefont {D.}~\bibnamefont {Gole\v{z}}}, \bibinfo
  {author} {\bibfnamefont {B.}~\bibnamefont {Remez}}, \bibinfo {author}
  {\bibfnamefont {P.}~\bibnamefont {Telang}}, \bibinfo {author} {\bibfnamefont
  {S.}~\bibnamefont {A}}, \bibinfo {author} {\bibfnamefont {L.}~\bibnamefont
  {Harnagea}}, \bibinfo {author} {\bibfnamefont {N.~R.}\ \bibnamefont
  {Cooper}}, \bibinfo {author} {\bibfnamefont {A.}~\bibnamefont {Millis}},
  \bibinfo {author} {\bibfnamefont {P.}~\bibnamefont {Werner}}, \bibinfo
  {author} {\bibfnamefont {A.~K.}\ \bibnamefont {Sood}}, \ and\ \bibinfo
  {author} {\bibfnamefont {A.}~\bibnamefont {Rao}},\ }\href@noop {} {\bibfield
  {journal} {\bibinfo  {journal} {arXiv:2003.10799}\ } (\bibinfo {year}
  {2020})}\BibitemShut {NoStop}%
\end{thebibliography}%

\end{document}